\documentclass[msnc]{informs3} 

\DoubleSpacedXI 

\usepackage[colorinlistoftodos,bordercolor=orange,backgroundcolor=orange!20,line
color=orange,textsize=scriptsize]{todonotes}
\catcode`\@=11
\catcode`\@=12
\usepackage{footnote}

\newcommand{\tbl}[1]{{\color{black} #1}}
\newcommand{\tb}[1]{{\color{black} #1}}

\usepackage{natbib}
 \bibpunct[, ]{(}{)}{,}{a}{}{,}%
 \def\bibfont{\small}%

 \usepackage{pdfpages}

 \usepackage{amsmath}
\usepackage{algorithm, algorithmicx, algpseudocode}
\usepackage{float}
\usepackage{subcaption}
\usepackage{bm}
\usepackage{placeins}
\usepackage{booktabs}
\newlength\mylenA
\newlength\mylenB

\newcommand{\R}{\mathbb{R}}

\newcommand{\A}{\mathcal{A}}

\newcommand{\PP}{\mathbb{P}}
\newcommand{\N}{\mathbb{N}}
\newcommand{\X}{\mathcal{S}}
\usepackage{bbm}

\newcommand{\I}{\mathcal{I}}
\newcommand{\cN}{\mathcal{N}}
\newcommand{\M}{\mathcal{M}}

\newcommand{\trp}{^{\top}}
\newcommand{\opt}{^{\star}}

\newcommand{\pieff}{\pi_{\sf eff}}
\newcommand{\pib}{\pi_{\sf base}}
\newcommand{\pia}{\pi_{\sf alg}}
\newcommand{\carda}{{\sf Card-Assort}}

\newcommand{\E}{\mathbb{E}}
\newcommand{\U}{\mathcal{U}}

\newcommand{\bs}{\bar{s}}

\TheoremsNumberedBySection  
\ECRepeatTheorems

\EquationsNumberedBySection 

\usepackage{natbib}

\RUNTITLE{Making Adherence-Aware Recommendations}

\TITLE{The Best Decisions Are Not the Best Advice: \\ Making Adherence-Aware Recommendations}

\ARTICLEAUTHORS{%
\AUTHOR{Julien Grand-Cl\'ement}
\AFF{Information Systems and Operations Management Department, Ecole des hautes etudes commerciales de Paris, \EMAIL{grand-clement@hec.fr}}
\AUTHOR{Jean Pauphilet}
\AFF{Management Science and Operations, London Business School, \EMAIL{jpauphilet@london.edu}} 
}

\ABSTRACT{%
 Many high-stake decisions follow an expert-in-loop structure in that a human operator receives recommendations from an algorithm but is the ultimate decision maker. 
Hence, the algorithm's recommendation may differ from the actual decision implemented in practice. 
However, most algorithmic recommendations are obtained by solving an optimization problem that assumes recommendations will be perfectly implemented. 
We propose an adherence-aware optimization framework to capture the dichotomy between the recommended and the implemented policy and analyze the impact of partial adherence on the optimal recommendation. 
Our framework provides useful tools to analyze the structure and to compute optimal recommendation policies that are naturally immune against such human deviations, and are guaranteed to improve upon the baseline policy.
}

 \HISTORY{.}

\begin{document}

\maketitle
\section{Introduction}
While some decisions can be automated and made directly by algorithms based on artificial intelligence (AI), many high-stake decisions follow an expert-in-loop structure in that an expert decision maker (e.g., a doctor) receives information, predictions, or even recommendations, and decides which course of action to follow. Consequently, the human decision maker (DM) does not systematically implement what the algorithm recommended. In other words, they may have a discretionary power to override/reject the recommendations from the algorithm, hence impacting the potential benefits from the AI tool. 
For instance, in a field experiment,
\citet{kesavan2020field} observed that merchants overrode the recommendations from a data-driven decision tool 71.24\% of the time, resulting in a 5.77\% reduction in profitability. 

To understand this phenomenon and its ultimate impact on the quality of the decision being made, a growing body of literature has investigated the mechanisms driving non- or partial adherence of humans to algorithmic recommendations. In this work, we ask a complementary question: Given the fact that the decision maker will partially implement recommendations made by an algorithm, should we adjust these recommendations in the first place and how? In other words, we investigate the impact of partial adherence on algorithm design and decision recommendation. 
\tbl{
Our main contributions are as follows.

\paragraph{A new model of partial adherence.}
We consider a model of sequential decision-making based on Markov decision processes (MDPs) and assume that the decision maker currently follows a {\em baseline} policy $\pib$ (or state of practice) \tbl{and is provided with a {\em recommendation} policy $\pia$ by an algorithm.}
We propose a framework, namely adherence-aware MDP, to compute recommendations that are immune against human deviations. 
\tbl{Our framework is behavioral in that it models the human switching behavior between their baseline policy and the algorithmic recommendations, but without specifying \emph{why} these deviations are undertaken by the DM.}
Despite its simplicity, we show that our model is consistent with \tb{five} different models for the DM's adherence decision, including random or adversarial adherence decisions. 
Furthermore, we provide examples where the co-existence of the human DM and the algorithmic recommendations performs either strictly worse or strictly better than any of the two policies alone, hence illustrating the ability of our model to capture the rich range of situations observed in practice.}
In particular, we show that (even rare) human deviations from algorithmic recommendations can lead to arbitrarily poor performance compared with both the expected performance of the algorithm and that of the current state of practice. In other words, we show that deploying a recommendation engine that was designed assuming its recommendations will be final decisions can have a dramatic impact on the effective performance. This set of negative results underscores the importance of accounting for the current baseline and the partial adherence phenomenon when building recommendation systems.

\tbl{
\paragraph{A tractable, structured, and flexible model.}
We study the appealing structural and computational properties of our adherence-aware MDP framework. In particular, we show that an optimal recommendation policy may be chosen stationary and deterministic, which is important from an implementation standpoint, and that it may be computed efficiently by a reduction to a classical MDP problem. We also show several structural properties, such as piecewise constant optimal recommendation policy and monotonicity of the optimal return (both as regards the adherence level). We identify classes of MDPs for which the decision maker may overlook the issue of partial adherence at some states (i.e., where the partial adherence phenomenon has no impact on the algorithmic recommendation to be made). We finally present extensions of our framework, including models where the adherence levels are state-dependent, action-dependent, uncertain, or where the baseline policy is not entirely known.
 }

\paragraph{Numerical study.}
We evaluate the practical impact of our model on a series of numerical experiments. Our simulations highlight the importance of accounting for the potential non-adherence of the decision maker, showing empirically that severe performance deteriorations can happen 
when partial adherence is overlooked in the search for an optimal policy. The magnitude of this performance deterioration depends both on the current baseline policy and on the level of adherence of the decision maker. Consequently, in addition to classical sensitivity and robustness analyses used in the literature, we encourage practitioners to conduct a systematic {\em adherence-robustness} analysis of their algorithms to  assess their effective performance prior to deployment.

\tbl{
The rest of the paper is organized as follows:
We present related work from the operations literature in Section \ref{sec:litreview}. 
Section \ref{sec:decision-adherence} introduces our framework for sequential decision-making under partial adherence, discusses its connection with various models for the DM's adherence decision, and provides examples of situations where the co-existence of human and algorithmic decisions leads to improved or, on the contrary, impaired system performance. 
In Section \ref{sec:adamdp.analysis}, we present algorithms to compute optimal recommendation policies, and we analyze their structural properties and their sensitivity to the adherence level.
We illustrate the practical impact of imperfect adherence and the value of our framework on numerical experiments in Section \ref{sec:simu}. Finally, we discuss extensions of our framework in Section \ref{sec:extension}.
}
\section{Literature review}\label{sec:litreview}
Our paper contributes to the rich literature of behavioral operations that studies the partial adherence of decision makers to machine recommendations. This phenomenon is also referred to in the literature as \emph{discretion}, \emph{overriding}, or \emph{deviation}.

Many field studies have documented this phenomenon in a wide range of tasks and industries such as demand forecasting  \citep{fildes2009effective,kremer2011demand,kesavan2020field}, warehouse operations \citep{sun2022predicting}, medical treatment adherence \citep{lin2021does}, or task sequencing  \citep{ibanez2018discretionary}. 
Actually, partial adherence also occurs when the recommendation does not come from a machine.  
In the context of chronic diseases, for instance, the World Health Organization (WHO) defines {\em adherence} as ``the extent to which a person’s behavior-taking medication, following a diet, and/or executing lifestyle changes corresponds with agreed recommendations from a health-care provider''~\citep{sabate2003adherence}. The WHO notes that adherence of the patients to therapy for chronic illnesses is as low as 50 \% in the long-term, and that this partial adherence leads to suboptimal clinical outcomes.
To anticipate its potential impact on operational performance, it is important to understand the drivers of partial adherence, such as information asymmetry or algorithmic aversion. 

In the context of operations, assuming that humans have more and better information than the machine, deviations due to information asymmetry can be beneficial to effective performance. In an inventory management setting, \citet{van2010ordering} conclude that providing store manager discretion may result in higher profits due to their superior information. In a field experiment with an automotive replacement parts retailer, \citet{kesavan2020field} evaluate that merchants overriding demand forecasts increases (resp. decreases) profitability for growth- (resp. decline-) stage products, suggesting that the information advantage of merchants increases when the machine has limited access to historical data on the product. However, on average, they observe a negative effect of human overriding power. Similarly, \citet{fildes2009effective} document the heterogeneous impact of human adjustment on prediction accuracy, depending on the company but also the magnitude and direction of the adjustment. In another context, \citet{sun2022predicting} study the box size recommendation algorithm of Alibaba. Since the algorithm ignores the foldability and compressibility of the items, they observe that warehouse workers are able to pack some orders in smaller boxes than the ones recommended. 

Partial adherence can also result from multiple conflicting objectives that are weighted differently by the human and the algorithm. In Alibaba's warehouses for instance, \citet{sun2022predicting} hypothesize that workers switching to larger boxes might do so to save packing effort at the expense of time and cost. In a healthcare setting, \citet{ibanez2018discretionary} observe that doctors tend to re-prioritize tasks so as to group similar tasks together and reduce mental switching costs, but that such prioritization may reduce long-term productivity. 

Another reason that could explain why humans fail to follow machine recommendation is algorithm aversion, as first documented by \citet{dietvorst2015algorithm}. Algorithm aversion refers to a general preference to rely on humans instead of algorithms. This general preference could be due to an inflated confidence in human performance. In a lab experiment, for instance, \citet{logg2019algorithm} observed that subjects (and in particular experts) were more prone to follow their own judgment over an algorithm's advice, or advice provided by another human. Alternatively, \citet{dietvorst2018overcoming} hypothesize that decision makers seek control over the output. In an empirical study, they successfully reduced algorithm aversion by offering decision makers some control over the machine's output. \citet{lin2021does} propose and empirically evaluate algorithm use determinants in algorithm aversion. 

In an effort to propose alternative explanations to algorithm aversion, \citet{devericourt2022your} develop a theoretical framework to study the evolution of the decision maker's belief about the performance of a machine and her overruling decisions over time. In their setting, decisions and recommendations are binary (to act or not to act, e.g., collect a biopsy or not) and the decision maker only collects performance data when choosing to act. Because of this verification bias, \citet{devericourt2022your} identify situations under which a \tbl{(rational)} decision maker fails to learn the true performance of the machine, and indefinitely overrules its recommendation with some non-zero probability.  

Understanding the drivers of partial adherence is useful to propose solutions and incorporate behavioral aspects into the algorithmic recommendations. 
\tbl{In a pricing setting, for example, \citet{caro2023believing} observe adherence patterns that are consistent with the fact that inventory and sales are more salient to managers and conduct two interventions aimed at increasing the salience of revenues.}
A growing literature has studied \emph{features} of the recommendation system or the recommended policy that could increase adoption, such as partial control over the output \citep[see discussion above and][]{dietvorst2018overcoming}, simplicity \citep{bastani2021improving}, or interpretability \citep[see, e.g.,][]{kallus2017recursive,bravo2020mining,ciocan2022interpretable,jacq2022lazy}. 
\tbl{
The underlying intuition is that policies that have simple structural forms are more likely to be adopted because of legal requirements for a `right to explanation' \citep{goodman2017european} and because decision makers and stake-holders value policy they can understand and audit \citep{bertsimas2013fairness,bertsimas2022predicting}. 
}    
Assuming that humans are more likely to adhere to recommendations that constitute small changes to their current practice, \citet{bastani2021improving} propose a reinforcement learning approach to compute optimal `tips', i.e., small changes in the current practice, and validate their approach in a controlled experiment. 
In an attempt to increase interpretability of reinforcement learning policies, \citet{jacq2022lazy} propose the 
{lazy-MDP} framework to learn and recommend {\em when to act} (i.e., in what states of the system), on the top  of the decisions. \citet{meresht2020learning} propose to learn when to switch control between machines and human decision makers. Nonetheless, these works assume that the simplicity or interpretability of the recommendation will not only increase adherence, but will lead to perfect adherence. In this paper, we complement this literature by challenging this assumption and investigating the impact of partial adherence directly on the actions to be recommended.
We develop a framework to incorporate the potential departure of the human decision maker {\em within the search for a good recommendation policy}.
\tbl{Our goal resembles that of robust optimization under implementation errors where there is a similar discrepancy between the computed solution and the implemented one  as in~\cite{bertsimas2010nonconvex,men2014fabrication}, except that their error model is purely adversarial and their decision problem static, and that our model accounts for the current baseline practices.}

In a similar vein, \citet{sun2022predicting} reduce non-adherence in Alibaba's warehouses by 19.3\% and packing time by 4.5\%, by modifying the box size recommendations for the ``at-risk'' orders (defined as having $>50\%$ chance of being overruled). 
In this paper, we have a similar objective of adjusting the recommendation of the algorithm to the expected adherence level. However, instead of an ad-hoc adjustment, we propose to account for the adherence level directly in the optimization problem which the recommendation is a solution of. Furthermore, our objective is not to increase adherence per se but to adjust the algorithm's recommendation to the adherence level, so as to increase the performance of the human-in-the-loop system.

 \section{Modeling partial adherence in a decision framework}\label{sec:decision-adherence}
In this section we formally introduce our model of decision under partial adherence. 

We consider a human decision maker (DM) which repeatedly interacts with an environment. The goal of the DM is to maximize a cumulative expected return, which captures both the instantaneous reward and the long-run objective. A policy of the DM is a map from the set of possible states of the environment to the set of actions. We assume that we have access to a {\em baseline} policy, called $\pi_{\sf base}$, which models the historical decisions of the DM. 
In a healthcare setting, for example, the DM is a medical practitioner, observes the health condition of a patient at each time period, and chooses a treatment to maximize the chances of survival, e.g., intravenous fluids and vasopressors for hospital patients with sepsis~\citep{komorowski2018artificial}, proactive transfers to the intensive care units for patients in the emergency room~\citep{grand2022robust}, or drug treatment decisions for heart disease in patients with type 2 diabetes~\citep{steimle2017markov}. The baseline policy $\pi_{\sf base}$ captures the current standard of care.

Classical methods from the operations management literature design models and algorithms to compute an alternative {\em recommendation policy} $\pi_{\sf alg}$ that leads to improved performance compared with the baseline. 
The underlying assumption is that the DM, convinced by the value of the algorithmic approach, will systematically follow $\pi_{\sf alg}$ and not revert to $\pi_{\sf base}$. However, in many practical problems, $\pi_{\sf alg}$ is only a \emph{recommendation}. The practitioner does not commit to implementing it. She has some discretionary power and the resulting policy is likely to be neither $\pi_{\sf base}$ nor $\pi_{\sf alg}$, but a mixture of the two. The main objective and contribution of our paper is to incorporate this partial adherence phenomenon within the optimization problem that defines $\pi_{\sf alg}$, i.e., adjust the recommended policy to the adherence level.

\subsection{Preliminaries on Markov decision process}\label{sec:mdp-preliminaries}
{\color{black} Formally, we adopt the framework of} Markov Decision Processes~\citep[MDPs;][]{puterman2014markov}. 
The system or environment is described via a set of possible states $\X$. At every decision period, the DM is at a given state $s \in \X$, chooses an action $a \in \A$, transitions to the next state $s' \in \X$ with a probability $P_{sas'} \in [0,1]$ and obtains a reward $r_{sas'} \in \R$. The future rewards are discounted by a factor $\lambda \in (0,1)$ and we assume that $\X$ and $\A$ are finite sets. An MDP instance $\M$ consists of a tuple $\M = \left(\X,\A,\bm{P},\bm{r},\bm{p}_{0},\lambda\right)$, with $\bm{r}=\left(r_{sas'}\right)_{s,a,s'} \in \R^{\X \times \A \times \X}$ and $\bm{P} = \left(P_{sas'}\right)_{sas'} \in \left(\Delta(\X)\right)^{\X \times \A}$, and $\bm{p}_{0} \in \Delta(\X)$ is an initial probability distribution over the set of states $\X$. Here, we denote $\Delta(\X)$ the simplex over $\X$, defined as
\[ \Delta(\X) = \left\{ \bm{p} \in \R^{\X} \; | \; p_{s} \geq 0, \forall \; s \in \X, \sum_{s \in \X} p_{s}=1\right\}.\] 
A policy $\pi$ maps, for each period $t \in \N$, the state-action history $(s_{0},a_{0},s_{1},a_{1},...,s_{t})$ to a probability distribution over the set of actions $\A$. A policy $\pi$ is Markovian if it only depends of the current state $s_{t}$, and stationary if it is Markovian and it does not depend on time. Therefore, a stationary policy is simply a map $\pi:\X \rightarrow \Delta(\A)$. We call $\Pi = \left(\Delta(\A)\right)^{\X}$ the set of stationary policies, $\Pi_{\sf M}$ the set of Markovian policies, and $\Pi_{\sf H}$ the set of all policies (possibly history-dependent).
In an MDP, the goal of the DM is to compute a policy $\pi$ to maximize the return $R(\pi)$, defined as
    \begin{equation}\label{eq:expected-reward}
     R(\pi)=\E^{\pi} \left[ \sum_{t=0}^{+\infty} \lambda^{t}r_{s_{t}a_{t}s_{t+1}} \right],
    \end{equation}
with $s_{t}$ the state visited at time period $t$, $a_{t}$ the action chosen with probability $\pi_{sa}$, and the expectation is as regards with the distribution  defined by the policy $\pi$ on the set of infinite-horizon trajectories. \tb{The return $R(\cdot)$ is sometimes called {\em expected reward}, and we use the term {\em return} to distinguish it from the instantaneous reward $r_{sa}$.} \tbl{The {\em value function} $\bm{v}^{\pi} \in \R^{\X}$ of a policy $\pi \in \Pi_{\sf H}$ represents the return obtained starting from any state: $v^{\pi}_{s} = \E^{\pi} \left[ \sum_{t=0}^{+\infty} \lambda^{t}r_{s_{t}a_{t}s_{t+1}} \; | \; s_{0} = s \right], \forall \; s \in \X$.} Note that in all generality, the return function $\pi \mapsto R(\pi)$ is neither convex nor concave on $\Pi$. An optimal policy can be chosen stationary and deterministic and can be computed efficiently \citep[see][chapter 6]{puterman2014markov}. We will say that a policy $\pi'$ is an {\em $\epsilon$-optimal policy} if its return is within $\epsilon>0$ of the optimal return: $R(\pi') + \epsilon \geq \max \{ R(\pi) \; | \; \pi \in \Pi\}$.
\begin{remark}[Finite-horizon setting]
In this paper, we only consider MDPs with infinite horizon. It is straightforward to extend our framework and results to the case of finite-horizon MDPs by adding an absorbing state with instantaneous reward $0$ after the last period.
\end{remark}

\subsection{Adherence-aware MDP}\label{sec:adherence-aware-mdps}
{\color{black} We now incorporate the phenomenon of partial adherence into an MDP framework.} 
Let $\M$ be an MDP instance, $\pi_{\sf base}$ a baseline policy, and $\pi_{\sf alg}$ a recommendation policy. We assume that $\pi_{\sf base}$ belongs to the set $\Pi$ of stationary policies. 
\tbl{To capture the fact that the DM does not systematically implement $\pi_{\sf alg}$, let us} 
introduce a parameter $\theta \in [0,1]$, which we call the {\em adherence level}. Intuitively, the adherence-level $\theta$ quantifies the compliance of the decision maker to follow the recommendation policy $\pia$ instead of the baseline policy $\pib$. 
\tbl{Therefore, the policy effectively implemented by the DM depends on $\pia$, $\pib$, and $\theta$. In particular, we consider an effective policy of the form: }
\begin{equation}\label{eq:expected effective policy}
\pieff(\pi_{\sf alg},\theta) = \theta \pi_{\sf alg} + (1-\theta)\pi_{\sf base}.
\end{equation}
\tbl{According to this model,} when $\theta = 0$, the DM always follows the baseline policy $\pi_{\sf base}$, and when $\theta = 1$, the DM always follows the recommendation policy $\pi_{\sf alg}$. When $\theta \in (0,1)$, the DM follows an effective policy $\pieff(\pi_{\sf alg},\theta)$, which is a mixture of $\pi_{\sf alg}$ and $\pi_{\sf base}$. 
Consequently,
the effective return for the DM is $R(\pieff(\pi_{\sf alg},\theta))$, with $\pieff(\pi_{\sf alg},\theta) = \theta \pi_{\sf alg} + (1-\theta)\pi_{\sf base}.$
For a fixed adherence level $\theta$, our objective is to compute an optimal recommendation policy such that the effective return $\pi_{\sf alg} \mapsto R(\pieff(\pi_{\sf alg},\theta))$ is maximized, i.e., our goal is to solve the following decision problem, called {\em Adherence-aware MDP} (\ref{eq:definition-ada-mdp}):
\begin{equation}\label{eq:definition-ada-mdp}\tag{{\sf AdaMDP}}
    \sup_{\pia \in \Pi_{\sf H}} \; R(\pieff(\pia,\theta)).
\end{equation}
\tb{When the supremum in the above optimization program is attained, we write $\pia\opt(\theta)$ for an optimal recommendation policy and} we write $\pieff^{\star}(\theta)$ for the resulting optimal effective policy, i.e., $\pieff\opt(\theta) = \pieff(\pi\opt_{\sf alg}(\theta),\theta)$. For simplicity, we assume for now that $\theta$ is the same for all states $s \in \X$, an assumption we will challenge in Section \ref{sec:extension}. We first note that an optimal policy $\pi^{\star}_{\sf alg}(\theta)$ for \ref{eq:definition-ada-mdp} can be chosen stationary and deterministic, two properties that are appealing from an implementation standpoint. 
\begin{proposition}\label{prop:pi alg star stationary deterministic}
The supremum in \ref{eq:definition-ada-mdp} is attained at an optimal recommendation policy $\pi^{\star}_{\sf alg}(\theta)$ that can be chosen stationary and deterministic:
\tb{
\[\sup_{\pia \in \Pi_{\sf H}} \; R(\pieff(\pia,\theta)) = \max_{\pia \in \Pi} \; R(\pieff(\pia,\theta)). \] }
\end{proposition}
The proof of Proposition \ref{prop:pi alg star stationary deterministic} uses some more advanced results that we will introduce in Section \ref{sec:algorithms}. We present the detailed proof in Appendix \ref{app:proof prop pi alg star stat deter}.

\begin{remark} Interestingly, a similar type of mixture policies have been studied in the online learning literature, yet with a different motivation. To address the exploration-exploitation trade-off, many policies obtained via reinforcement learning are implemented together with an ad-hoc exploration mechanism. Instead, \citet{shani2019exploration} propose to compute ``exploration-conscious'' policies that are designed for a particular exploration policy (e.g., choosing actions uniformly at random) and exploration rate, which play a similar role as $\pi_{\sf base}$ and $1-\theta$ in our framework. However, they view the exploration policy and exploration rate as additional parameters one can tune to mitigate the exploration-exploitation tradeoff, while we consider $\pi_{\sf base}$ and $\theta$ as uncontrolled inputs (arising from potential human deviations) and study their impact on actual performance. 
\end{remark}

\subsection{Discussion: Mechanisms for partial adherence and effective policy}\label{sec:discussion-adherence-aware-mdps}
Our adherence-aware MDP framework posits that the effective policy can be simply expressed as a convex combination of the algorithmic and the baseline policies, as presented in \eqref{eq:expected effective policy}. In this section, we further justify the practical relevance of our framework by discussing how different models for the DM's adherence  decision connects with our framework. 

To model the DM's decision to adhere, we introduce a variable ${u}_{s,t} \in [0,1]$ indicating, in state $s$, at time $t$, whether she follows the recommended policy $\pia$  (the case $u_{s,t}=1$) or whether she follows $\pib$ (the case $u_{s,t} = 0$). We call $u_{s,t}$ the {\em adherence decision} at state $s$ and period $t$, and we write $u := \left(u_{s,t}\right)_{s \in \X, t \in \N}$. With this notation, the effective policy at state $s$ at time $t$ is given as 
\begin{equation}\label{eq:def effective policy}
        \pieff(\pia,u)_{s,t} = u_{s,t} {\pia}_{s,t} + (1-u_{s,t}){\pib}_{s,t},
\end{equation}
and specifying an adherence mechanism is equivalent to specifying how the DM chooses $u$. 

\paragraph{Random model.}
For example, the DM could sample $u_{s,t}$ following {\em any distribution} with support included in $[0,1]$ and with a mean $\theta$. For instance, in the case of a Bernoulli distribution with parameter $\theta$, at each time period, the decision maker follows $\pi_{\sf alg}$ with probability $\theta$ and $\pi_{\sf base}$ with probability $1-\theta$. 
In practice, this random model of adherence decisions can be interpreted as being agnostic to the reasons for partial adherence. Whatever the cause (e.g., algorithm aversion, information asymmetry), they are inaccessible to the algorithm, hence are perceived by the algorithm as random deviations from the recommended policy. In other words, this model mimics the observed behavior of DM but does not capture from first principles why she sometimes decides to deviate from the recommendations.
For example, in a stylized setting with a rational DM trying to learn whether a machine is more accurate than her, \citet{devericourt2022your} identify regimes where the DM's belief oscillates permanently, hence justifying models like this one, where the DM's adherence decisions $u_{s,t}$ and $u_{s,t'}$ may be different for $t \neq t'$, even though the state is the same. In the next theorem, we show that this model with {random} adherence decision $u$ is exactly equivalent to \ref{eq:definition-ada-mdp}.
\tb{
\begin{theorem}\label{th:random tv is equivalent to adamdp}
Consider the following model of {\bf random} adherence decisions, where each $\left(u_{s,t}\right)_{s \in \X}$ is sampled from a distribution with mean $\left(\theta,...,\theta\right) \in [0,1]^{\X}$, independently across $t \in \N$.
Then \[\sup_{\pia \in \Pi_{\sf H}} \E_{u} \left[  \: R(\pieff(\pia,u)) \right] = \max_{\pia \in \Pi}  \: R(\pieff(\pia,\theta)) \]
and an optimal recommendation may be chosen stationary and deterministic in the left-hand side of the above equation.
\end{theorem}
}

\tbl{
We present a detailed proof in Appendix \ref{app:proof random tv}. \tb{Note that under the assumption of Theorem \ref{th:random tv is equivalent to adamdp}, the random variables $u_{s,t}$ and $u_{s',t}$ may be dependent for $s \neq s'$.} In fact, the proof relies on showing that  $\E_{u} \left[  \: R(\pieff(\pia,u)) \right] =  R\left(\E_{u} \left[ \pieff(\pia,u)\right]\right)$, despite the return $R\left(\cdot\right)$ being non-linear. This follows from the properties that $u_{s,t}$ and $u_{s',t'}$ are independent across pairs $(s,t),(s',t')$ such that $t \neq t'$. Noting that $\E_{u} \left[ \pieff(\pia,u)\right] = \pieff(\pia,\theta)$ concludes the proof. 
}

\paragraph{Adversarial model.}
Alternatively, as discussed in the literature review in Section \ref{sec:litreview}, partial adherence can be driven by information asymmetry or conflicting objectives between the algorithm and the DM. In other words, the decision maker could choose to follow the recommendation policy $\pia$ or the baseline policy $\pib$ according to a different MDP instance $\M'$ than the MDP instance $\M$ that parametrized the algorithm. Adopting a conservative view, one can assume the DM picks each $u_{s,t} \in [\theta, 1]$ {\em adversarially} in a set $B \subseteq [\theta,1]^{\X \times \N}$:
\begin{equation}\label{eq:adversarial-model-general}
\sup_{\pia \in \Pi_{\sf H}} \; \min_{u \in B} R\left(\pieff(\pia,u)\right).
\end{equation}
Without any restrictions, i.e., in the case $B = [\theta,1]^{\X \times \N}$, the DM could decide to follow the algorithm in state $s$ at time $t$ and, when visiting the same state $s$ at a later stage, decide to override it.
Hence, we can enrich the set $B$ with several {consistency} constraints to model more realistic situations. 
In some settings, for instance, it might be more realistic to assume a {\em time-invariant adversarial} model, i.e., to assume that the DM's adherence behavior depends on the state but is consistent over time. For example, one could assume that she
chooses an adherence decision $u_s \in [\theta,1]$ adversarially for each state $s$ and adopts this policy throughout, i.e., $u_{s,t} = u_s, \forall \; t \in \N$. 
Note that the time-invariant adversarial model assumes that the decision maker has some discretionary power at the beginning but commits to one policy for the rest of the trajectory, which can be seen as contradictory. \tb{Another realistic model consists of {\em state-invariant} adherence decisions, i.e., $u_{s,t}=u_{t} \in [0,1]$ across all pairs $(s,t) \in \X \times \N$. }
 A fourth model could assume that the adherence decisions are {\em time- and state-invariant}, i.e., that $u_{s,t}=u \in [0,1]$ across all pairs $(s,t) \in \X \times \N$. 
Fortunately, as stated (informally) in Theorem \ref{th:model.equiv.informal}, studying our effective policy \eqref{eq:expected effective policy} is equivalent to studying any of these three adherence mechanisms:
\begin{theorem}\label{th:model.equiv.informal} (Informal statement) An optimal algorithmic recommendation $ \pi^{\star}_{\sf alg}(\theta)$, solution to \ref{eq:definition-ada-mdp}, is an optimal solution of the decision problem \eqref{eq:adversarial-model-general}, whenever the adherence decision $u$ is chosen according to one of the following adversarial models:  for all $(s,t) \in \X \times \N$,
\begin{itemize}
\item {\bf (Unconstrained Adversarial)} $u_{s,t}$ chosen independently and adversarially in  $[\theta,1]$.
\item {\bf (Time-invariant Adversarial)} $u_{s,t} = u_s$ with $u_s$ chosen independently and adversarially in  $[\theta,1]$.
\item \tb{ {\bf (State-invariant Adversarial)} $u_{s,t} = u_{t}$ with $u_t$ chosen independently and adversarially in  $[\theta,1]$.}
\item {\bf (Time- and State-invariant Adversarial)} $u_{s,t} = u$ with $u$ chosen adversarially in  $[\theta,1]$. 
\end{itemize}
\tb{Additionally, strong duality holds for these models of adversarial adherence decisions.}
\end{theorem}
We defer a formal statement and proof of Theorem \ref{th:model.equiv.informal} to Appendix \ref{app:model equivalence}. 
Theorem \ref{th:model.equiv.informal} shows that \ref{eq:definition-ada-mdp} can be interpreted as the robust counterpart of the aforementioned adversarial models, and perhaps surprisingly, that these robust models yield the same worst-case return, and from the proof of Theorem \ref{th:model.equiv.informal}, the same optimal policy as well. \tb{The strong duality results show that the case where $\pia$ is chosen {\em before} the adherence decisions $\bm{u}$ and the case where $\pia$ is chosen {\em after} the adherence decisions $\bm u$ are equivalent.}  We should emphasize, however, that Theorem \ref{th:model.equiv.informal} only claims an equivalence in terms of {\em optimal} effective return. For a given (sub-optimal) policy, its effective return under each model (\ref{eq:definition-ada-mdp} or one of the adversarial models) can differ. 

\tb{
\begin{remark}
The proof of Theorem \ref{th:model.equiv.informal} shows that for these adversarial models, a worst-case $u_{s,t}$ can be chosen as $u_{s,t} = \theta, \forall (s,t) \in \X \times \N$. Therefore, when $\theta = 0$, we recover the fact that the agent never follows the algorithmic recommendation $\pia$.
\end{remark}
}
Overall, Theorems \ref{th:random tv is equivalent to adamdp} and \ref{th:model.equiv.informal} show that our simple proposal for adherence-aware MDPs subsumes a collection of DM-level models of partial adherence, hence justifying our subsequent analysis of the effective policy \eqref{eq:expected effective policy} and the optimal recommendation problem \eqref{eq:definition-ada-mdp}.  

\tb{We summarize the equivalences obtained in this section in Table \ref{tab:random-adverasarial-summary}. For the adversarial model, {\em time-invariance} and {\em state-invariance} are described in Theorem \ref{th:model.equiv.informal}. For the random model of adherence decisions, {\em time-invariance} corresponds to a model where there exist two periods $t \neq t'$ for which the random variables $u_{s,t}$ and $u_{s',t'}$ are dependent for some states $s,s' \in \X$, and {\em state-invariance} corresponds to the case where there exist $s \neq s'$ and $t \in \N$ for which $u_{s,t}$ and $u_{s',t}$ are dependent random variables. The assumption in Theorem \ref{th:random tv is equivalent to adamdp} corresponds to random models that are {\em not} time-invariant. We provide more discussion on these  time-invariant and state-invariant random models at the end of Appendix \ref{app:proof random tv}. 
\begin{table}[htb]
\settowidth\mylenA{Model of adherence decisions} 
\setlength\mylenB{(\mylenA-2\tabcolsep)/2}

\begin{center}
    \begin{tabular}{c|c |*{2}{|wc{\mylenB}}}

\multicolumn{2}{c}{Constraints} & \multicolumn{2}{c}{Model of adherence decisions} \\
\cline{1-2} \cline{3-4}
 Time-invariance & State-invariance & Random & Adversarial \\
\midrule
       $\times$ &  $\times$ & \ref{eq:definition-ada-mdp} & \ref{eq:definition-ada-mdp} \\ 
        $\times$ & $\checkmark$ & \ref{eq:definition-ada-mdp} &  \ref{eq:definition-ada-mdp} \\ 
     $\checkmark$ & $\times$     & unknown &  \ref{eq:definition-ada-mdp} \\ 
       $\checkmark$  & $\checkmark$   & unknown &  \ref{eq:definition-ada-mdp} \\ 
\bottomrule
    \end{tabular}
    \end{center}
    \caption{\tb{Summary of the adherence decision models considered in this paper and their relations with  \ref{eq:definition-ada-mdp}. }}\label{tab:random-adverasarial-summary}
    
\end{table}
}
\paragraph{Cardinality-constrained model.}
\tbl{
Under an adversarial lens, one could model the DM's unwillingness to implement a large number of changes to her current practice by, e.g., imposing a limit on the number of states where she adheres. For example, let us assume that adherence decisions are time-invariant and let us model the DM's adherence problem as that of finding up to $k$ states where she follows the algorithmic recommendation, with $k \in \N$:
\begin{equation}\label{eq:constrained-independent-varying-adversarial}\tag{{\sf Constrained-AdaMDP}}
\min_{u \in \{0,1\}^{\X}, \sum_{s \in \X} u_{s} \leq k} R(\pieff(\pia,u)).
\end{equation}
The evaluation problem above (let alone the problem of then optimizing for $\pia$) is hard, as we characterize in the following result:
\begin{theorem}\label{th:hardness-apx-ada-mdp}
\ref{eq:constrained-independent-varying-adversarial} is APX-hard, i.e., there exists a constant $\alpha>0$, for which it is NP-hard to approximate \ref{eq:constrained-independent-varying-adversarial} within a factor smaller than $1+\alpha$.
\end{theorem}
Our proof of Theorem \ref{th:hardness-apx-ada-mdp} is based on a reduction from the constrained assortment optimization under the Markov Chain-based choice model~\citep{desir2020constrained} and we provide the details in Appendix \ref{app:proof hardness}. This shows that adding a simple cardinality constraint to \ref{eq:definition-ada-mdp} makes the decision problem intractable. \tb{For the sake of completeness, and since \ref{eq:constrained-independent-varying-adversarial} may be of independent interest, we provide a mixed-integer optimization formulation for solving \ref{eq:constrained-independent-varying-adversarial} in Appendix \ref{app:MIP-constrained-adamdp}.  }
}

\subsection{Examples of competition/complementarity between the human and the algorithm} \label{sec:bad-mdp-instance}
{\color{black} Before turning to a more formal analysis of our framework, we demonstrate the implications of the effective policy \eqref{eq:expected effective policy} on a simple MDP instance, to provide some intuition on the interactions at play between $\pia$ and $\pib$ as well as illustrate the rich range of situations that can arise in our framework. Indeed, we provide an example where the co-existence of the algorithmic and baseline policies can lead to arbitrarily bad performance and another example where, on the contrary, they complement each other. 

We consider the MDP instance from Figure \ref{fig:mdp-example.tot}. There are $5$ states, the rewards are independent from the chosen action and only depend on the current state. We assume that the transitions are deterministic and are represented with dashed arcs in Figure \ref{fig:mdp-instance.tot}, along with the rewards above the states. The actions consist in choosing the possible next states. The MDP starts in State $1$, and State $4$ and State $5$ are absorbing. The MDP instance is parametrized by $\epsilon \in \{-1,1\}$, which impacts the reward of State $5$.

The current policy $\pi_{\sf base}$ is represented in Figure \ref{fig:pi-b.tot}. 
Observe that $\pi_{\sf base}$ prescribes to transition from State 2 to State 5 but that, according to $\pi_{\sf base}$, State 2 should not be visited in the first place. For example, in a healthcare setting, State 2 could correspond to a newly introduced treatment, which the practitioner is not used to prescribing. The expected return of $\pi_{\sf base}$ is
\begin{align*}
    R\left(\pi_{\sf base}\right) & =  \frac{\lambda^2}{1-\lambda},
\end{align*}
where $\lambda \in (0,1)$ is the discount factor. Note that, by definition of the effective policy $\pieff$, for any $\theta \in [0,1]$, $\pi_{\sf base}  = \pieff(\pi_{\sf base},\theta)$. In other words, for any adherence level $\theta \in [0,1]$, recommending $\pi_{\sf base}$ leads exactly to the implementation of $\pi_{\sf base}$. 
We further consider that the algorithm prescribes the policy $\pia$ represented in Figure \ref{fig:pi-r-star-one.tot}, whose expected return is 
\begin{align*}
    R\left(\pia\right) & = 0.1 \lambda + \frac{\lambda^2}{1-\lambda} > R\left(\pi_{\sf base}\right).
\end{align*}
Detailed computations of policy returns reported in this section are presented in Appendix \ref{app:bad-mdp-instance}.

\begin{figure}[h]
\begin{center}
\begin{subfigure}{0.3\textwidth}
  \includegraphics[width=1.0\linewidth]{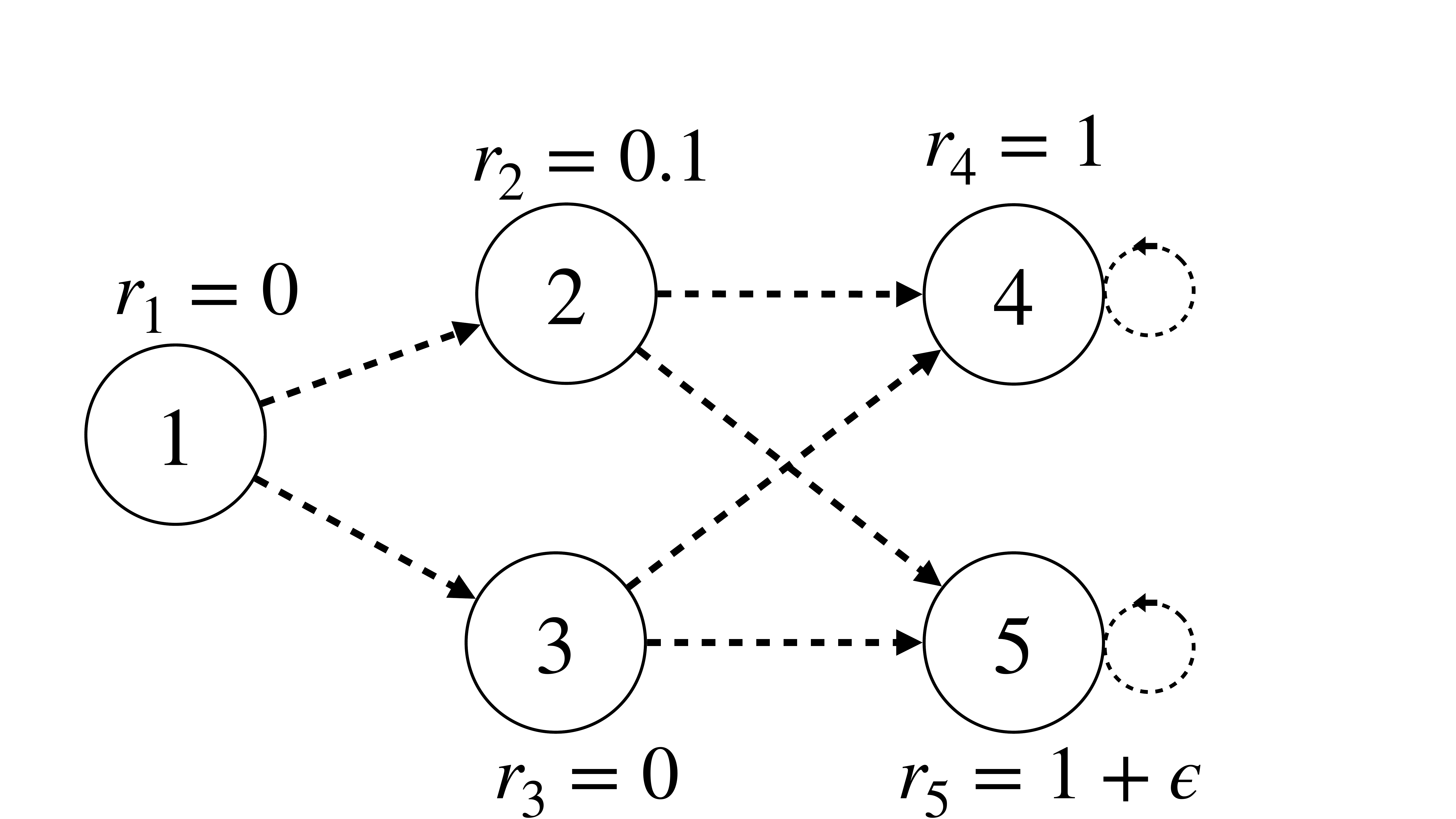}
\caption{MDP instance.}
\label{fig:mdp-instance.tot}
\end{subfigure}
\begin{subfigure}{0.3\textwidth}
  \includegraphics[width=1.0\linewidth]{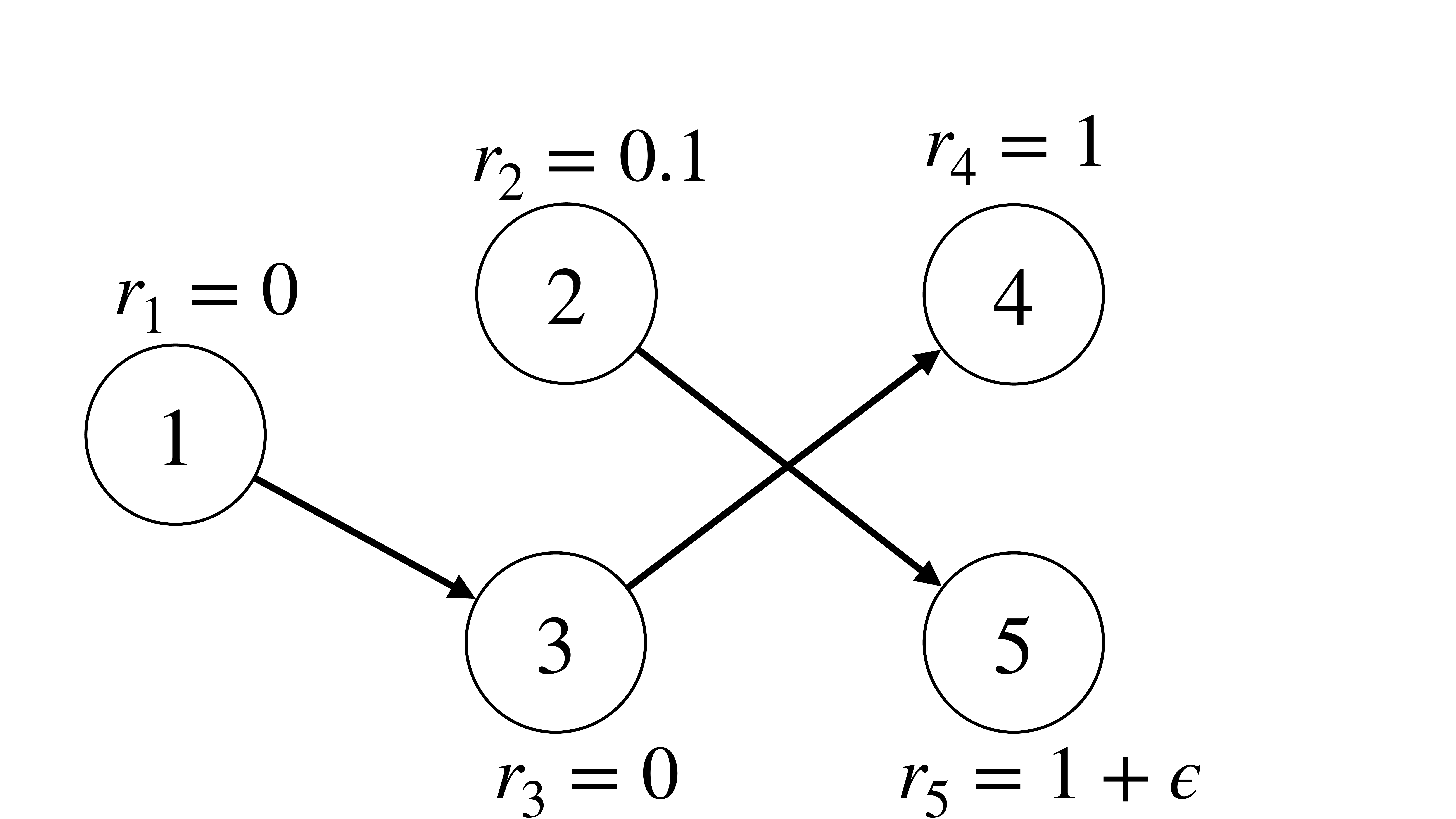}
\caption{Baseline policy $\pi_{\sf base}$.}
\label{fig:pi-b.tot}
\end{subfigure}
\begin{subfigure}{0.3\textwidth}
  \includegraphics[width=1.0\linewidth]{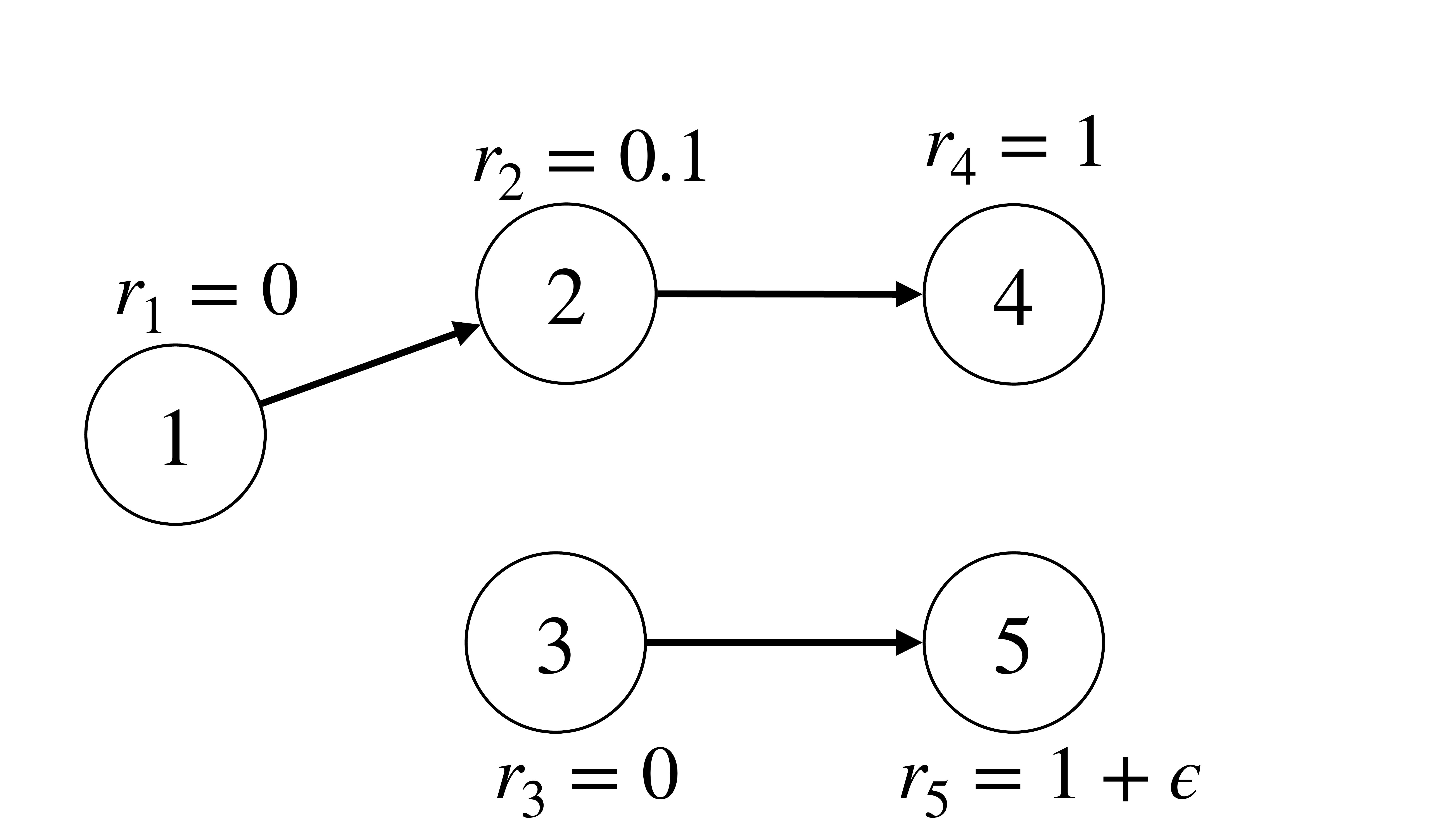}
\caption{Representation of $\pia$.}
\label{fig:pi-r-star-one.tot}
\end{subfigure}
\end{center}
\caption{Details on the transitions and rewards of our MDP instance.}
\label{fig:mdp-example.tot}
\end{figure}

\paragraph{Case 1: partial adherence hurts.} We first assume that $\epsilon=-1$. 
In this case, it is easy to verify that $\pia$ is optimal under perfect adherence ($\theta = 1$). If adherence is not perfect, however, continuing to recommend $\pia$ can lead to sub-optimal performance.
Indeed, $\pi_{\sf base}$ chooses suboptimal actions in State 2, which $\pi_{\sf alg}$  recommends to visit (unlike $\pi_{\sf base}$). So, the mixture policy $\pieff(\pia,\theta)$ can lead to worse performance than either $\pia$ or $\pi_{\sf base}$.
 Formally,  the return of the effective policy $\pieff(\pia,\theta)$ is equal to
 \[R\left(\pieff(\pia,\theta)\right) =  R(\pib) + 2 \theta \frac{\lambda^2}{1-\lambda} \left(\theta - \tilde{\theta}\right)\]
with $\Tilde{\theta} := 1 - 0.1 \dfrac{1-\lambda}{2\lambda} \leq 1$.  If $\Tilde{\theta} \leq 0$, the behavior of the effective return function is intuitive: In this case, we observe that $\theta \mapsto R\left(\pieff(\pia,\theta)\right)$ is increasing. In particular, $R\left(\pia\right) = R\left(\pieff(\pia,1)\right) \geq R\left(\pieff(\pia,\theta)\right)$, i.e., partial adherence degrades the effective return obtained by recommending $\pia$ compared with the perfect adherence case. Furthermore, $R\left(\pieff(\pia,\theta)\right) \geq  R\left(\pi_{\sf base}\right)$, i.e., recommending $\pia$ improves over the current standard of practice, $\pi_{\sf base}$. 

However, the analytic expression above reveals surprising behaviors when $\Tilde{\theta} > 0$. In this case, the function $\theta \mapsto R\left(\pieff(\pia,\theta)\right)$ is non-monotone (see Figure \ref{fig:mdp-instance-dominance}, obtained with $\lambda=0.5$, hence $\tilde{\theta} = 0.95$): It decreases on $[0,\Tilde{\theta}/2]$ and increases on  $[\Tilde{\theta}/2, 1]$. 
Since the effective policy is a convex combination of $\pia$ and $\pi_{\sf base}$, it is intuitive to believe that its performance will be bounded above and below by $R(\pia)$ and $R(\pi_{\sf base})$ respectively. This example disproves this intuition.
In particular, we have $R\left(\pieff(\pia,\Tilde{\theta})\right) <  R\left(\pi_{\sf base}\right)$. In other words, overlooking the adherence level $\theta$ and recommending the same policy $\pia$ may lead to lower return than the baseline policy itself! 
Actually, as we formally prove in the next section, this sub-optimality gap can be made arbitrarily large. 

Finally, via backward induction, we can find an optimal recommendation policy $\pi\opt_{\sf alg}(\theta)$ for any value of $\theta \in [0,1]$. In particular, we find an optimal recommendation policy of the following form (see derivations in Appendix \ref{app:bad-mdp-instance}): $\pi^\star_{\sf alg}(\theta) = \pi\opt$ if $\theta > \max ( 0, \bar{\theta})$ for $\pi\opt$ that chooses $1 \rightarrow 2, 2 \rightarrow 4, 3 \rightarrow 4$ and $\bar{\theta} = 1 - 0.1 (1-\lambda)/\lambda$; and $\pi^\star_{\sf alg}(\theta) = \pib$ if $\theta \leq \max(0,\bar{\theta})$. 
Note that by varying $\lambda$, the breakpoint  $\max(0,\bar{\theta})$ can be made arbitrarily close to $1$.
In the following section, we show that, for any MDP instance, the optimal recommendation policy $\pi^\star_{\sf alg}(\theta)$ enjoys such piecewise constant structure. 

\paragraph{Case 2: partial adherence helps (complementarity).} We now consider the case where $\epsilon = 1$ so that neither $\pia$ nor $\pib$ are optimal and there is room for improvement. Actually, we show in this example that partial adherence improves upon {\em both} policies, illustrating {\em complementarity benefits} between the human DM and the algorithm.
We now compute the expected return of the effective policy $\pieff(\pi_{\sf alg},\theta) = \theta \pi_{\sf alg} + (1-\theta) \pi_{\sf base}$. In particular, we obtain that
\[ R\left(\pieff(\pi_{\sf alg},\theta)\right) = R(\pi_{\sf alg}) + 2 R(\pi_{\sf base}) (1-\theta)(\theta-(1-\tilde{\theta})),\]
with $\tilde{\theta}$ previously defined. Thus, if $1-\tilde{\theta}<1$, we observe that $R\left(\pieff(\pi_{\sf alg},\theta)\right)  > \max \left\{ R(\pi_{\sf alg}), R(\pi_{\sf base}) \right\}$ for any $\theta \in (1-\Tilde{\theta},1)$. In other words, there exists a regime where the partial implementation of $\pia$ leads to greater performance than $\pia$ or $\pib$ alone. 

These examples show that, despite its simple form, the class of effective policies defined in \eqref{eq:expected effective policy} can capture many realistic situations where the co-existence of the algorithm and the DM hurts or benefits the overall system performance. 
Because our objective is prescriptive and we are interested in informing the design of the algorithmic recommendations $\pia$, we assume in the rest of the paper that recommendations are optimal for the true MDP parameter $\bm{r},\bm{P},\lambda$ and the adherence level $\theta$, i.e., where $\pia = \pia\opt(\theta)$ with $\pia\opt(\theta)$ an optimal solution to the optimization problem \eqref{eq:definition-ada-mdp}. This corresponds to the case where there is no model misspecification, and where $\theta$ is known. In particular, under this assumption, algorithmic recommendations that ignore the issue of partial adherence correspond to $\pia = \pia\opt(1)$, and Case 1 in this section shows that $R(\pieff\opt(\theta))$ may be much greater than $R(\pieff(\pia,\theta))$.
  Given an estimate of the adherence level $\theta$, our objective is thus to compute an optimal recommendation $\pia\opt(\theta)$ as a solution of an optimization problem, enabling us to prove important structural properties and tractability results in the next sections. We should emphasize that diverting from the assumption that the algorithmic recommendation is the solution of an optimization model leaves open the question of how to define (and compute) the algorithmic recommendation in practice.

\tb{
\begin{remark}
In our MDP instance for the second case (complementarity), neither $\pia$ nor $\pib$ are optimal. Indeed, by definition, if $\pia$ or $\pib$ is an optimal policy for the nominal MDP, then it is impossible that $R\left(\pieff(\pi_{\sf alg},\theta)\right)  > \max \left\{ R(\pi_{\sf alg}), R(\pi_{\sf base}) \right\}$, i.e., complementarity cannot occur. More complex models of partial adherence could lead to interesting human-machine complementarity, for instance in the case where both the algorithm and the human only have access to {\em partial information} on the state or action sets or have different objectives. Our agnostic model may adequately complement these cases where more is known (or assumed) about the rational behind partial adherence.
Because decision models are necessarily a simplification of real-life decisions, integrating more complex behavioural models behind partial adherence is an important direction for future work.
\end{remark}
}
}

\begin{figure}[h]
\begin{center}
\begin{subfigure}{0.4\textwidth}
  \includegraphics[width=1.0\linewidth]{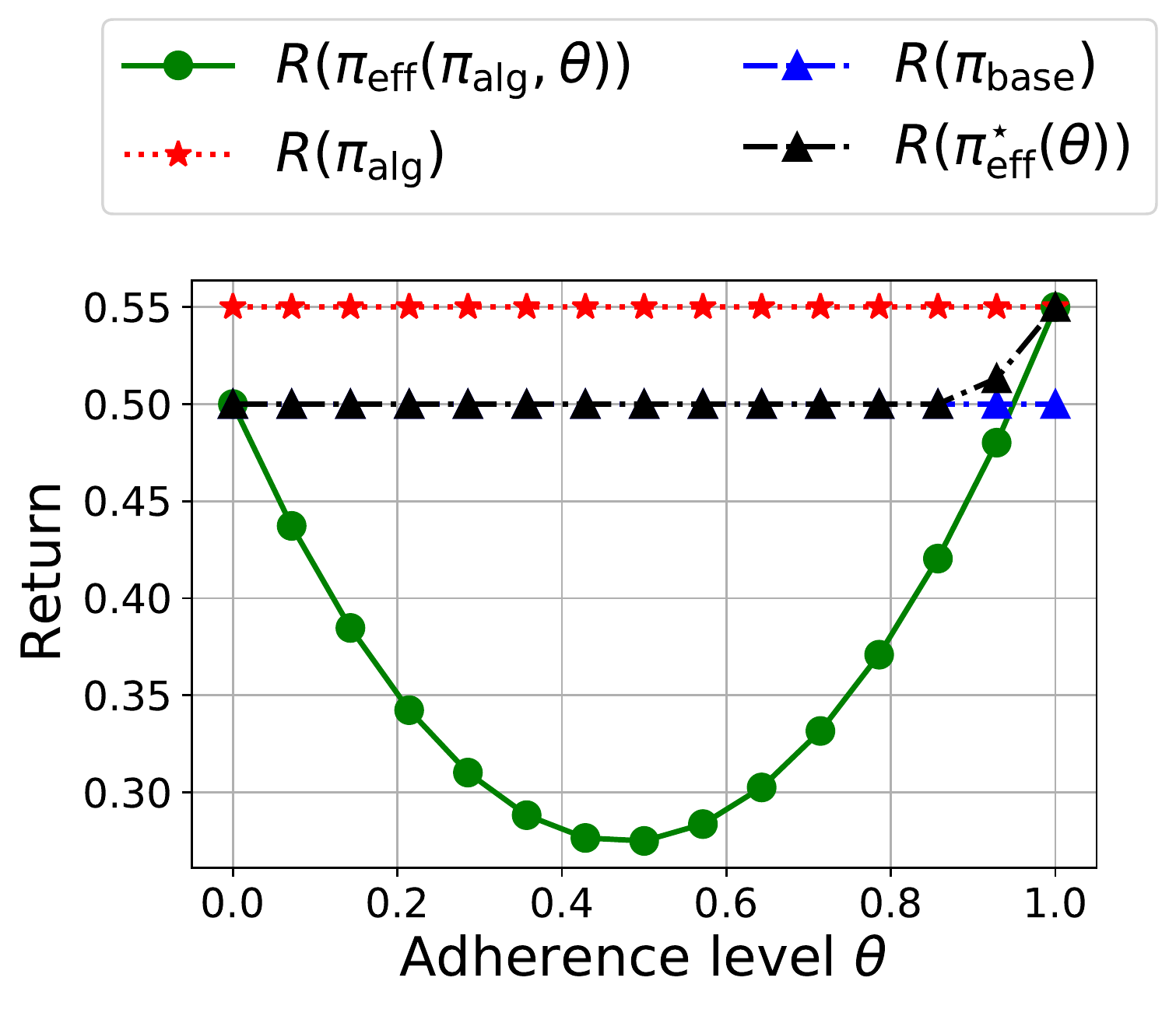}
\caption{Co-existence hurts ($\epsilon=-1$)}
\label{fig:mdp-instance-dominance}
\end{subfigure}
\begin{subfigure}{0.4\textwidth}
  \includegraphics[width=1.0\linewidth]{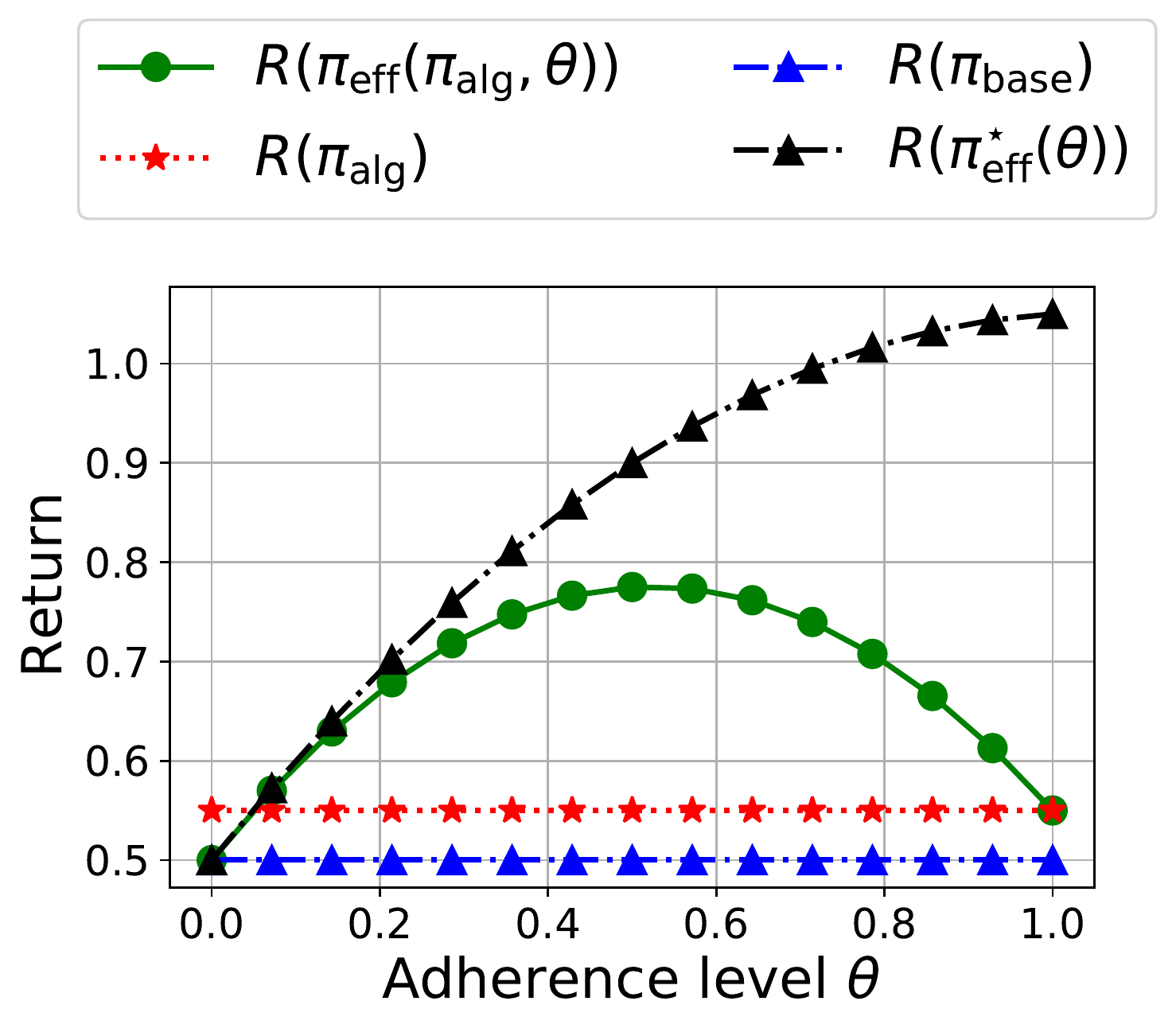}
\caption{Co-existence helps ($\epsilon=1$)}
\label{fig:mdp-instance-complementarity}
\end{subfigure}
\end{center}
\caption{Illustrating the impact of the partial adherence phenomenon (hence the coexistence of a baseline and algorithmic policy) in the MDP instance from Figure \ref{fig:mdp-instance.tot}. We choose $\lambda=0.5$ in our simulations.}
\label{fig:bad-instance-simu}
\end{figure}

\tbl{
\section{Analyzing adherence-aware MDPs}\label{sec:adamdp.analysis}
We now theoretically analyze the class of adherence-aware MDPs we introduced in the previous section. As a motivation, we first provide negative results showing the worst-case performance deterioration that can be experienced by overlooking the partial adherence phenomenon, i.e., by recommending $\pia^\star(1)$ instead of $\pia^\star(\theta)$. We then show how to compute optimal adherence-aware recommendations efficiently and investigate how they depend structurally on $\theta$.
}
\subsection{Worst-case analysis of the performance of $\pia^\star(1)$}\label{sec:structural-properties}
As the example in Section \ref{sec:bad-mdp-instance} shows, an optimal recommendation policy $\pia\opt(\theta)$ may be different from an optimal nominal policy $\pia\opt(1)$, which itself can lead to worse performance than the baseline policy $\pib$ alone. 
We now formalize these observations.

First, we analyze the performance of $\pieff(\pi^\star_{\sf alg}(1), \theta)$ for $\pia\opt(1)$ an optimal nominal policy and show that recommending $\pi_{\sf alg}\opt(1)$ (i.e., ignoring the partial adherence effect) can lead to arbitrarily worse  returns than the baseline policy.
\begin{proposition}\label{prop:naive-approach-negative}
 For any scalar $M \geq 0$, for any adherence level $\theta \in (0,1)$, there exists an MDP instance $\M$ such that 
 $R\left(\pi_{\sf base}\right) \geq M + R\left(\pieff(\pi_{\sf alg}^{\star}(1),\theta)\right),$ where $\pia\opt(1)$ is an optimal policy for the nominal MDP instance $\M$.
\end{proposition}
\proof{Proof of Proposition \ref{prop:naive-approach-negative}} Fix $M \geq 0$ and $\theta \in (0,1)$ and consider the MDP instance of Section \ref{sec:bad-mdp-instance} with $\epsilon=-1$, with $\pia\opt(1)$ as in Figure \ref{fig:pi-r-star-one.tot}. In the limit where $\lambda \rightarrow 1$, we have $R\left(\pieff(\pi_{\sf alg}^{\star}(1),\theta)\right) - R\left(\pi_{\sf base}\right) \sim 2 \theta \frac{\lambda^2}{1-\lambda} (\theta - \tilde{\theta})  \rightarrow -\infty$ since $\theta < 1$. 
Hence,  we can have $R\left(\pieff(\pi_{\sf alg}^{\star}(1),\theta)\right) - R\left(\pi_{\sf base}\right) \leq -M$ for $\lambda$ close to 1. \hfill \halmos
\endproof
Proposition \ref{prop:naive-approach-negative} generalizes the observation that $\pieff(\pi_{\sf alg}^{\star}(1),\theta) $ can lead to \emph{arbitrarily worse} performance than the current baseline policy itself (e.g., the current state of practice). As elicited in the example from Section \ref{sec:bad-mdp-instance}, this phenomenon happens when the baseline policy $\pi_{\sf base}$ chooses sub-optimal actions in some states. As a result, the effective policy $\pieff(\pi_{\sf alg}^{\star}(1),\theta)$ can also end up in these bad states that are overlooked by $\pi_{\sf alg}^{\star}(1)$, which assumes that the actions are always chosen from $\pi_{\sf alg}^{\star}(1)$. 
Consequently, for any value of $\theta \in (0,1)$, the policy $\pi_{\sf alg}\opt(1)$ can be arbitrarily sub-optimal. 
\begin{corollary} \label{cor:naive-approach-negative}
 For any scalar $M \geq 0$, for any adherence level $\theta \in (0,1)$, there exists an MDP instance $\M$ such that 
 $R\left(\pieff^\star(\theta)\right) \geq M + R\left(\pieff(\pi_{\sf alg}^{\star}(1),\theta)\right).$
\end{corollary} 
\proof{Proof of Corollary \ref{cor:naive-approach-negative}} The result follows from Proposition \ref{prop:naive-approach-negative} since  $R\left(\pi_{\sf base}\right) = R\left(\pieff(\pi_{\sf base},\theta)\right) \leq R\left(\pieff^\star(\theta)\right)$.
\hfill \halmos
\endproof

While Proposition \ref{prop:naive-approach-negative} and Corollary \ref{cor:naive-approach-negative} show that ignoring the adherence level $\theta$ can lead to arbitrarily large losses in performance, there are worst-case statements where, for each value of $\theta \in [0,1)$, a particular MDP instance $\M$ 
is constructed. In practice, one might be interested in a single MDP instance and the impact of varying $\theta \in [0,1]$ on this instance in particular, which is the focus of the rest of this section. 
\subsection{Solving adherence-aware MDPs} \label{sec:algorithms}
We now show how to efficiently compute an optimal policy $\pieff\opt(\theta)$ for adherence-aware MDPs. Note that when $\theta=1$, the DM is simply solving a classical MDP problem, which can be done efficiently with various algorithms such as value iteration, policy iteration, and linear programming \citep[see chapter 6 in][]{puterman2014markov}. \tbl{ Additionally, for the classical MDP problem, it is well-known that an optimal policy can be chosen stationary and deterministic without loss of optimality, which greatly simplifies implementation and interpretation of such policies in practice.  We show that the same holds for the adherence-aware MDP problem in the next proposition.
\begin{proposition}\label{prop:solving-adaMDP}
There exists a unique vector $\bm{v}^{\infty} \in \R^{\X}$ defined as
\begin{equation}\label{eq:v-infinity}
     v^{\infty}_{s} = \max_{\bm{\pi}_{s} \in \Delta(\A)}  \theta \cdot  \sum_{a \in \A} \pi_{sa} \bm{P}_{sa}^{\top} \left( \bm{r}_{sa} + \lambda \bm{v}^{\infty} \right) + (1-\theta) \cdot \sum_{a \in \A} \pi_{{\sf base},sa} \bm{P}_{sa}^{\top} \left( \bm{r}_{sa} + \lambda \bm{v}^{\infty} \right), \forall \; s \in \X,
\end{equation}
and an optimal recommendation policy $\pia^{\star}(\theta)$ can be computed as a stationary deterministic policy attaining the $\arg \max$ of Equation \eqref{eq:v-infinity} for each $s \in \X$.
\end{proposition}
The proof of Proposition \ref{prop:solving-adaMDP} is akin to our proof of Proposition \ref{prop:pi alg star stationary deterministic}, presented in Appendix \ref{app:proof prop pi alg star stat deter}, and we omit it for conciseness.
We note that we can rewrite Equation \eqref{eq:v-infinity} as
\begin{equation}\label{eq:v-infinity-prime}
     v^{\infty}_{s} = \max_{\bm{\pi}_{s} \in \Delta(\A)}  \sum_{a \in \A} \pi_{sa}\left(r'_{sa} + \lambda\bm{P}'^{\top}_{sa} \bm{v}^{\infty}\right), \forall \; s \in \X,
\end{equation}
with $\bm{P}' \in \left(\Delta(\X)\right)^{\X \times \A},\bm{r}' \in \R^{\X \times \A }$ defined as
\begin{equation}\label{eq:surrogate-mdp}
\begin{aligned}
 \bm{P}_{sa}' & := \theta \cdot \bm{P}_{sa} + (1-\theta) \cdot \sum_{a' \in \A} \pi_{{\sf base},sa'}\bm{P}_{sa'}, \\
r'_{sa} & :=  \theta \cdot \bm{P}_{sa}^{\top}\bm{r}_{sa} + (1-\theta) \cdot \sum_{a' \in \A} \pi_{{\sf base},sa'}\bm{P}^{\top}_{sa'}\bm{r}_{sa'},
\end{aligned}
\end{equation}
for all $(s,a) \in \X \times \A$.
This shows  that for any $\theta \in [0,1]$, an optimal recommendation $\pi_{\sf alg}\opt(\theta)$ can be viewed as the optimal policy for another MDP instance $\M'=\left(\X,\A,\bm{P}',\bm{r}',\bm{p}_{0},\lambda\right)$, where the new transition probabilities $\bm{P}'$ and the new rewards $\bm{r}'$ are defined as \eqref{eq:surrogate-mdp}, and, interestingly, where the instantaneous rewards only depend on the current state-action pair $(s,a)$ but not on the subsequent state $s'$.
In the context of ``exploration-conscious'' reinforcement learning and in the simpler case where $r_{sas'} = r_{sa}, \forall \; (s,a,s') \in \X \times \A \times \X$ in the MDP instance $\M$, \citet{shani2019exploration} refer to the MDP instance $\M'$ as the {\em surrogate MDP}. This shows that we can efficiently compute an optimal recommendation policy by computing an optimal policy of the surrogate MDP.
}
Note that even though $\pi_{\sf alg}\opt(\theta)$ can be chosen deterministic since it is an optimal policy to the surrogate MDPs, the effective policy $\pieff\opt(\theta)$ may be randomized, since by definition $\pieff\opt(\theta) = \theta \pi_{\sf alg}\opt(\theta)  +(1-\theta)\pi_{\sf base}$. 

\tbl{For the sake of completeness}, we now describe two efficient methods to compute $\bm{v}^{\infty}$.
\paragraph{Iterative method: value iteration.} Let us define the operator $f:\R^{\X} \rightarrow \R^{\X}$ as
\begin{equation}\label{eq:bellman-like-operator}
f_{s}(\bm{v}) = \max_{\bm{\pi}_{s} \in \Delta(\A)}  \theta \cdot  \sum_{a \in \A} \pi_{sa} \bm{P}_{sa}^{\top} \left( \bm{r}_{sa} + \lambda \bm{v} \right) + (1-\theta)  \sum_{a \in \A} \pi_{{\sf base},sa} \bm{P}_{sa}^{\top} \left( \bm{r}_{sa} + \lambda \bm{v} \right), \forall \; s \in \X.
\end{equation}
Note that when $\theta=1$, this is the classical Bellman operator.
The operator $f$ is a contraction for $\ell_{\infty}$: for any $\bm{v},\bm{w} \in \R^{\X}$, we have $
\| f(\bm{v}) - f(\bm{w}) \|_{\infty} \leq \lambda \| \bm{v} - \bm{w} \|_{\infty}.$
Therefore, as for classical MDPs, the fixed-point $\bm{v}^{\infty}$ can be computed efficiently via value iteration (VI): $\bm{v}^{0} =\bm{0}, \bm{v}^{t+1} = f(\bm{v}^{t}), \forall \; t \in \N.$
To obtain an $\epsilon$-optimal recommendation policy, we can stop as soon as $\| \bm{v}^{t} - f\left(\bm{v}^{t}\right) \|_{\infty} \leq \epsilon (1-\lambda)(2\lambda)^{-1}$, which is satisfied after $O\left(\log\left(\epsilon^{-1}\right)\right)$ iterations~\citep[][theorem 6.3.3]{puterman2014markov}. 
\paragraph{Linear programming formulation.}
The optimal value function $\bm{v}^{\infty} \in \R^{\X}$ can also be computed with linear programming~\citep[][section 6.9]{puterman2014markov}. In particular, $\bm{v}^{\infty}$ is the unique solution to the optimization problem $\min \left\{ \sum_{s \in \X} v_{s} \; | \; v_{s} \geq f_{s}(\bm{v}), \forall \; s \in \X \right\}$, which can reformulated in the following linear program with $|\X|$ decision variables and $|\X| \times |\A|$ linear constraints:
\[ \min \; \{ \bm{p}_{0}\trp\bm{v} \; | \;  v_{s}  \geq \theta \bm{P}_{sa}^{\top} \left(\bm{r}_{sa} + \lambda \bm{v}\right) + (1-\theta) \sum_{a' \in \A} \pi_{{\sf base},sa'} \bm{P}_{sa'}^{\top}\left( \bm{r}_{sa'} + \lambda \bm{v}\right), \forall \; (s,a) \in \X \times \A\}.\] 

\tbl{
\subsection{Structure and sensitivity of $\pia^\star(\theta)$ with respect to the adherence level}
We now investigate how the optimal recommendation $\pia^\star(\theta)$ and its performance $R(\pieff(\pia^\star(\theta),\theta))$ depend on the adherence level $\theta$.}

First, the example from Section \ref{sec:bad-mdp-instance} illustrates that the mapping $\theta \mapsto R(\pieff(\pi,\theta))$, for a fixed policy $\pi$, is not necessarily monotone. Still, we can recover monotonicity when considering $\pi_{\sf alg}^\star(\theta)$ instead, as shown in the next proposition.
\begin{proposition}\label{prop:monotonicity}
For any MDP instance $\M$, the map $\theta \mapsto R(\pieff\opt(\theta))$ is non-decreasing on $[0,1]$.
\end{proposition}
\proof{Proof.} \tbl{This is straightforward from the equivalence of \ref{eq:definition-ada-mdp} and the models of adversarial adherence decisions from Theorem \ref{th:model.equiv.informal}. We provide a simple, more direct proof below.}
Let $\theta_{1},\theta_{2} \in [0,1]$ with $\theta_{1} \leq \theta_{2}$. We will show that $R(\pieff\opt(\theta_{1})) \leq R(\pieff\opt(\theta_{2}))$. 
Following the definition of $\pieff\opt(\theta_{1})$, we have $\pieff^{\star}(\theta_{1}) = \theta_{1} \pi^{\star}_{\sf alg}(\theta_{1}) + (1-\theta_{1})\pi_{\sf base}$.
We can rewrite this as
\[\pieff^{\star}(\theta_{1}) = \theta_{2} \left( \frac{\theta_{1}}{\theta_{2}} \pi^{\star}_{\sf alg}(\theta_{1}) + \frac{\theta_{2}-\theta_{1}}{\theta_{2}} \pi_{\sf base} \right) + (1-\theta_{2})\pi_{\sf base},\]
and $\hat{\pi} := \frac{\theta_{1}}{\theta_{2}} \pi^{\star}_{\sf alg}(\theta_{1}) + \frac{\theta_{2}-\theta_{1}}{\theta_{2}}\pi_{\sf base}$ is a policy since $0 \leq\theta_{1} \leq \theta_{2} \leq 1$. Overall, we conclude that 
$R(\pieff\opt(\theta_{1})) = R(\pieff(\hat{\pi},\theta_2)) \leq R(\pieff\opt(\theta_{2}))$, by optimality of $\pi^\star_{\sf alg}(\theta_2)$.
\hfill \halmos
\endproof
Proposition \ref{prop:monotonicity} shows that as the DM deviates more and more from the recommendation policy (i.e., as $\theta$ decreases), the optimal effective return decreases. Note that this result holds because we consider $\pi^\star_{\sf alg}(\theta)$, in other words because we adjust our recommended policy as the adherence level varies. Since $\pieff\opt(0) = \pi_{\sf base}$, Proposition \ref{prop:monotonicity} also implies that $R(\pieff\opt(\theta)) \geq R(\pi_{\sf base})$: recommending $\pieff\opt(\theta)$ can only improve performance compared with the current baseline, which may not be the case when recommending $\pi_{\sf alg}\opt(1)$, as highlighted in Proposition \ref{prop:naive-approach-negative}. Overall, Proposition \ref{prop:monotonicity} also suggests that it is always beneficial to try to increase the compliance of the decision maker (i.e., increase the value of $\theta$), as this leads to more returns for the optimal effective policy $\pieff\opt(\theta)$. 

\tbl{Actually, we now show that the optimal recommendation $\pia^\star(\theta)$ does not vary continuously in $\theta$ but rather enjoys a piecewise constant structure:}
\begin{proposition}\label{prop:piece-constant-policy} For any MDP instance $\M$:
\begin{enumerate}
\item There exists $\bar{\theta} \in [0,1)$, such that $\pi_{\sf alg}^{\star}(\theta) = \pi_{\sf alg}^{\star}(1)$ for any $\theta \in [\bar{\theta},1]$. \label{prop:piecewise-statement-1}
\item There exists $n \in \mathbb{N}$ and $0=\theta_1 < \theta_2 < \dots < \theta_{n}=1$ such that, for any $i \in \{1,...,n-1\}$, $\pi_{\sf alg}^{\star}(\theta)$ can be chosen constant over the interval $[\theta_i, \theta_{i+1}]$. 
\label{prop:piecewise-statement-2}
\item If $\pi_{\sf base} = \pi_{\sf alg}\opt(\underline{\theta})$ for some $\underline{\theta} \in [0,1]$, then $\pi_{\sf alg}\opt(\theta)=\pi_{\sf base}$ for any $\theta \in [0,\underline{\theta}]$. 
\label{prop:piecewise-statement-3}
\end{enumerate}
\end{proposition}
Combined with the fact that $\pi_{\sf alg}^\star(1)$ is an optimal recommendation for $\theta=1$, Statement \ref{prop:piecewise-statement-1} shows that, when the adherence level is sufficiently close to $1$, we can overlook the issue of partial adherence and output the same recommendation as when $\theta=1$, which \tbl{reduces} to the classical MDP model. More generally, Statement \ref{prop:piecewise-statement-2} in Proposition \ref{prop:piece-constant-policy} shows that, in general, $\pi_{\sf alg}^{\star}(\theta)$ has a piecewise constant structure. 
The piecewise constant structure of $\pi^\star_{\sf alg}(\theta)$ combined with the fact that $\pi_{\sf base}$ is an optimal recommendation for $\theta=0$ also ensures that $\pi_{\sf base}$ is an optimal recommendation in a neighborhood of $0$. 
Statement \ref{prop:piecewise-statement-3} generalizes this observation and states that if the baseline policy is an optimal recommendation policy for an adherence level $\underline{\theta}$, then it is optimal for any lower adherence level. A trivial example is the case where $\underline{\theta}=1$, i.e., when $\pi_{\sf base}$ is optimal in the classical MDP model, then we should systematically recommend the baseline. To motivate our study, we implicitly assumed that $R(\pi_{\sf base}) < R(\pi^\star_{\sf alg}(1))$, i.e., that the baseline policy could be improved.

\tb{Lastly, we uncover two conditions on the MDP instance under which the partial adherence phenomenon can be ignored by the decision-maker.
}
We start with a simple example where the optimal recommendation $\pia^\star(\theta)$ does not depend on $\theta$ and $\pib$. We observe that when the transitions $\bm{P}_{sa} \in \Delta(\X)$ do not depend on the action but only on the current state: $\bm{P}_{sa} = \bm{P}_s \in \Delta(\X)$ and when $r_{sas'} = r_{sa}$ for all $(s,a,s') \in \X \times \A \times \X$, then the optimality equation \eqref{eq:v-infinity} becomes
\begin{equation*}
     v^{\infty}_{s} =    \theta \cdot  \max_{\bm{\pi}_{s} \in \Delta(\A)}  \left\{     \bm{\pi}_{s}^\top \bm{r}_{s} \right\} + \theta \cdot \lambda \bm{P}_{s}^{\top}\bm{v}^{\infty} + (1-\theta) \cdot \bm{\pi}_{{\sf base},s}^\top \bm{r}_{s} + (1-\theta) \cdot \lambda \bm{P}_{s}^{\top}\bm{v}^{\infty}, \forall \; s \in \X,
 \end{equation*}
and we can choose an optimal recommendation policy $\pi_{\sf alg}\opt(\theta)$ that is {\em independent} from $\theta$ and $\pi_{\sf base}$. In other words, partial adherence only impacts the effective return but {\em it does not change the optimal recommendation}. This special case occurs, for example, when the DM faces a sequence of independent single-stage decision problems (e.g., patients arriving independently to be treated) where each decision provides an immediate reward but does not impact the next decision problem, see~\cite{devericourt2022your} for a detailed study of this case in a learning setting.

\tb{
We now describe a condition under which the decision-maker may ignore partial adherence at a given state. Inspecting the surrogate MDP defined in Equation \eqref{eq:surrogate-mdp}, we note that the new pair of rewards and transitions $(\bm{r}',\bm{P}')$ is a convex combination of the nominal parameters $(\bm{r},\bm{P})$ and the rewards and transitions induced by $\pib$. Therefore, if $\pib$ chooses an optimal action at a state $\bs \in \X$, we may expect that the algorithmic recommendation coincides with $\pib$ at $\bs$. We show that this intuition is true in the next proposition.
\begin{proposition}\label{prop:value-similar}
Let $\bs \in \X$ such that $v^{\pia\opt(1)}_{\bs} = v^{\pib}_{\bs}$.  Then for any $\theta \in [0,1]$, we have $ v^{\pieff\opt(\theta)}_{\bs} = v^{\pib}_{\bs}$ and we can choose $\pia\opt(\theta)_{\bs} = \pi_{{\sf base},\bs}$.
\end{proposition}
We provide the proof of Proposition \ref{prop:value-similar} in Appendix \ref{app:proof prop value similar}. Proposition \ref{prop:value-similar} shows that if the baseline policy obtains the optimal nominal value at a given state $\bs \in \X$, then the decision-maker can guarantee this same value at $\bs$ for any value of the adherence level $\theta \in [0,1]$ by recommending the same action as the baseline policy.
\tb{We conclude this section by noting that obtaining a meaningful bound on the suboptimality of a policy $\pia$ against $\pia\opt(\theta)$ for a given value $\theta \in [0,1]$ of the adherence level is an interesting direction for future work. We derive a bound in Appendix \ref{app:suboptimality bound}, noting that it may be hard to interpret, due to the piece-wise constant structure of the optimal recommendation policies (Proposition \ref{prop:piece-constant-policy}).
}
}

\section{Numerical experiments}\label{sec:simu}
In this section, we numerically study the impact of the adherence level and of the baseline policy on two decision-making examples, in machine replacement and healthcare respectively, that have been studied in the MDP literature. We solve all the decision problems using the value iteration algorithm presented in Section \ref{sec:algorithms}. Among others, these numerical results illustrate the importance of taking into account the current state of practice and the adherence level when designing algorithmic recommendations. In particular, the adherence-aware optimization framework we develop in this paper provides simple tools to evaluate the robustness of a policy with respect to the adherence level and to obtain improved solutions in situations where the performance is the most impacted. 

\subsection{Machine replacement problem}\label{sec:simu-machine}
\begin{figure}[htb]
\begin{subfigure}{0.5\textwidth}
  \includegraphics[width=0.9\linewidth]{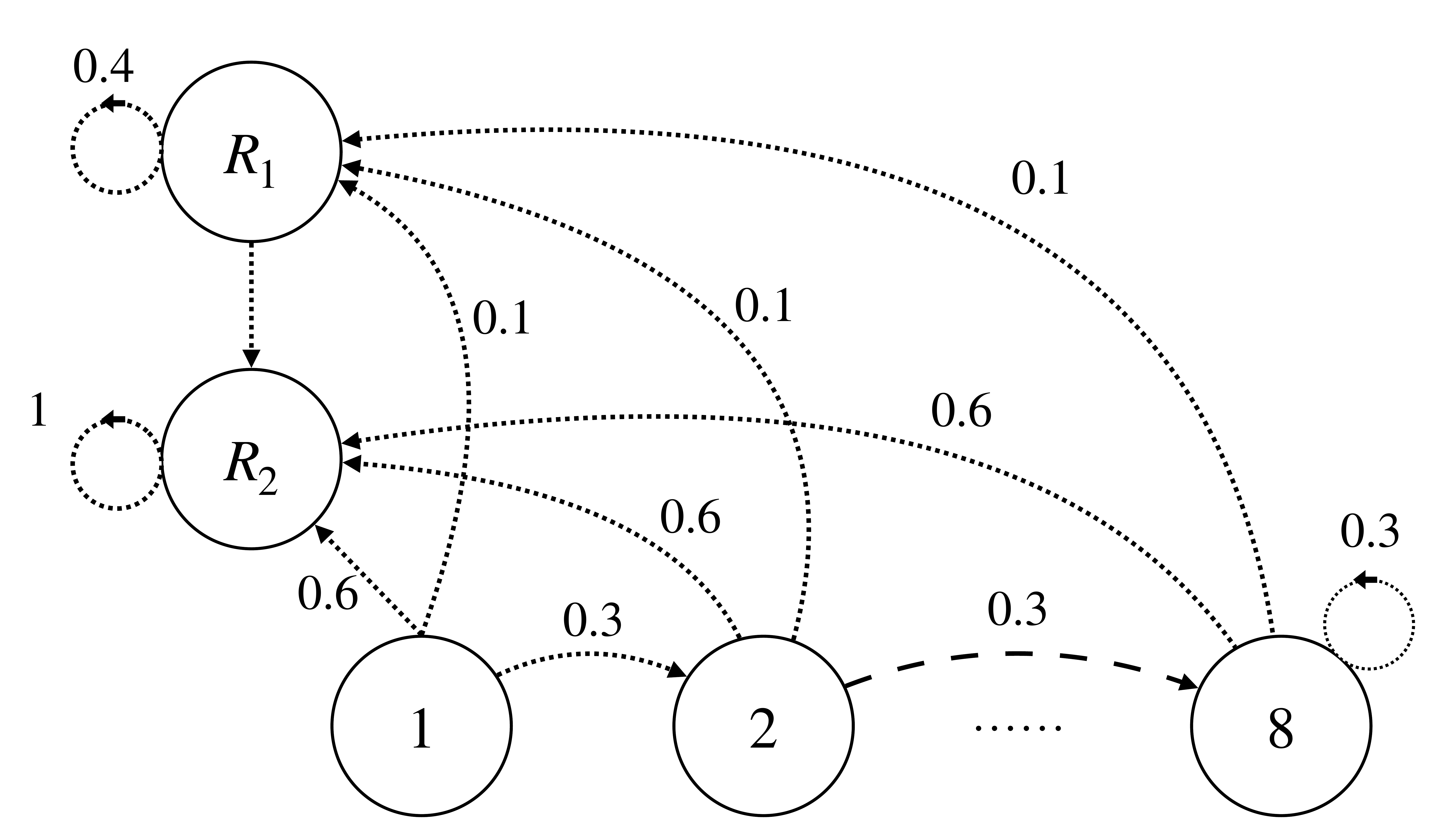}
\caption{Transition probabilities for \\ action {\sf repair}.}
\label{fig:machine-mdp-repair}
\end{subfigure}
\begin{subfigure}{0.5\textwidth}
  \includegraphics[width=0.9\linewidth]{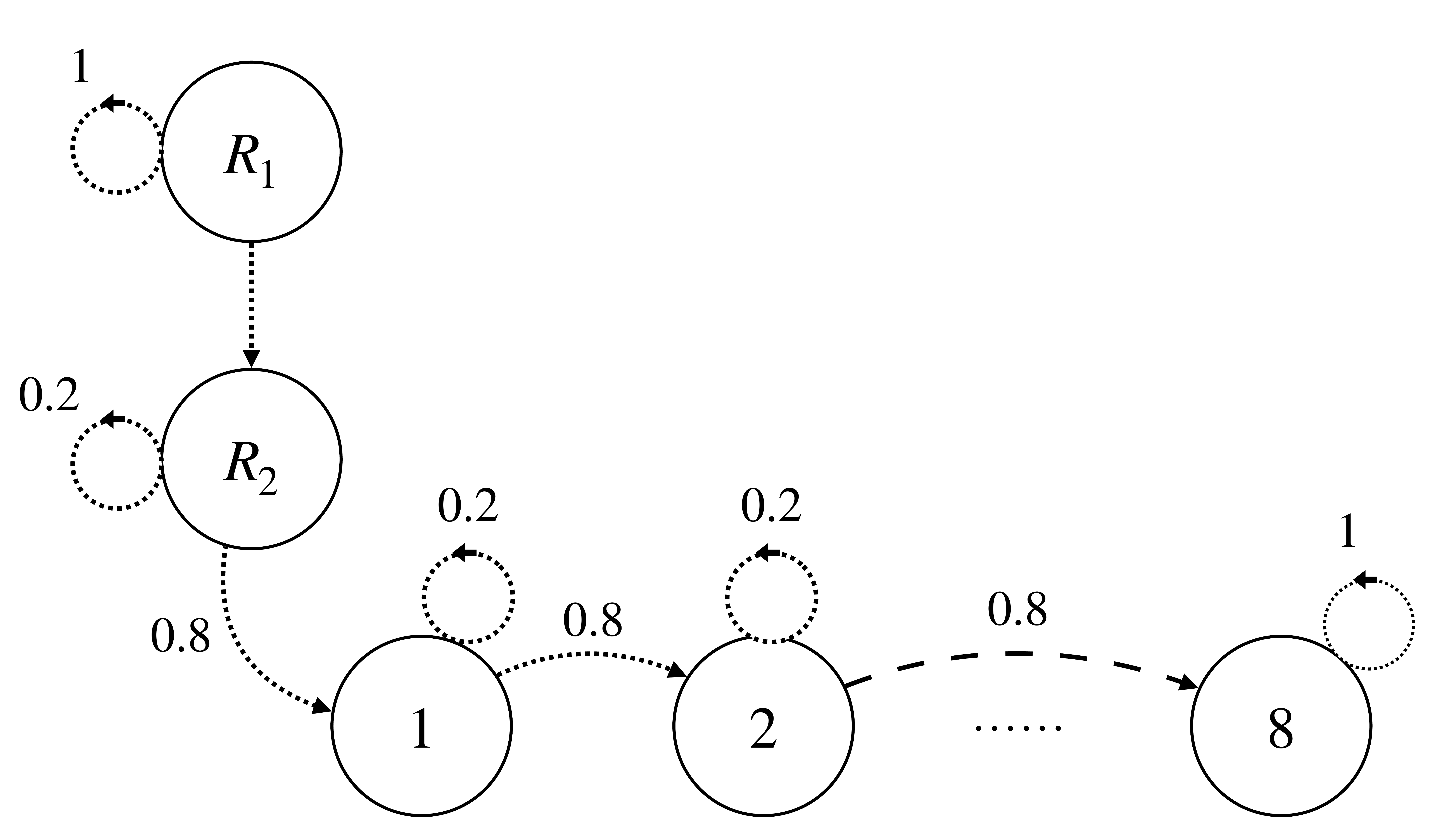}
\caption{Transition probabilities for \\ action {\sf wait}.}
\label{fig:machine-mdp-wait}
\end{subfigure}
\caption{Transition probabilities for the machine replacement MDP. There is a reward of $18$ in state $R1$, of $10$ in state $R2$ and of $0$ in state $8$. All others states have a reward of $20$.}
\label{fig:machine-mdp}
\end{figure}

\begin{figure}[h]
\begin{center}
\begin{subfigure}{0.3\textwidth}
  \includegraphics[width=1.0\linewidth]{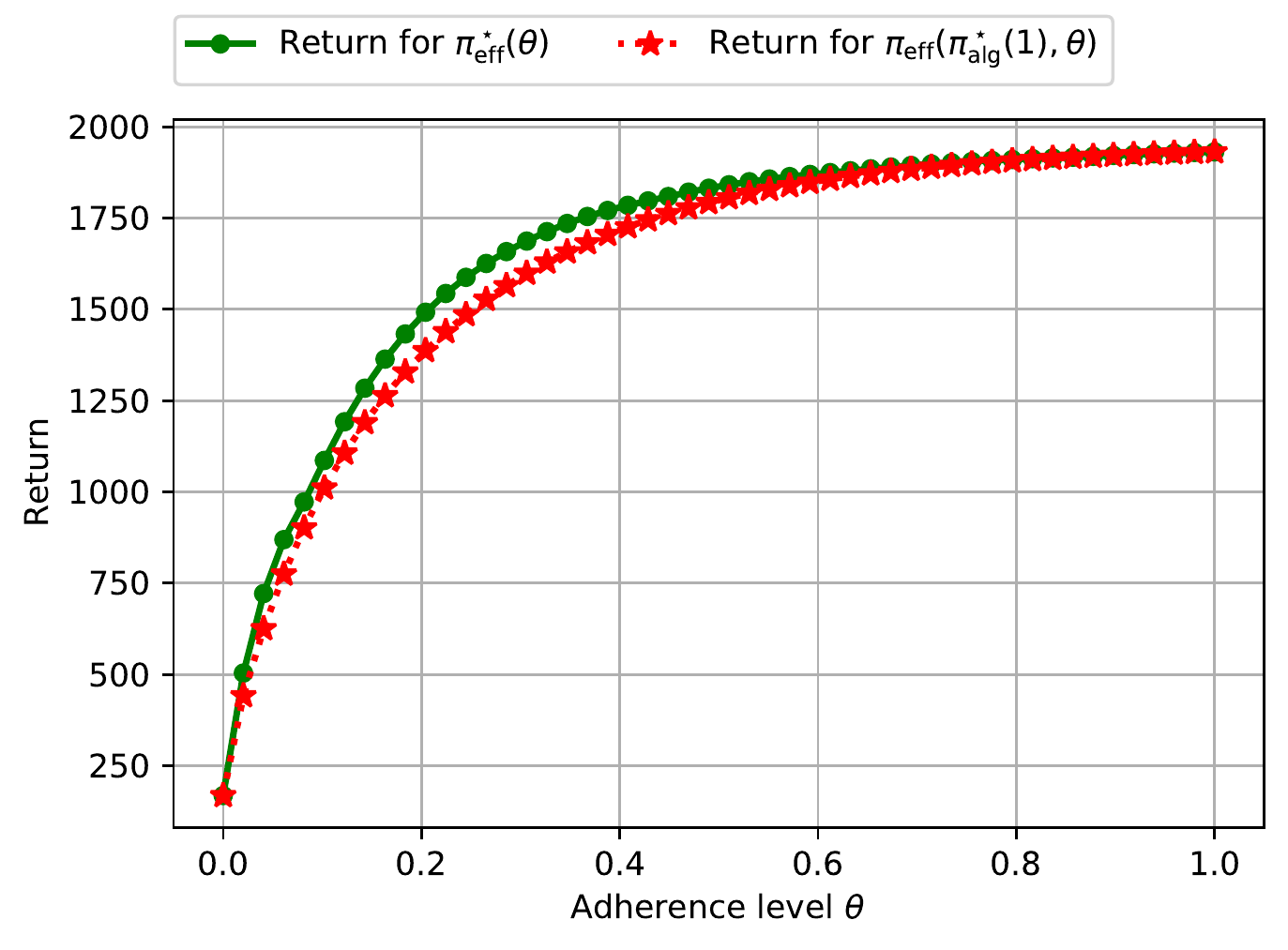}
\caption{Returns for recommending $\pi\opt_{\sf alg}(\theta)$ and $\pi\opt_{\sf alg}(1)$.}
\label{fig:machine-wait-return}
\end{subfigure}
\begin{subfigure}{0.3\textwidth}
  \includegraphics[width=1.0\linewidth]{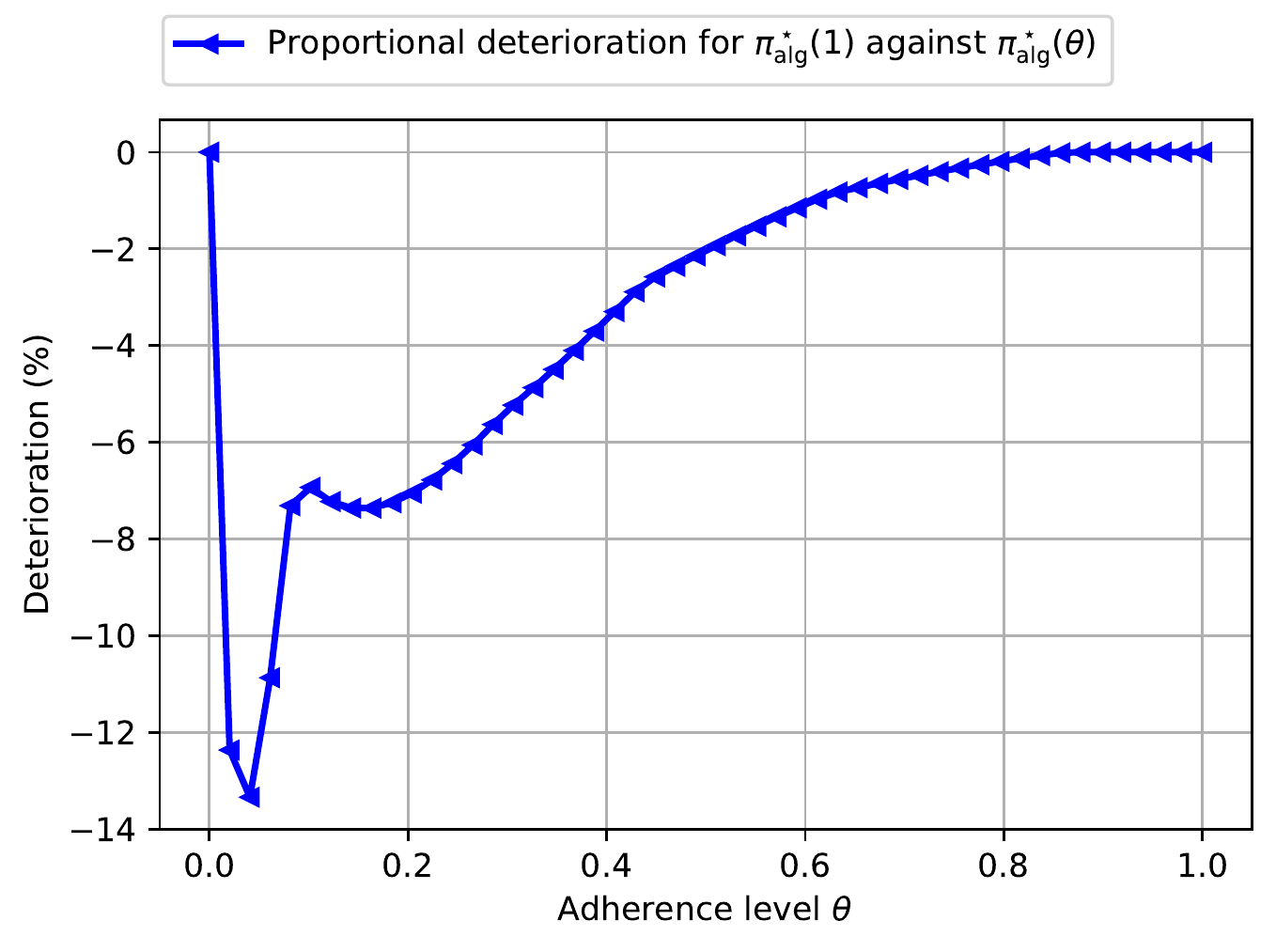}
\caption{Proportional deterioration for $\pi\opt_{\sf alg}(1)$ against $\pi\opt_{\sf alg}(\theta)$.}
\label{fig:machine-wait-decrease}
\end{subfigure}
\begin{subfigure}{0.3\textwidth}
  \includegraphics[width=1.0\linewidth]{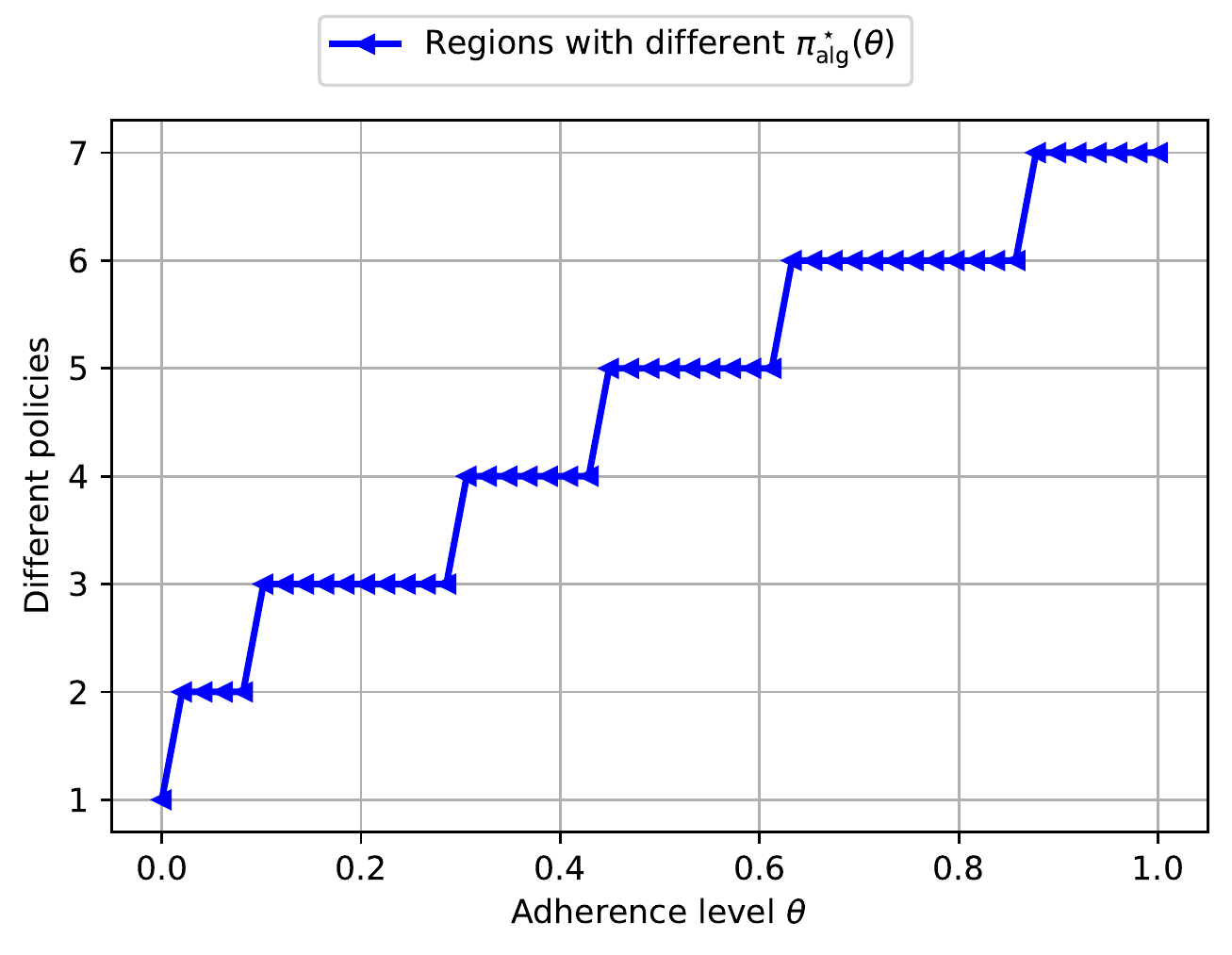}
\caption{Subregions with constant recommendation policies.}
\label{fig:machine-wait-regions}
\end{subfigure}
\end{center}
\caption{Numerical results for the machine replacement MDP with $\pi_{\sf base}$ always choosing action {\em wait}.}
\label{fig:machine-wait-comparison}
\end{figure}

We start with the a {machine replacement problem} introduced in \cite{delage2010percentile} and studied in \cite{wiesemann2013robust,goyal2023robust}.
\paragraph{MDP instance.} We represent the machine replacement MDP in Figure \ref{fig:machine-mdp}. The set of states is $\{1,2,3,4,5,6,7,8,R_{1},R_{2}\}$ and the set of actions is $\{${\sf repair, wait}$\}$. Each state models the condition of the same machine. In State $8$ the machine is broken, while State $R_{1}$ and State $R_{2}$ model some ongoing reparations. State $R_{1}$ is a normal repair while State $R_{2}$ is a long repair. We use the same rewards and transitions as in \cite{delage2010percentile}. In particular, there is  a reward of $0$ in State $8$, a reward of $18$ in State $R_{1}$, a reward of $10$ in State $R_{2}$, and a reward of $20$ in the remaining states. We set a discount factor of $\lambda=0.99$ and the DM starts in State $1$.
\paragraph{Numerical results.} Assuming $\theta=1$, an optimal policy $\pi_{\sf alg}\opt(1)$ is to choose action {\sf wait} in States $1,2,3,4,R_{2}$ and action {\sf repair} in States $5,6,7,8,R_{1}$. We now compare the effective return of $\pi_{\sf alg}\opt(1)$ with that of the best recommendation $\pi^\star_{\sf alg}(\theta)$, for varying values of the adherence level $\theta$.
We first consider the case where $\pi_{\sf base}$ chooses to always wait instead of repairing the machine. 
We present the results of our empirical study in Figure \ref{fig:machine-wait-comparison}. 
In Figure \ref{fig:machine-wait-return}, we report the effective return of both policies, namely $R(\pieff\opt(\theta))$ and $R(\pieff(\pi_{\sf alg}\opt(1),\theta))$, for varying $\theta \in [0,1]$.
We also compute the proportional deterioration in performance, $\left(R(\pieff\opt(\theta))-R(\pieff(\pi_{\sf alg}\opt(1),\theta))\right)/R(\pieff\opt(\theta))$ in Figure \ref{fig:machine-wait-decrease}. As expected from Proposition \ref{prop:piece-constant-policy}, when $\theta$ is sufficiently close to $1$ (here, for $\theta \geq 0.88$), we have $\pieff\opt(\theta)=\pieff\opt(1)$ and there is no deterioration in performance. However, as the value of $\theta$ decreases towards $0$, overlooking the adherence level and recommending $\pi_{\sf alg}\opt(1)$ can lead to as much as $13.34\%$ proportional deterioration compared with the optimal return $R(\pieff\opt(\theta))$. We also note in Figure \ref{fig:machine-wait-decrease} that small changes in $\theta$ can lead to very severe deterioration, for instance in the region $\theta \in [0,0.20]$, i.e., for very low adherence from the human decision maker. The different regions over which the optimal decision $\theta \mapsto \pi_{\sf alg}\opt(\theta)$ is constant are shown in Figure \ref{fig:machine-wait-regions}, which highlights that the optimal recommendation policy may change many times as the adherence level decreases.

\begin{figure}[h]
\begin{center}
\begin{subfigure}{0.3\textwidth}
  \includegraphics[width=1.0\linewidth]{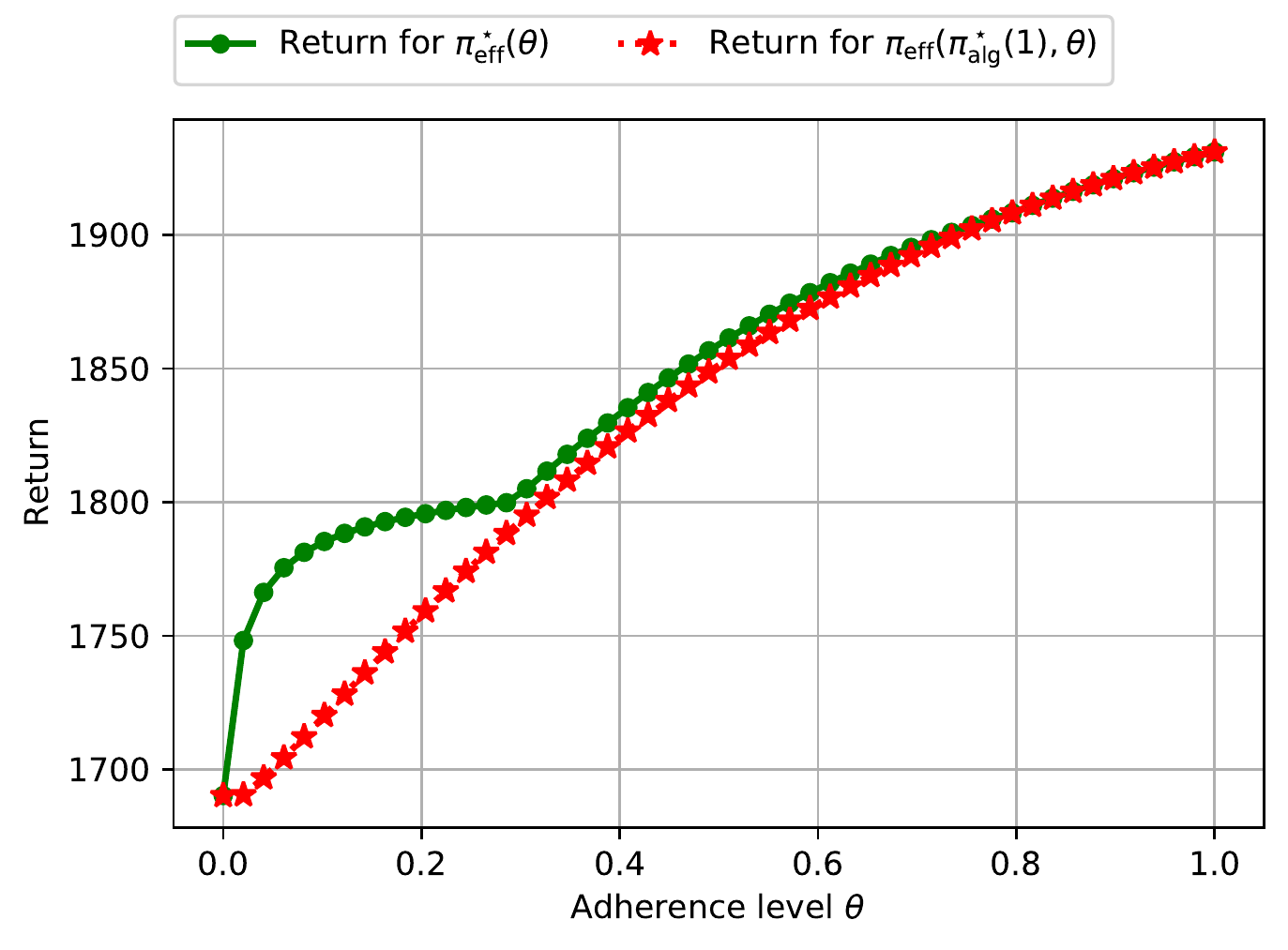}
\caption{Returns for recommending $\pi\opt_{\sf alg}(\theta)$ and $\pi\opt_{\sf alg}(1)$.}
\label{fig:machine-pi-b-2-return}
\end{subfigure}
\begin{subfigure}{0.3\textwidth}
  \includegraphics[width=1.0\linewidth]{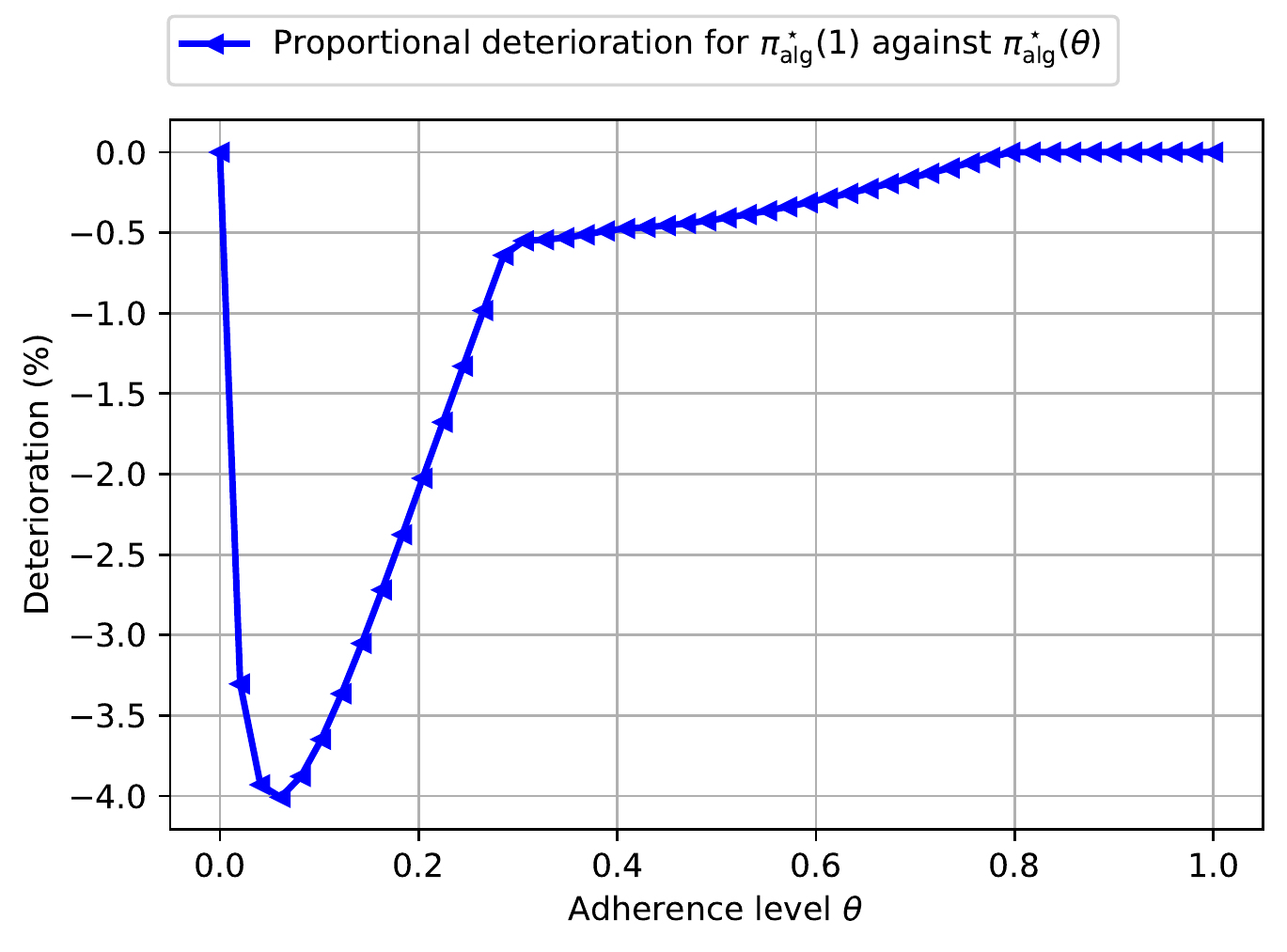}
\caption{Proportional deterioration for $\pi\opt_{\sf alg}(1)$ against $\pi\opt_{\sf alg}(\theta)$.}
\label{fig:machine-pi-b-2-decrease}
\end{subfigure}
\begin{subfigure}{0.3\textwidth}
  \includegraphics[width=1.0\linewidth]{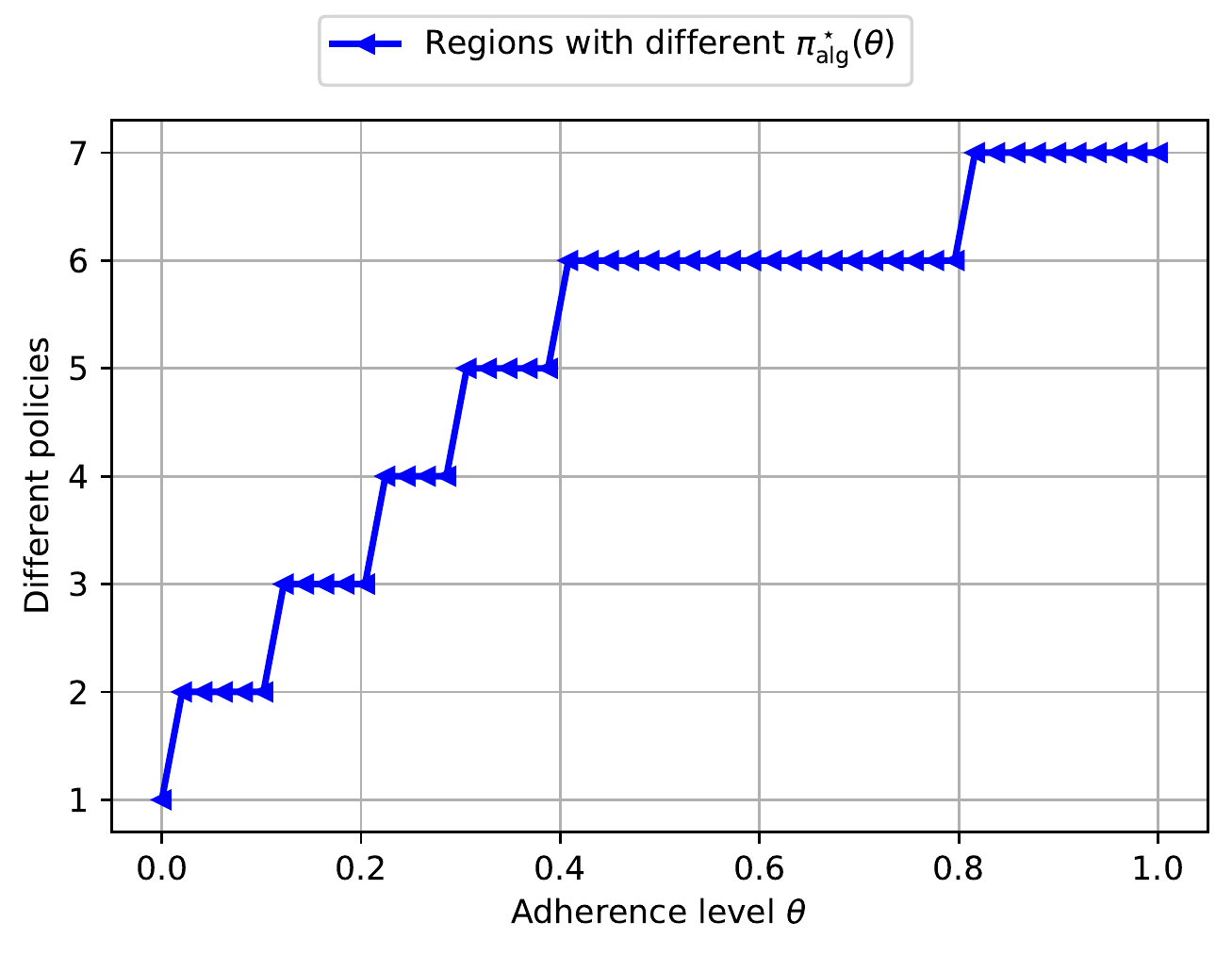}
\caption{Subregions with constant recommendation policies.}
\label{fig:machine-pi-b-2-regions}
\end{subfigure}
\end{center}
\caption{Numerical results for the machine replacement MDP with $\pi_{\sf base}$ repairing in the absorbing states $8,R_{1}$ and waiting in the other states.}
\label{fig:machine-pi-b-2-comparison}
\end{figure}

We also study the impact of the adherence level when $\pi_{\sf base}$ is the policy that avoids being trapped in the ``bad'' states (States $8,R_{1},R_{2}$). In particular, let us consider a policy $\pi_{\sf base}$ that always waits when the machine is not broken (State $1$ to State $7$) or in the normal repair state (State $R_{2}$), but chooses to repair in State $8$ and in the long repair state (State $R_{1}$). The numerical results are presented in Figure \ref{fig:machine-pi-b-2-comparison}. In this case, we see that the performance of $\pi_{\sf alg}\opt(1)$ are robust for $\theta \geq 0.35$, with a proportional deterioration of only $0.5 \%$ compared to the return of the optimal recommendation policy $\pi_{\sf alg}\opt(\theta)$ (Figure \ref{fig:machine-pi-b-2-decrease}). However, for $\theta \leq 0.35$, there is a significant drop in performance, leading to a  $4.01 \%$ reduction in effective return. 
\subsection{Stylized healthcare decision problem}\label{sec:simu-healthcare}
\begin{figure}[h]
\begin{center}
\begin{subfigure}{0.32\textwidth}
  \includegraphics[width=0.9\linewidth]{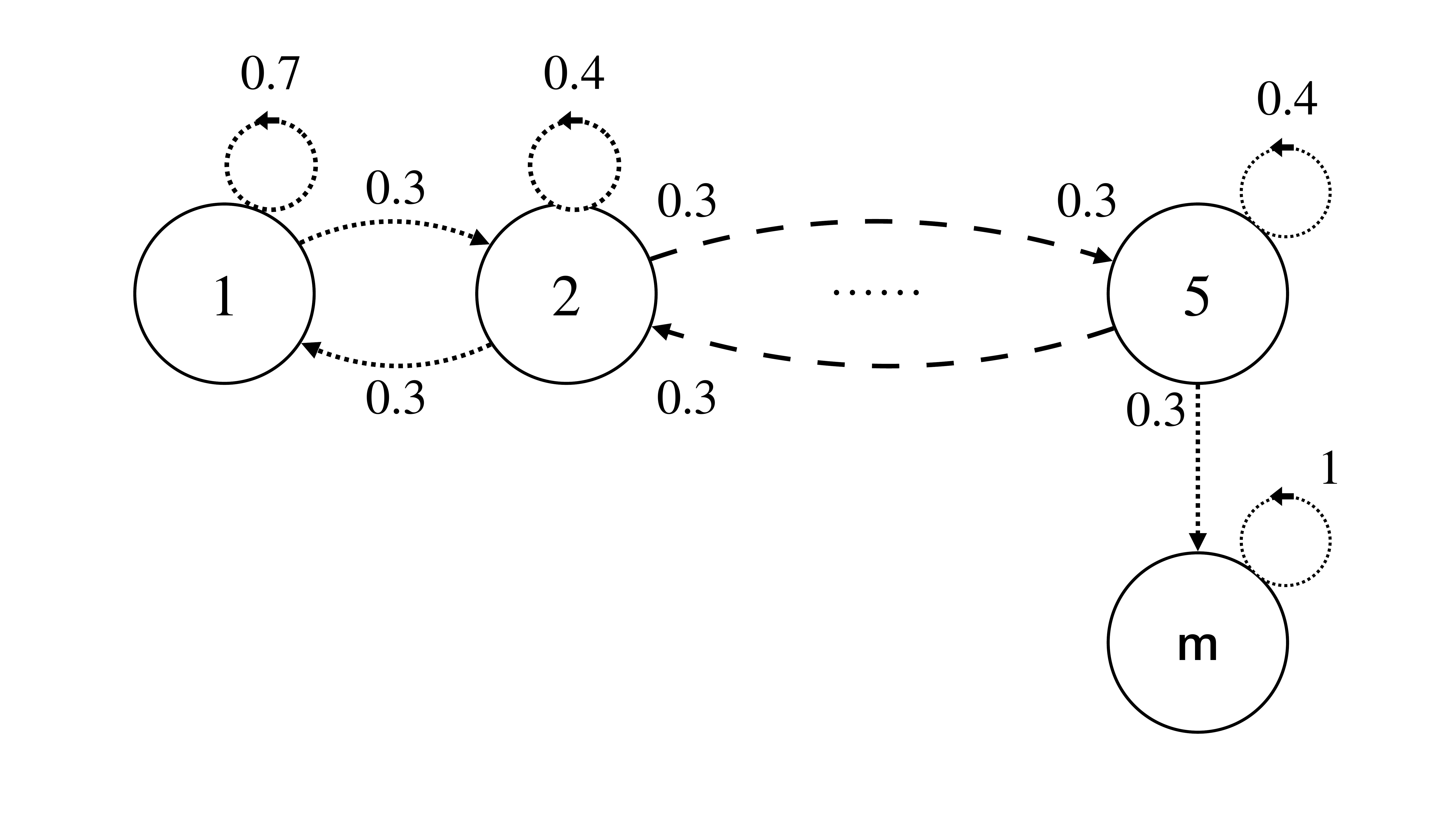}
\caption{Transition probabilities for \\ action {\sf low}.}
\label{fig:healthcare-mdp-wait}
\end{subfigure}
\begin{subfigure}{0.32\textwidth}
  \includegraphics[width=0.90\linewidth]{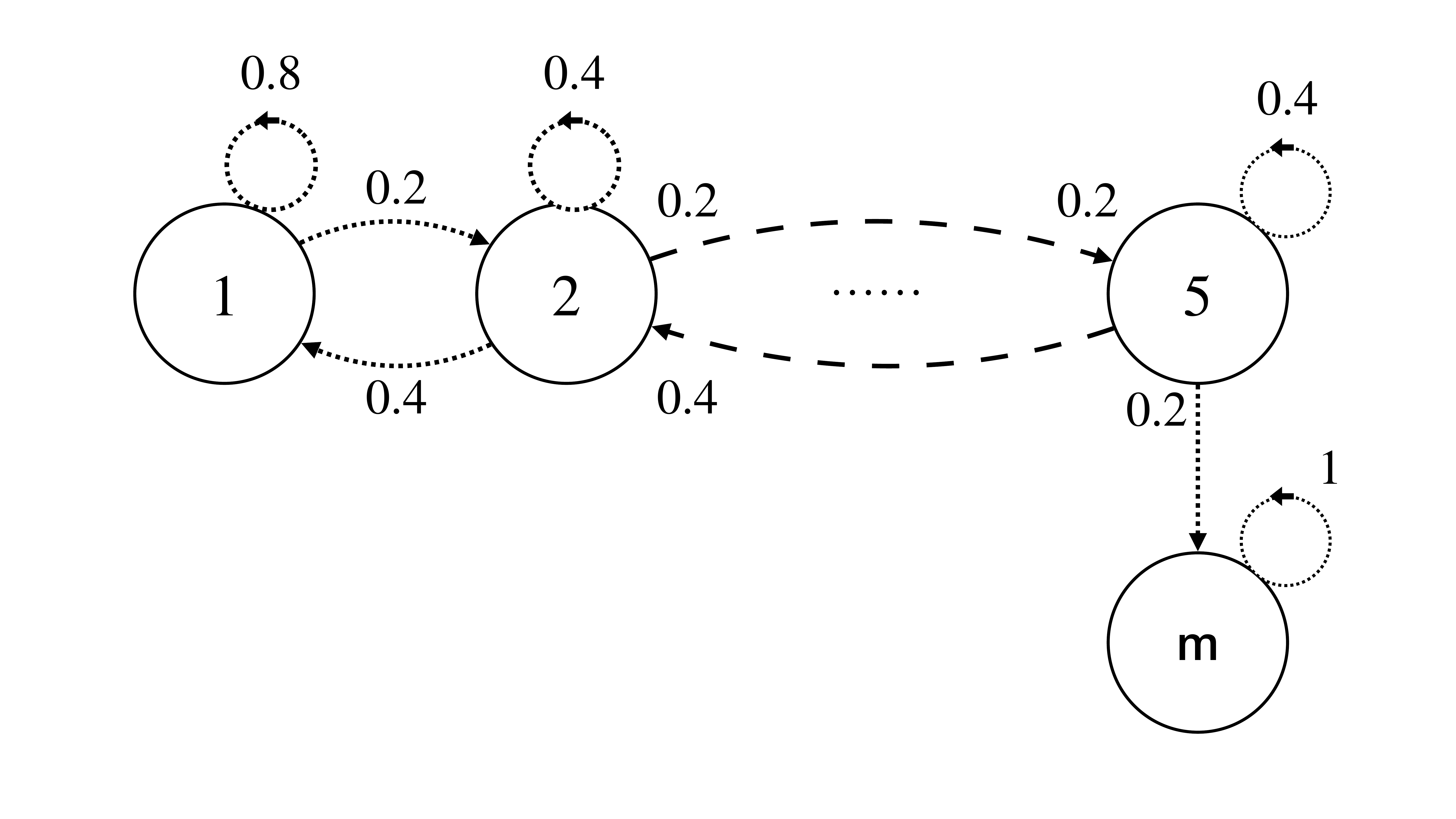}
\caption{Transition probabilities for \\ action {\sf medium}.}
\label{fig:healthcare-mdp-low}
\end{subfigure}
\begin{subfigure}{0.32\textwidth}
  \includegraphics[width=0.9\linewidth]{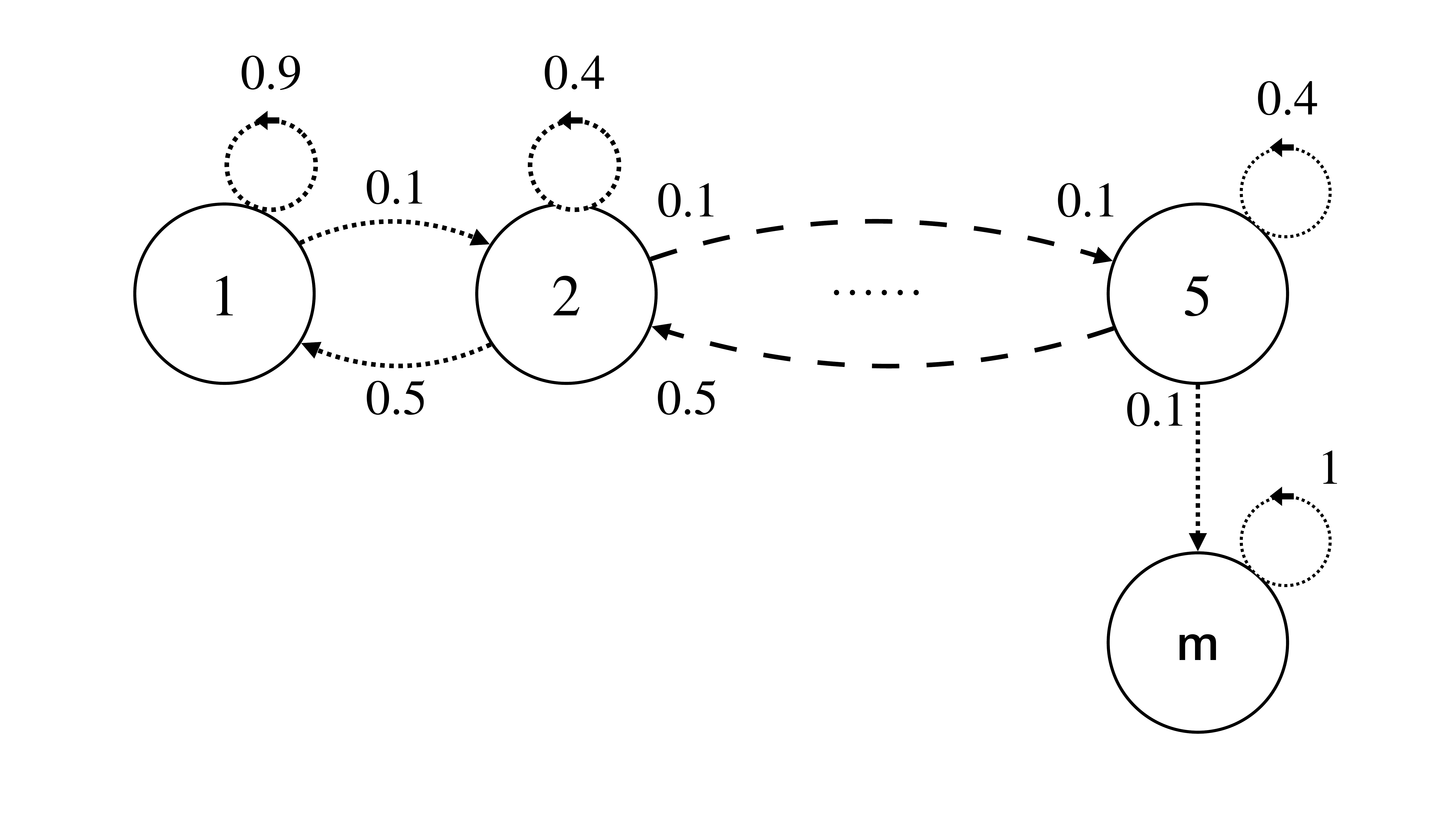}
\caption{Transition probabilities for \\ action {\sf high}.}
\label{fig:healthcare-high}
\end{subfigure}
\end{center}
\caption{Transition probabilities for the healthcare MDP instance.}
\label{fig:healthcare-mdp}
\end{figure}

\begin{figure}[h]
\begin{center}
\begin{subfigure}{0.3\textwidth}
  \includegraphics[width=1.0\linewidth]{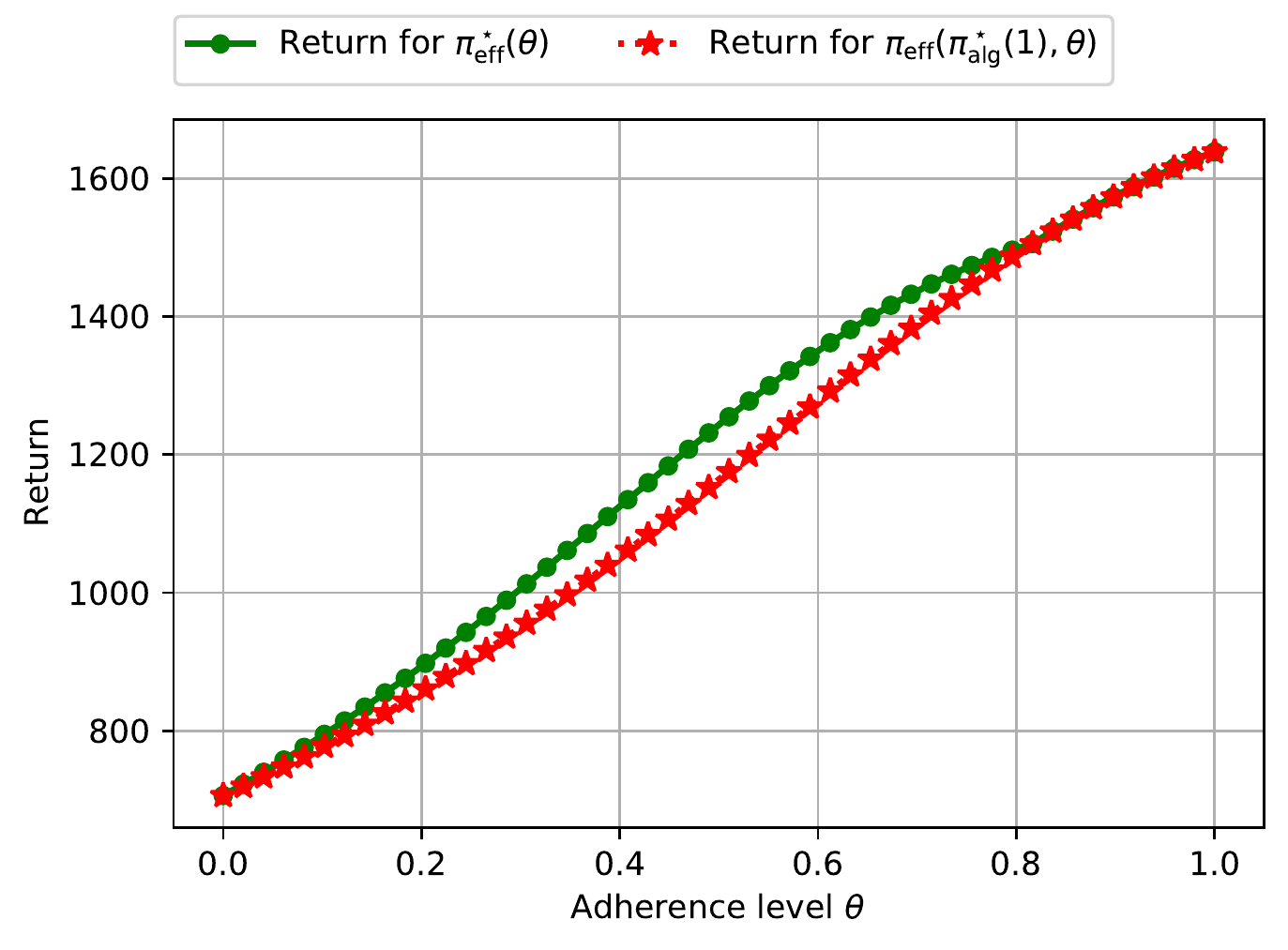}
\caption{Returns for recommending $\pi\opt_{\sf alg}(\theta)$ and $\pi\opt_{\sf alg}(1)$.}
\label{fig:healthcare-pi-b-low-return}
\end{subfigure}
\begin{subfigure}{0.3\textwidth}
  \includegraphics[width=1.0\linewidth]{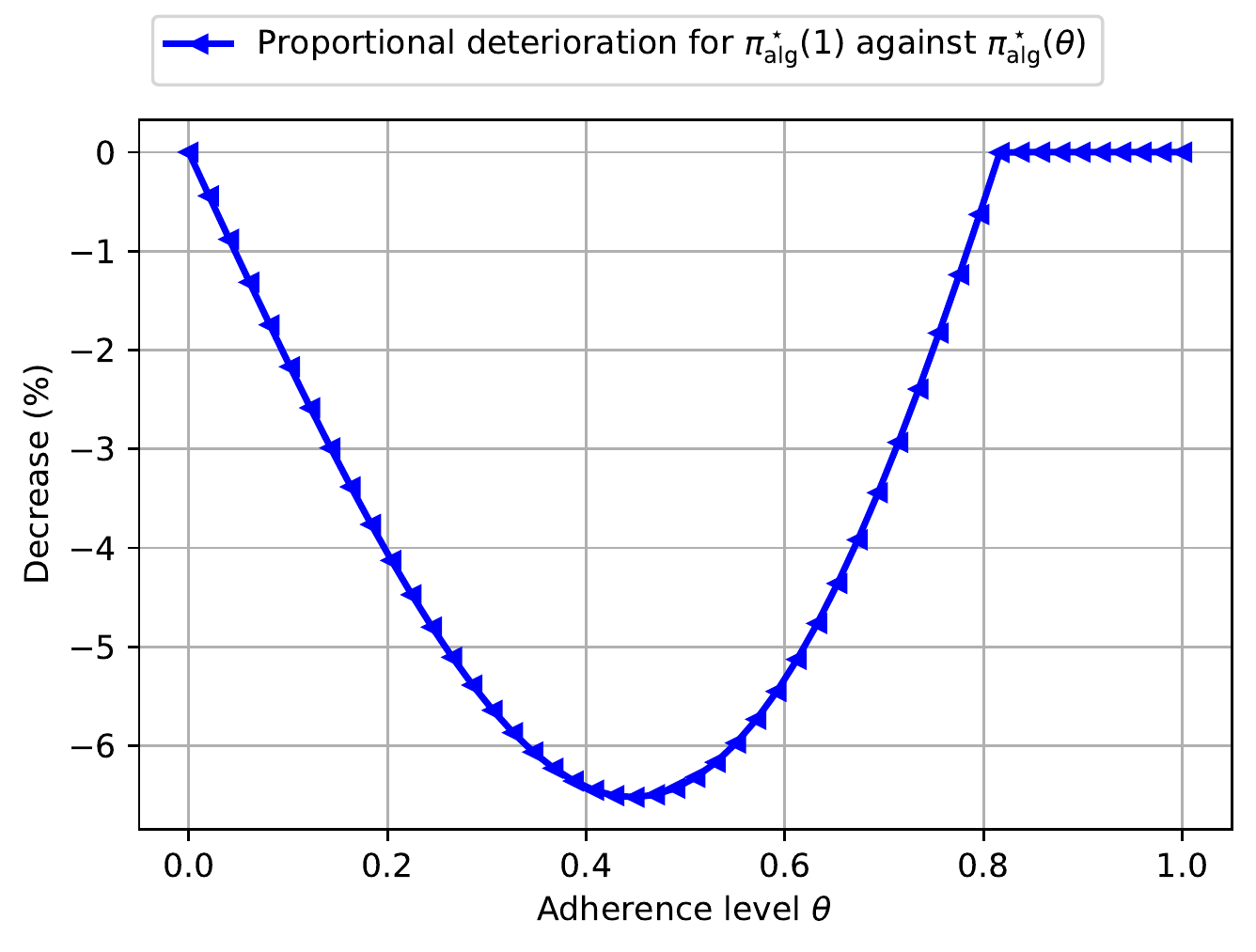}
\caption{Proportional deterioration for $\pi\opt_{\sf alg}(1)$ against $\pi\opt_{\sf alg}(\theta)$.}
\label{fig:healthcare-pi-b-low-decrease}
\end{subfigure}
\begin{subfigure}{0.3\textwidth}
  \includegraphics[width=1.0\linewidth]{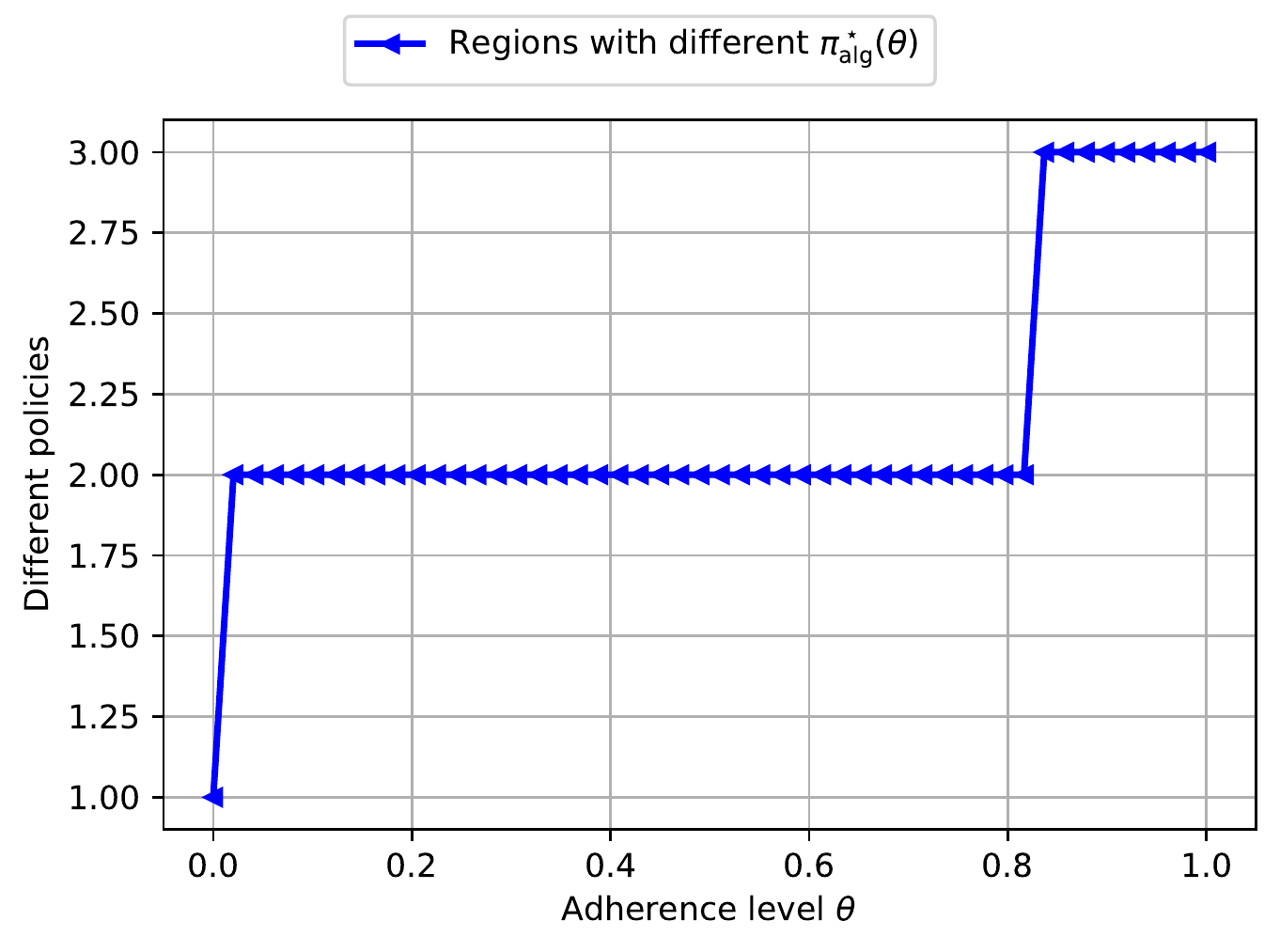}
\caption{Subregions with constant recommendation policies.}
\label{fig:healthcare-pi-b-low-regions}
\end{subfigure}
\end{center}
\caption{Numerical results for the healthcare MDP with $\pi_{\sf base}$ choosing action {\sf low} in all states.}
\label{fig:healthcare-mdp-pi-b-low}
\end{figure}

\begin{figure}[h]
\begin{center}
\begin{subfigure}{0.3\textwidth}
  \includegraphics[width=1.0\linewidth]{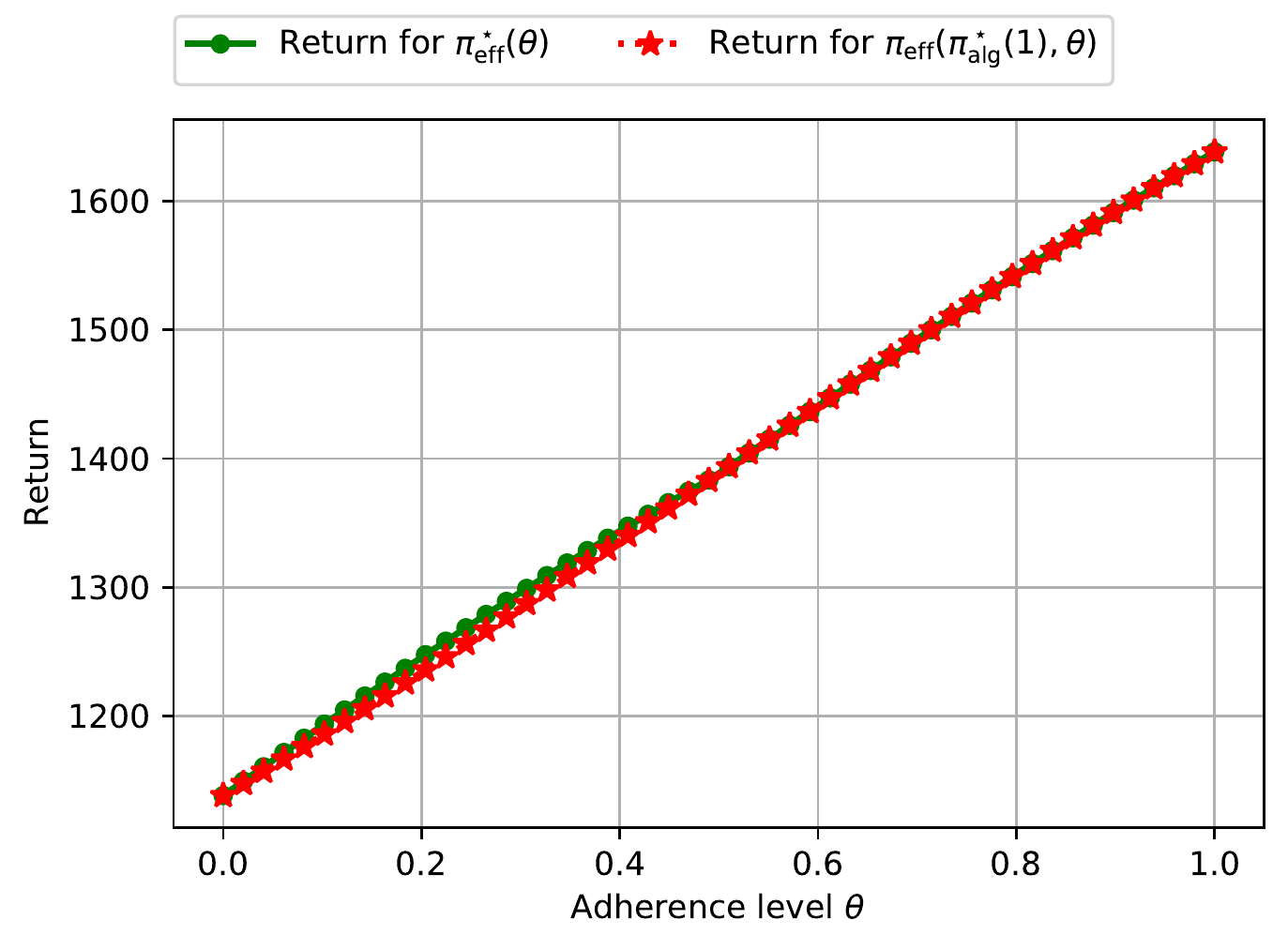}
\caption{Returns for recommending $\pi\opt_{\sf alg}(\theta)$ and $\pi\opt_{\sf alg}(1)$.}
\label{fig:healthcare-pi-b-med-return}
\end{subfigure}
\begin{subfigure}{0.3\textwidth}
  \includegraphics[width=1.0\linewidth]{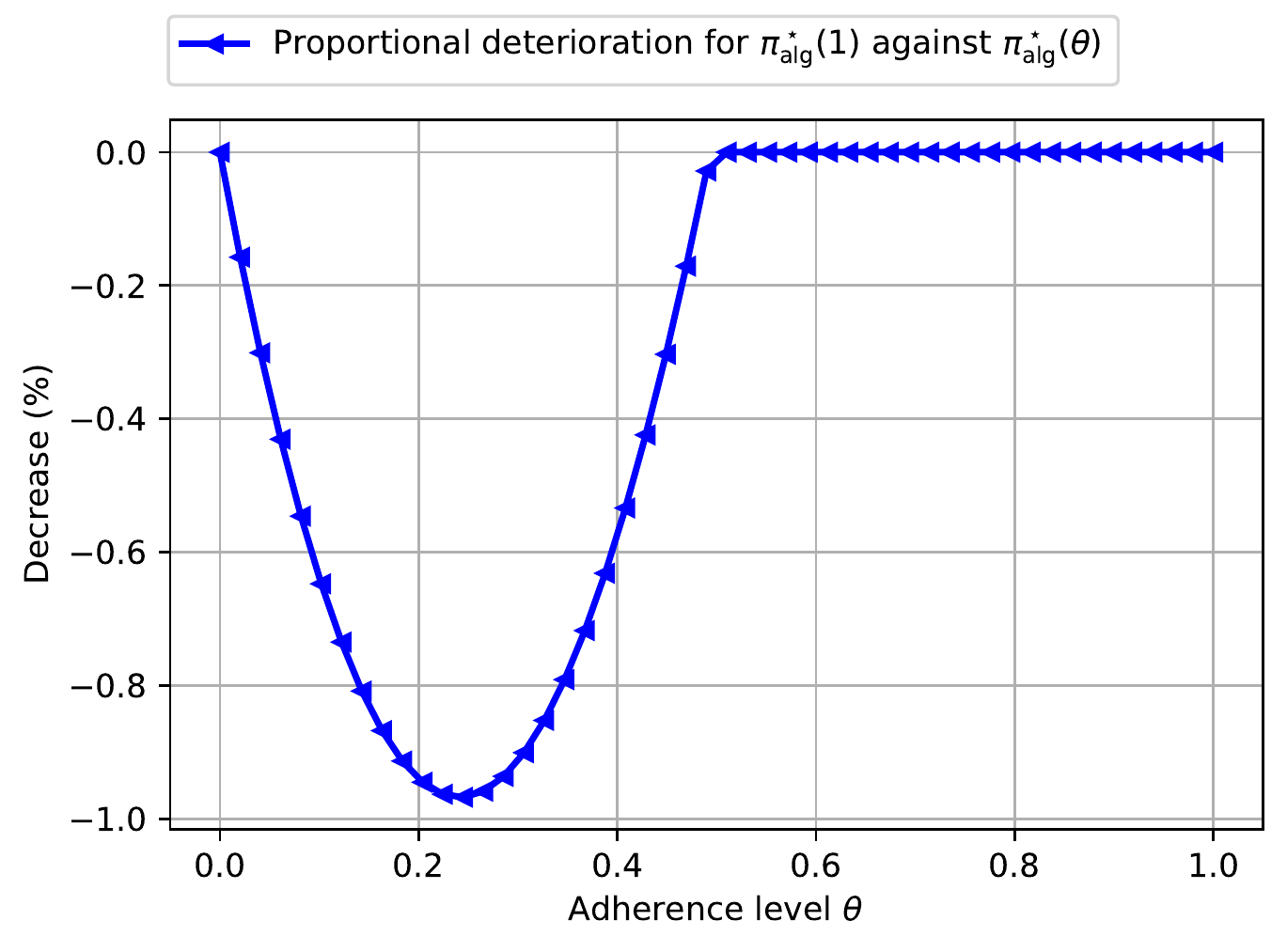}
\caption{Proportional deterioration for $\pi\opt_{\sf alg}(1)$ against $\pi\opt_{\sf alg}(\theta)$.}
\label{fig:healthcare-pi-b-med-decrease}
\end{subfigure}
\begin{subfigure}{0.3\textwidth}
  \includegraphics[width=1.0\linewidth]{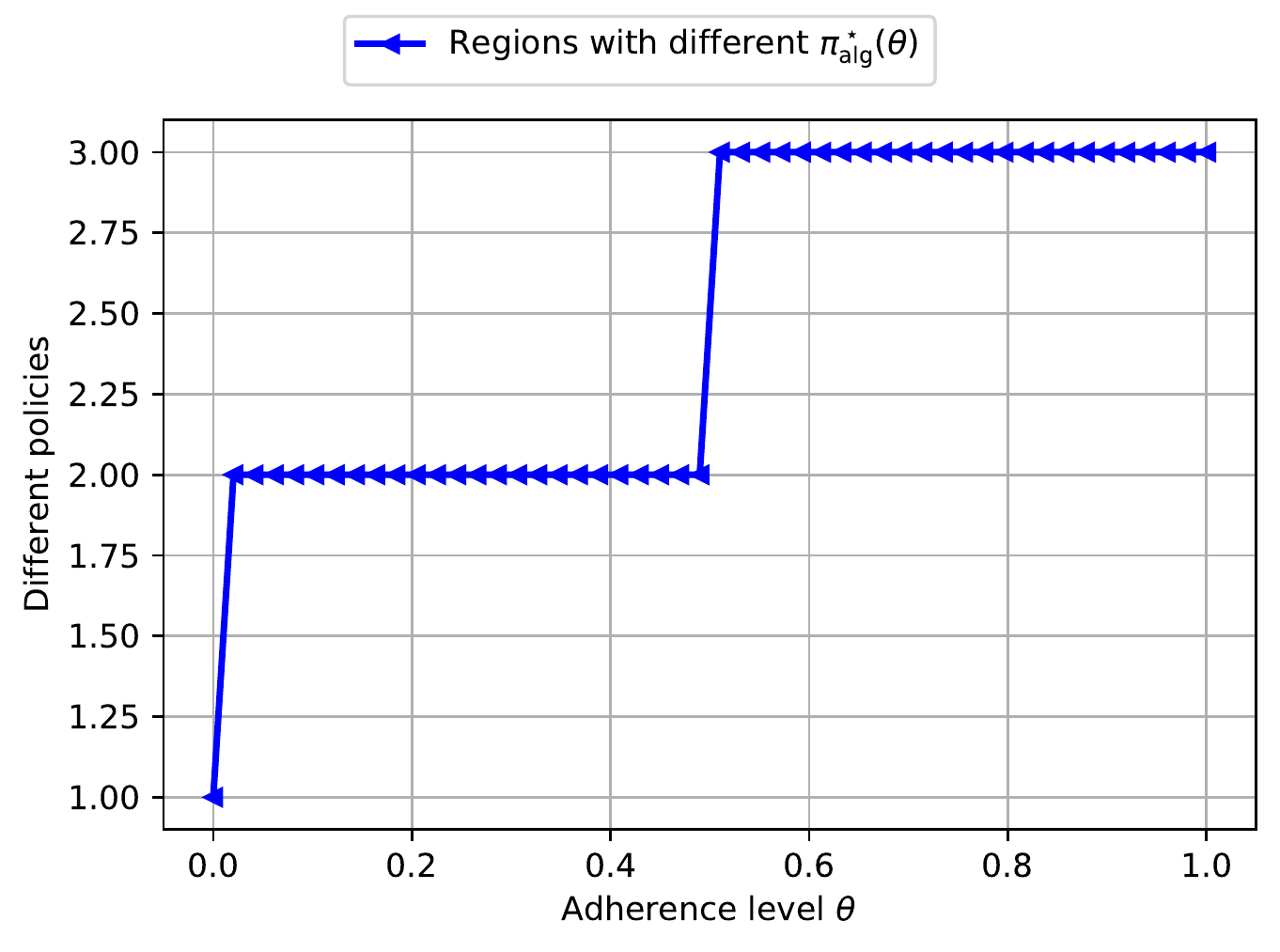}
\caption{Subregions with constant recommendation policies.}
\label{fig:healthcare-pi-b-med-regions}
\end{subfigure}
\end{center}
\caption{Numerical results for the healthcare MDP with $\pi_{\sf base}$ choosing action {\sf medium} in all states.}
\label{fig:healthcare-mdp-pi-b-med}
\end{figure}

\begin{figure}[h]
\begin{center}
\begin{subfigure}{0.3\textwidth}
  \includegraphics[width=1.0\linewidth]{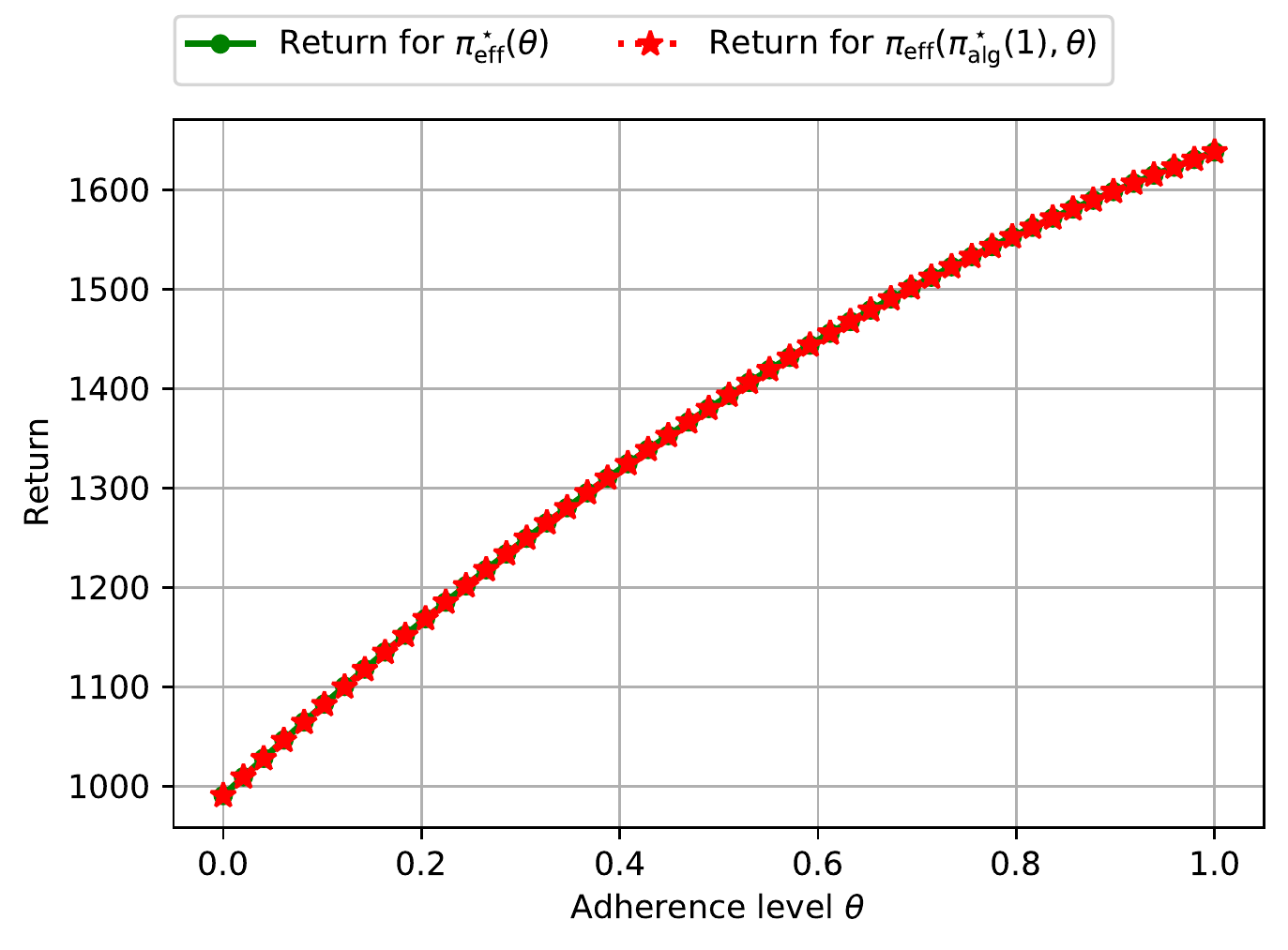}
\caption{Returns for recommending $\pi\opt_{\sf alg}(\theta)$ and $\pi\opt_{\sf alg}(1)$.}
\label{fig:healthcare-pi-b-high-return}
\end{subfigure}
\begin{subfigure}{0.3\textwidth}
  \includegraphics[width=1.0\linewidth]{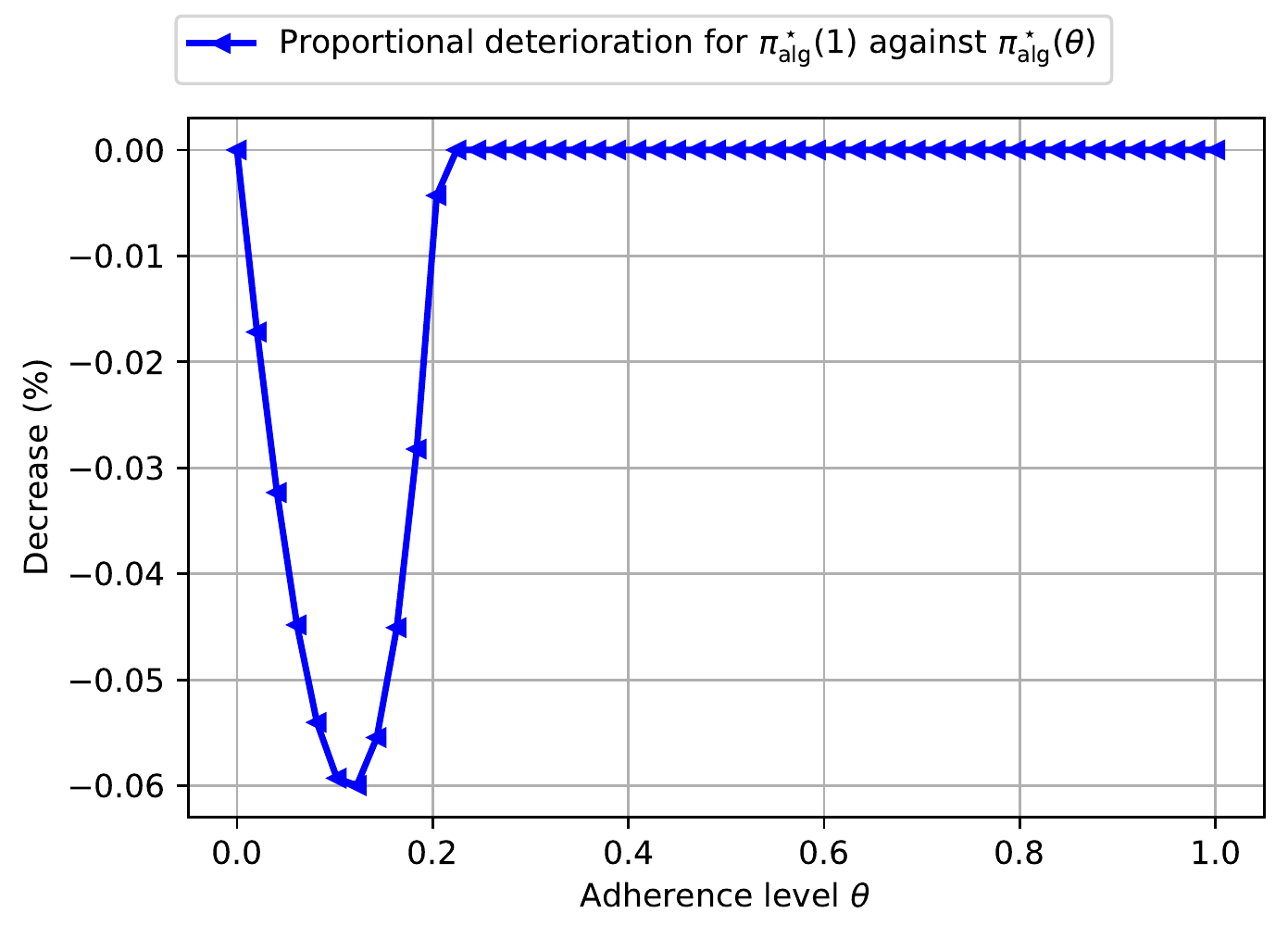}
\caption{Proportional deterioration for $\pi\opt_{\sf alg}(1)$ against $\pi\opt_{\sf alg}(\theta)$.}
\label{fig:healthcare-pi-b-high-decrease}
\end{subfigure}
\begin{subfigure}{0.3\textwidth}
  \includegraphics[width=1.0\linewidth]{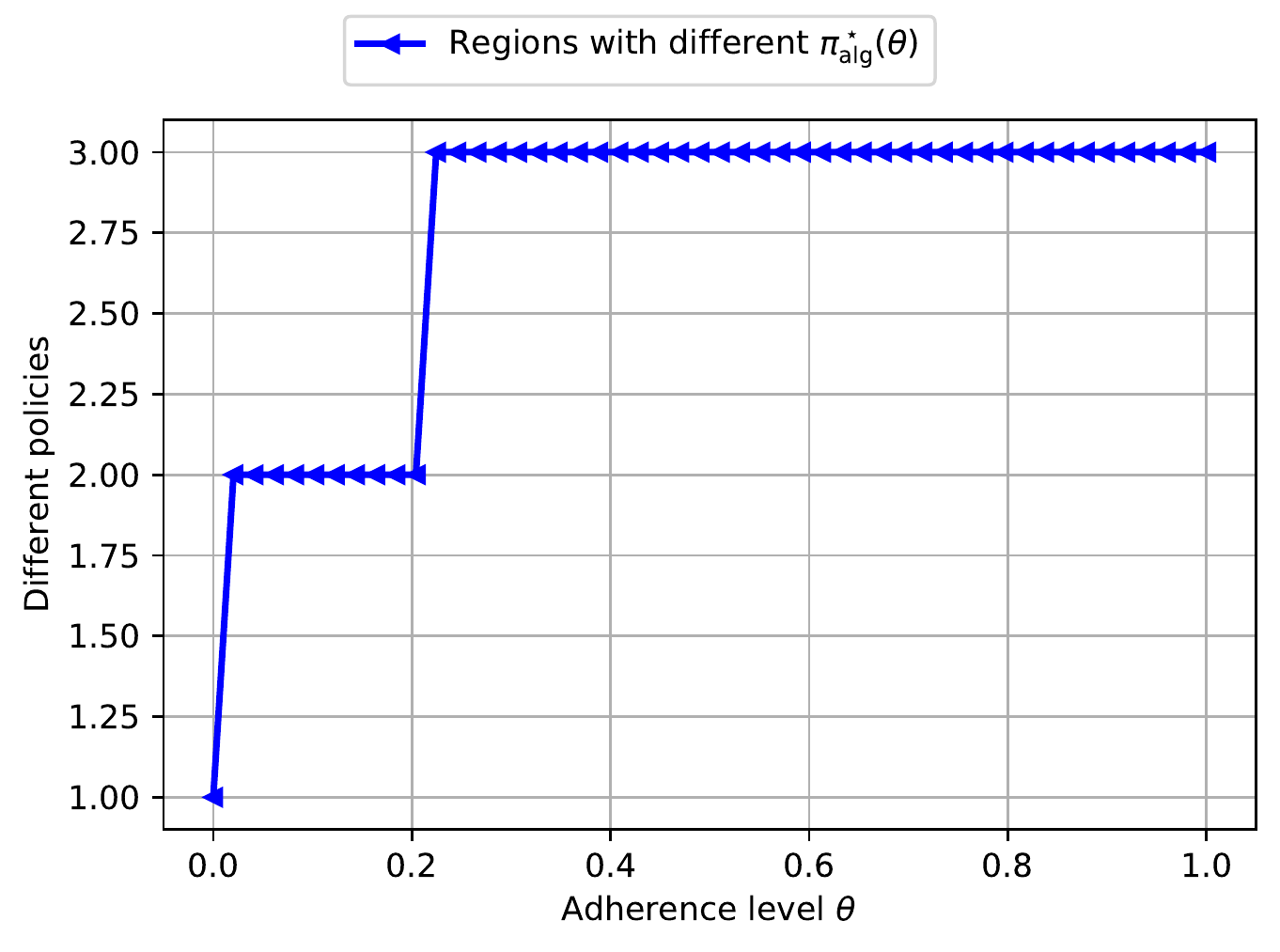}
\caption{Subregions with constant recommendation policies.}
\label{fig:healthcare-pi-b-high-regions}
\end{subfigure}
\end{center}
\caption{Numerical results for the healthcare MDP with $\pi_{\sf base}$ choosing action {\sf high} in all states.}
\label{fig:healthcare-mdp-pi-b-high}
\end{figure}

We consider an MDP instance inspired from sequential decision-making in healthcare. In particular, we approximate the evolution of the patient's health dynamics using a Markov chain, using a simplification of the models in \cite{goh2018data} and \cite{grand2022robust}. 
\paragraph{MDP instance.} The dynamics of the MDP is represented in Figure \ref{fig:healthcare-mdp}. There are $5$ states representing the severity of the health condition of the patient, and an absorbing {\em mortality} state {\sf m}. State $1$ represents a healthy condition for the patient while State $5$ is more likely to lead to mortality. There are three actions $\{${\sf low, medium, high}$\}$, corresponding to prescription of a given drug dosage at every state. In any given state (except mortality), there is a reward of $20$ for choosing action {\sf low}, a reward of $15$ for choosing action {\sf medium}, and a reward of $10$ for choosing action {\sf high}. There is a reward of $0$ in the mortality state {\sf m}. The goal of the decision maker is to choose a policy to keep the patient alive (by avoiding the mortality state {\sf m}) while minimizing the invasiveness of the \tbl{treatment}. We choose a discount factor of $\lambda=0.99$ and the patient starts in State $1$.
\paragraph{Numerical results.} An optimal policy $\pi_{\sf alg}\opt(1)$ is to choose action {\sf low} in States $1,2$, and to choose action {\sf high} in States $3,4,5$. We now test the robustness of $\pi_{\sf alg}\opt(1)$ to partial adherence of the patient. In particular, we consider three different baseline policies $\pi_{\sf base}$. In Figure \ref{fig:healthcare-mdp-pi-b-low}, Figure \ref{fig:healthcare-mdp-pi-b-med} and Figure \ref{fig:healthcare-mdp-pi-b-high}, we consider baseline policies $\pi_{\sf base}$ that always chooses action {\sf low}, {\sf medium} or {\sf high} in every health states, respectively. Our simulations highlights the sensitivity of the effective performance of $\pi_{\sf alg}\opt(1)$, with respect to both the baseline policy and the adherence level. In particular, while $\pi_{\sf alg}\opt(1)$ may loose up to $6.52 \%$ of the optimal effective return when the baseline policy always chooses low dosage (Figure \ref{fig:healthcare-pi-b-low-decrease}), it only \tbl{loses} a maximum of $0.97 \%$ of the optimal effective return when the baseline policy always chooses medium dosage (Figure \ref{fig:healthcare-pi-b-med-decrease}), and \tbl{loses} close to $0 \%$ of the optimal effective return when the baseline policy always chooses high dosage (Figure \ref{fig:healthcare-pi-b-high-decrease}). In addition, we observe that the range of the $\theta$-values for which $\pi_{\sf alg}\opt(1)$ is optimal differs greatly from one baseline policy to another (Figures \ref{fig:healthcare-pi-b-low-regions}-\ref{fig:healthcare-pi-b-med-regions}-\ref{fig:healthcare-pi-b-high-regions}): when $\pi_{\sf base}$ always chooses low dosage, $\pi_{\sf alg}\opt(1)$ is optimal for $\theta \geq 0.82$, whereas when $\pi_{\sf base}$ always chooses medium dosage, $\pi_{\sf alg}\opt(1)$ is optimal for $\theta \geq 0.51$, and when $\pi_{\sf base}$ always chooses high dosage, $\pi_{\sf alg}\opt(1)$ is optimal for $\theta \geq 0.20$.

\FloatBarrier

\section{Extensions and discussion}\label{sec:extension}
Finally, we discuss additional properties and potential extensions of our adherence-aware decision framework. 
\subsection{Heterogeneous adherence levels \color{black} across states} \label{sec:extension-heterogeneous}
We have restricted our previous analysis to the case of a {\em homogeneous} adherence level $\theta \in [0,1]$, common to all states $s \in \X$. However, in practice, it is possible that the adherence level differs across states. For instance, in a healthcare setting, practitioners may be more prone to overlook the algorithms' recommendations when the patient is in a critical health condition because any error may have life-threatening consequences. To model this practical consideration, we can extend our model to {\em heterogeneous adherence levels}, $\theta_{s} \in [0,1]$ for each state  $s \in \X$. In this model, at every decision period $t \in \N$ and visited state $s_{t}$, the decision maker decides to follow the recommendation policy $\pi_{\sf alg}$ (with probability $\theta_{s}$) or the baseline policy $\pi_{\sf base}$ (with probability $1-\theta_{s}$).  The effective policy $\pieff(\pi_{\sf alg},\bm{\theta})$ is now defined as 
\begin{equation}\label{eq:state-dependent effective policy}
\pieff(\pi_{\sf alg},\bm{\theta})_{s} = \theta_{s} \bm{\pi}_{{\sf alg},s} + (1-\theta_{s})\bm{\pi}_{{\sf base},s},\forall \; s \in \X.
\end{equation}
All the structural results from Section \ref{sec:structural-properties} would generalize to this simple extension. 
In particular, Proposition \ref{prop:monotonicity} still holds provided the non-decreasing property of $\theta \mapsto R(\pieff\opt(\theta))$ is replaced with an order-preserving property:
\[\theta_{s} \leq \theta'_{s}, \forall \; s \in \X \Rightarrow R(\pieff\opt(\bm{\theta})) \leq R(\pieff\opt(\bm{\theta}')).\]
Importantly, we can still efficiently find an optimal recommendation policy $\pi\opt_{\sf alg}(\bm{\theta})$ for any adherence level $\bm{\theta} \in [0,1]^{\X}$, by adapting the value iteration and the linear programming formulation to the map $f_{\bm{\theta}}:\R^{\X} \rightarrow \R^{\X}$, defined as
\[f_{\bm{\theta},s}(\bm{v}) = \max_{\bm{\pi}_{s} \in \Delta(\A)}  \theta_{s} \cdot  \sum_{a \in \A} \pi_{sa} \bm{P}_{sa}^{\top} \left( \bm{r}_{sa} + \lambda \bm{v} \right) + (1-\theta_{s}) \cdot \sum_{a \in \A} \pi_{{\sf base},sa} \bm{P}_{sa}^{\top} \left( \bm{r}_{sa} + \lambda \bm{v} \right), \forall \; s \in \X.\]

\tbl{
\subsection{Heterogeneous adherence levels across states and actions} \label{sec:action-dependent-adherence}
Furthermore, it is plausible in practice that recommendations that are close to the baseline actions are more likely to be followed than drastically different ones, e.g., in a healthcare setting where the actions correspond to drug dosages.  
To model this situation, we can extend our framework further to involve an adherence level that depends on each state $s \in \X$ and each action in $a \in \A$. Formally, we could study policies of the form 
\begin{equation*}\label{eq:wrong state-action pieff}
\pieff(\pi_{\sf alg},\bm{\theta})_{sa} = \theta_{sa} \pi_{{\sf alg},sa} + (1-\theta_{sa})\pi_{{\sf base},sa},\forall \; (s,a) \in \X \times \A.
\end{equation*}
However, for every state $s \in \X$, we need $\pieff(\pi_{\sf alg},\bm{\theta})_{s} \in \Delta(\A)$, which imposes some non-trivial restrictions on the values of $\theta_{sa}$ (which would depend on the probability of playing each action according to $\pia$ and $\pib$).

To circumvent this issue, we propose an alternative model where $\pieff(\pi_{\sf alg},\bm{\theta})_{s} \in \Delta(\A)$ by design. For the sake of simplicity, in this section, we assume that $\pib$ is a deterministic stationary policy: for each state $s \in \X$ we write $\pib(s) \in \A$ for the action chosen by the policy $\pib$. 
At a state $s \in \X$, a recommended action $a$ is sampled from the probability distribution $\bm{\pi}_{{\sf alg},s} \in \Delta(\A)$. Then with probability $\theta_{sa} \in [0,1]$ the DM follows the recommendation (action $a$), otherwise the action selected by the DM is $\pib(s)$. With this model, the effective policy $\pieff(\pia,\bm{\theta})$ for some $\left(\theta_{sa}\right)_{(s,a) \in \X \times \A} \in [0,1]^{\X \times \A}$ is such that
\begin{align*}
\pieff(\pia,\bm{\theta})_{sa} = \begin{cases} \pi_{{\sf alg},sa}\theta_{sa} & \mbox{ if } a \neq \pib(s), \\ 1-\sum_{a' \in \A \setminus \{\pib(s)\}} \pi_{{\sf alg},sa'}\theta_{sa'} & \mbox{ if }   a = \pib(s). \end{cases}
\end{align*}
Note that the expression for the case $a = \pib(s)$ simply follows from 
\begin{equation}\label{eq:proba-pieff-reformulation}
1-\sum_{a' \in \A \setminus \{\pib(s)\}} \pi_{{\sf alg},sa'}\theta_{sa'} = \pi_{{\sf alg}, s\pib(s)} + \sum_{a' \in \A \setminus \{\pib(s)\}} \pi_{{\sf alg},sa'}(1-\theta_{sa'}),
\end{equation}
i.e., action $\pib(s)$ is chosen either because it has been sampled following $\bm{\pi}_{{\sf alg},s}$ or because another action $a'$ was sampled but the decision maker chose to follow $\pib$, which happens with probability $1-\theta_{sa'}$.
We can now write the value function of a policy $\pieff(\pia,\bm{\theta})$. For any $s \in \X$, we obtain,
 using \eqref{eq:proba-pieff-reformulation}:
\[v^{\pieff(\pia,\bm{\theta})}_{s} = \sum_{a \in \A } \pi_{sa} \left( \theta_{sa} \bm{P}_{sa}^{\top}\left(\bm{r}_{sa} + \lambda \bm{v}^{\pieff(\pia,\bm{\theta})}\right) + (1-\theta_{sa})\bm{P}_{s\pib(s)}^{\top}\left(\bm{r}_{s\pib(s)} + \lambda \bm{v}^{\pieff(\pia,\bm{\theta})}\right)\right).\]
Overall, we have obtained that the value function $\bm{v}^{\pieff(\pia,\bm{\theta})}$ satisfies
\[v^{\pieff(\pia,\bm{\theta})}_{s} = \sum_{a \in \A} \pi_{sa}\left(r'_{sa} + \lambda\bm{P}'^{\top}_{sa} \bm{v}^{\pieff(\pia,\bm{\theta})}\right), \forall \; s \in \X\]
with $\bm{P}' \in \left(\Delta(\X)\right)^{\X \times \A},\bm{r}' \in \R^{\X \times \A }$ the transition probabilities and the instantaneous rewards of another surrogate MDP $\M'$ with transitions and rewards defined as
$ \bm{P}_{sa}' := \theta_{sa} \cdot \bm{P}_{sa} + (1-\theta_{sa}) \cdot \bm{P}_{s\pib(s)},
r'_{sa} :=  \theta_{sa} \cdot \bm{P}_{sa}^{\top}\bm{r}_{sa} + (1-\theta_{sa}) \cdot \bm{P}^{\top}_{s\pib(a)}\bm{r}_{s\pib(a)},$
for all $(s,a) \in \X \times \A$. This shows that for this model of state-action-dependent adherence level, we can efficiently find an optimal recommendation policy by computing an optimal (nominal) policy for the surrogate MDP $\M'$. 
}

\subsection{Uncertain adherence level}\label{sec:extension-uncertain}
In our framework, we have assumed that the adherence level $\theta \in [0,1]$ was known and used as an input to design the recommendation policy $\pi_{\sf alg}$. This assumption is likely violated in practice, where $\theta$ is not perfectly known. Instead, we can assume that the true adherence level $\theta$
is uncertain but belongs to an interval $[ \underline{\theta},\bar{\theta}]$. 
 Under this assumption, we take a  {\em robust optimization} approach~\citep{bertsimas2004price,ben2009robust} and model the uncertainty in the value of $\theta$ as an adversarial choice from the set $[ \underline{\theta},\bar{\theta}]$ of all possible realizations. The goal is to compute an optimal {\em robust} recommendation policy, that optimizes the worst-case objective over all plausible values of the adherence levels:
\begin{equation}\label{eq:max-min-MDP}
\sup_{\pia \in \Pi_{\sf H}} \: \min_{\theta \in [ \underline{\theta},\bar{\theta}]} \, R(\pieff(\pi_{\sf alg},\theta)).
\end{equation}
The optimization problem \eqref{eq:max-min-MDP} is reminiscent to {\em robust MDPs}, which consider the case where the rewards and/or the transition probabilities are unknown~\citep{iyengar2005robust,wiesemann2013robust}, but in our setting the same adherence level $\theta$ has an impact on the transition probabilities out of every states $s \in \X$ in the surrogate MDP, which contradicts the classical rectangularity assumption for robust MDPs.
 However,  thanks to the structural properties highlighted in Section \ref{sec:structural-properties}, the optimization problem \eqref{eq:max-min-MDP} can be solved as efficiently as \ref{eq:definition-ada-mdp}, the adherence-aware decision-making problem with known adherence level $\theta$. Crucially, an optimal recommendation policy can still be chosen stationary (i.e., in the set $\Pi$) instead of history-dependent (i.e., in the set $\Pi_{\sf H}$), and deterministic. 
Formally, we have the following theorem (proof detailed in Appendix \ref{app:proof-th-robust-theta}):
\begin{theorem}\label{th:robust-to-nominal}
An optimal robust recommendation policy in \eqref{eq:max-min-MDP} may be chosen stationary:
\[\sup_{\pia \in \Pi_{\sf H}} \:  \min_{\theta \in [ \underline{\theta},\bar{\theta}]} \, R(\pieff(\pi_{\sf alg},\theta)) = \max_{\pi_{\sf alg} \in \Pi} \:  \min_{\theta \in [ \underline{\theta},\bar{\theta}]} \, R(\pieff(\pi_{\sf alg},\theta)).\]
Additionally, the pair $\left(\pi_{\sf alg}\opt(\underline{\theta}),\underline{\theta}\right)$ with $\pi_{\sf alg}\opt(\underline{\theta})$ a deterministic policy is an optimal solution to \eqref{eq:max-min-MDP}.
\end{theorem}
Theorem \ref{th:robust-to-nominal} is remarkable in that it shows that the same value of $\theta$ (in particular, the most pessimistic value $\underline{\theta}$) is attaining the worst-case return for all policies. In practice, it reduces the problem of estimating the true adherence level to the (admittedly easier) task of obtaining a valid lower bound only. Furthermore, Theorem \ref{th:robust-to-nominal}  also has significant computational impact since it shows that solving \eqref{eq:max-min-MDP} can be done by applying the same algorithms as the one described in Section \ref{sec:algorithms} with $\theta = \underline{\theta}$. The resulting recommendation will also be a deterministic policy, which is desirable in practice. The proof is very similar to the case of time- and state-invariant adversarial adherence decision in Theorem \ref{th:model.equiv.informal} and we present it in Appendix \ref{app:proof-th-robust-theta}.
\tbl{
\subsection{Uncertain baseline policy}\label{sec:uncertain-baseline-policy}
Similarly, the baseline policy $\pib$ is currently a known input to our adherence-aware MDP framework. However, in practice it is possible that the algorithm only has access to an estimation $\hat{\pi}_{{\sf base}}$ of the baseline policy, learned from a finite dataset, and that the true baseline policy differs from $\hat{\pi}_{{\sf base}}$. We consider a robust approach where the recommendation policy optimizes over the worst-case baseline policy $\pib \in \Gamma$, where the set $\Gamma \subseteq \left(\Delta(\A)\right)^{\X}$ represents feasible baseline policies that are close to the estimation $\hat{\pi}_{{\sf base}}$, i.e., we consider 
\begin{equation}\label{eq:robust-ada-mdp}
\sup_{\pia \in \Pi_{\sf H}} \min_{\pib \in \Gamma} R(\theta \pia + (1-\theta) \pib).
\end{equation}
The following theorem shows that \eqref{eq:robust-ada-mdp} is still a tractable optimization problem under some mild assumption on $\Gamma$. We provide the detailed proof in Appendix \ref{app:proof-robust-ada=robust-mdp}.
\begin{theorem}\label{th:robust-ada-mdp=robust-mdp}
Assume that the set of feasible baseline policies $\Gamma$ satisfies the following {\em rectangularity assumption}: $
\Gamma = \times_{s \in \A} \Gamma_{s}$  where $\Gamma_{s} \subseteq \Delta(\A)$ is a convex, compact set for each $s \in \X$. Then an optimal solution to \eqref{eq:robust-ada-mdp} exists and can be chosen stationary. Additionally, if the set $\Gamma$ is a polytope or defined with conic constraints, then an optimal solution to \eqref{eq:robust-ada-mdp} can be computed efficiently.
\end{theorem}
Our proof is based on showing that the optimization problem \eqref{eq:robust-ada-mdp} can be reformulated as an s-rectangular robust MDP~\citep{wiesemann2013robust} with uncertain pair $(\bm{r},\bm{P})$ of instantaneous rewards and transition probabilities. This follows from the interpretation of \ref{eq:definition-ada-mdp} as solving a surrogate MDP, where the rewards and transitions, defined in \eqref{eq:surrogate-mdp}, are dependent on $\pib$.
}
\subsection{Varying adherence level}\label{sec:extension-varying}
The adherence level $\theta$ may also vary over time. As the DM observes the recommendation made by the algorithm over time, her trust in the recommendation, hence her adherence, may increase (or decrease). 

One could endogeneize these dynamics by making $\theta$ explicitly dependent on the recommended policy $\pi_{\sf alg}$. However, the works of \citet{boyaci2020human,devericourt2022your} highlight how complex these dynamics can be, even for highly stylized decision problems, because of cognitive limitations and asymmetric performance evaluation. Therefore, we conjecture that such game-theoretic approaches (where $\pi_{\sf alg}$ and $\theta$ are updated at each step) would be intractable for the type of complex multi-stage decision problems we consider in this paper. Furthermore, as discussed in Section \ref{sec:litreview}, many mechanisms could explain partial adherence. Consequently, any method that restricts the reasons for non-adherence (e.g., information asymmetry, algorithm aversion, cognitive limitations) and derives update rules for the adherence level $\theta$ based on these mechanisms could suffer from model misspecification. 

Alternatively, one could capture the dynamic nature of $\theta$ by estimating it from past observations in an online fashion. At a high-level, the optimization problem to which $\pi_{\sf alg}^\star(\theta)$ is a solution resembles that of an MDP whose transition probabilities depend on $\theta$ (and $\pi_{\sf base}$). Hence, a varying adherence level would lead to non-stationary transition probabilities. In the multi-armed bandit literature, two types of assumptions are used to address non-stationarity. \citet{garivier2011upper} introduced a piecewise stationary assumption, where the parameters are constant over certain time periods and change at unknown time steps. Alternatively, \citet{besbes2014stochastic,besbes2015non} considered a slowly varying setting where the absolute difference between parameters at two consecutive time-steps are bounded (by a so-called variation budget). 
Although originally derived for multi-armed bandit problems, both these frameworks have been extended and used to solve non-stationary MDPs (or non-stationary reinforcement learning problems) as well. We refer to \citet{auer2008near} and \citet{cheung2019non} for an analysis of non-stationary MDPs under the piecewise stationary and slowly varying assumptions respectively. \tbl{Beyond the technical difficulties addressed by the aforementioned works, learning $\theta$ from past historical data also suffers from a censorship issue: if both $\pi_{\sf alg}$ and $\pi_{\sf base}$ recommend the same action at a given  state $s_{t}$, then it is impossible to distinguish adherence from non-adherence. 

We see our model based on partial adherence in {\em offline} sequential decision-making as a first step towards a better understanding of the phenomena arising in expert-in-loop systems and a better design of algorithmic recommendations. \tb{The online extension of our framework, where the adherence level (and potentially the baseline policy $\pi_{\sf base}$) needs to be continuously learned from past observations constitutes an interesting future direction, as well as the case where the real MDP parameters $(\bm{r},\bm{P})$ themselves are only partially known to the human agent and the algorithm and must be learned over time.}
}
\section*{Acknowledgements}
We would like to thank the three anonymous reviewers and the associate editor for their insightful comments that lead to a more complete version of the paper.
\bibliographystyle{plainnat}
\bibliography{ref}
\newpage
\begin{APPENDICES}

\tbl{

\section{Proof of Theorem \ref{th:random tv is equivalent to adamdp}}\label{app:proof random tv}
\proof{Proof of Theorem \ref{th:random tv is equivalent to adamdp}.}
\tb{Let us assume that the random variables $(u_{s,t})_{s,t}$ are such that $u_{s,t}$ an $u_{s',t'}$ are independent for any $t \neq t'$, and that for any $t \in \N$, $\E_{u}\left[ \left(u_{s,t}\right)_{s \in \X} \right] = (\theta,...,\theta) \in [0,1]^{\X}$.
We prove that $\max_{\pia \in \Pi} \:  R\left( \theta \pia + (1-\theta)\pib \right) 
= \sup_{\pia \in \Pi_{\sf H}}  \: \E_{u} \left[ R(\pieff(\pia,u)) \right]$.}
The proof proceeds in three steps.
\paragraph{Step 1.}
We first show that we can restrict ourselves to Markovian policies: $\sup_{\pia \in \Pi_{\sf H}}  \: \E_{u} \left[ R(\pieff(\pia,u)) \right] = \sup_{\pia \in \Pi_{\sf M}}  \: \E_{u} \left[ R(\pieff(\pia,u)) \right]$.
Note that for some fixed values of $u \in [0,1]^{\X \times \N}$, the map $\pia \mapsto  R(\pieff(\pia,u))$ is a function of the values of $\PP^{\pia}\left(s_{t}=s,a_{t}=a\right)$ for $(s,a) \in \X \times \A$ and $t \in \N$. 
Following \citet[corollary 5.5.2,][]{puterman2014markov}, for any history-dependent policy $\pia \in \Pi_{\sf H}$, there exists a Markovian policy $\pia' \in \Pi_{\sf M}$, potentially randomized, such that for any pair $(s,a) \in \X \times \A$ and any time $t \in \N$, we have $\PP^{\pia}\left(s_{t}=s,a_{t}=a\right) = \PP^{\pia'}\left(s_{t}=s,a_{t}=a\right).$
Therefore, for any history-dependent policy $\pia \in \Pi_{\sf H}$, we can find a Markovian policy $\pia'$ such that $  \E_{u} \left[ R(\pieff(\pia,u)) \right] =  \E_{u} \left[ R(\pieff(\pia',u)) \right].$ From this we conclude that $\sup_{\pia \in \Pi_{\sf H}}  \: \E_{u} \left[ R(\pieff(\pia,u)) \right] = \sup_{\pia \in \Pi_{\sf M}}  \: \E_{u} \left[ R(\pieff(\pia,u)) \right].$ 
\paragraph{Step 2.}
We now show that for any $\pia \in \Pi_{\sf M}$, we have $ \E_{u} \left[R(\pieff(\pia,u))\right] = R(\pieff(\pia,\theta))$. Indeed, let us define, for $\pi \in \left(\Delta(\A)\right)^{\X}$, the transition matrix $ \bm{P}^{\pi} \in \R^{\X \times \X} $ as $ P^{\pi}_{ss'} = \sum_{a \in \A} \pi_{sa}P_{sas'}, \forall \; (s,s') \in \X \times \X$ and $\bm{r}_{\pi} \in \R^{\X}, r_{\pi,s} = \sum_{a \in \A} \pi_{sa}\bm{P}_{sa}^{\top}\bm{r}_{sa}.$ 
Note that $\bm{P}^{\pi} $ and $\bm{r}_\pi$ depend linearly on $\pi$.
Then by definition we have, for a Markovian policy $\pi = \left(\pi_{t}\right)_{t \in \N}$ with $\pi_{t} \in \left(\Delta(\A)\right)^{\X}$ for $t \in \N$,
\[R(\pi) = \E^{\pi} \left[ \sum_{t=0}^{+\infty} \lambda^{t}r_{s_{t}a_{t}s_{t+1}} \right] = \bm{p}_{0}^{\top}\left( \sum_{t=0}^{+\infty} \lambda^{t} \prod_{t'=0}^{t-1} \bm{P}^{\pi_{t'}}\bm{r}_{\pi_{t}} \right).\]
We have
\begin{align}
\E_{u} \left[ R(\pieff(\pia,u))\right] & = \bm{p}_{0}^{\top}\left( \E_{u} \left[ \sum_{t=0}^{+\infty} \lambda^{t} \prod_{t'=0}^{t-1} \bm{P}^{(\pieff(\pia,u))_{t'}}\bm{r}_{(\pieff(\pia,u))_{t}} \right] \right) \nonumber \\
& = \bm{p}_{0}^{\top}\left( \sum_{t=0}^{+\infty} \lambda^{t} \E_{u} \left[ \prod_{t'=0}^{t-1} \bm{P}^{(\pieff(\pia,u))_{t'}}\bm{r}_{(\pieff(\pia,u))_{t}} \right] \right) \label{eq:rdv-prf-1}  \\
& = \bm{p}_{0}^{\top}\left( \sum_{t=0}^{+\infty} \lambda^{t}  \prod_{t'=0}^{t-1} \E_{u} \left[\bm{P}^{(\pieff(\pia,u))_{t'}}\right]\E_{u} \left[\bm{r}_{(\pieff(\pia,u))_{t}} \right] \right) \label{eq:rdv-prf-2} \\
& = \bm{p}_{0}^{\top}\left( \sum_{t=0}^{+\infty} \lambda^{t}  \prod_{t'=0}^{t-1} \bm{P}^{\E_{u} \left[(\pieff(\pia,u))_{t'}\right]}\bm{r}_{\E_{u} \left[(\pieff(\pia,u))_{t}\right]} \right)  \label{eq:rdv-prf-3} \\
& = \bm{p}_{0}^{\top}\left( \sum_{t=0}^{+\infty} \lambda^{t}  \prod_{t'=0}^{t-1} \bm{P}^{(\pieff(\pia,\theta))}\bm{r}_{(\pieff(\pia,\theta))} \right) \label{eq:rdv-prf-4} \\
& = R\left( \pieff(\pia,\theta) \right) \nonumber \\
& = R\left(\theta \pia + (1-\theta)\pib\right)\label{eq:rdv-prf-6}
\end{align}
where \eqref{eq:rdv-prf-1} follows from the dominated convergence theorem,
\tb{ \eqref{eq:rdv-prf-2} follows from the adherence decisions being independent across time, \eqref{eq:rdv-prf-3} follows from linearity of the expectation and the definition of $\bm{P}^{\pi}$ and $\bm{r}_{\pi}$, \eqref{eq:rdv-prf-4} follows from $\E_{u} \left[\pieff(\pia,u)\right] = \pieff(\pia,\theta)$,} and finally \eqref{eq:rdv-prf-6} follows from the definition of $R\left(\theta \pia + (1-\theta)\pib\right)$.
\paragraph{Step 3.} In Step 1, we have shown $\sup_{\pia \in \Pi_{\sf H}}  \: \E_{u} \left[ R(\pieff(\pia,u)) \right] = \sup_{\pia \in \Pi_{\sf M}}  \: \E_{u} \left[ R(\pieff(\pia,u)) \right]$. In Step 2, we have shown $\sup_{\pia \in \Pi_{\sf M}}  \: \E_{u} \left[ R(\pieff(\pia,u)) \right] = \sup_{\pia \in \Pi_{\sf M}}  \: R(\pieff(\pia,\theta))$. Proposition \ref{prop:pi alg star stationary deterministic} shows that $\sup_{\pia \in \Pi_{\sf M}}  \: R(\pieff(\pia,\theta)) = \max_{\pia \in \Pi}  \: R(\pieff(\pia,\theta))$, which concludes our proof.
\hfill \halmos
\endproof
}
\tbl{\paragraph{Other random models of adherence decisions.}
We briefly discuss here the viability of {\em Time-invariant random} adherence decision models, where there are some correlations across the adherence decisions across times. One possible {\em time-invariant} random models corresponds to
\[ \pieff(\pia,u)_{s,t} = u_{s} {\pia}_{s,t} + (1-u_{s}){\pib}_{s,t}\]
with $u_{s}$ sampled following a distribution with mean $\theta$ independently across all $s \in \X$. Another random model of adherence decisions corresponds to {\em Time- and State-invariant random} adherence decisions, where
\[ \pieff(\pia,u)_{s,t} = u {\pia}_{s,t} + (1-u){\pib}_{s,t},\]
with $u$ sampled following a distribution with mean $\theta$ and support in $[0,1]$. These models appear harder to analyze than the random models of deviation from Theorem \ref{th:random tv is equivalent to adamdp}, where the fact that the decisions are {\em independent} over time plays a crucial role in our proof. We simply note that an interesting property arises when the adherence decisions $u_{s,t}$ is common across all states and times: $u_{s,t} = u, \forall \; (s,t) \in \X \times \N$, and chosen at random following a distribution {\em supported in $\{0,1\}$}, with mean $\theta \in [0,1]$. In this case, the decision maker chooses either to follow $\pia$ (in every state and at every period) with probability $\theta$, or to follow $\pib$ with probability $1-\theta$. This situation may occur in the case where the decision maker is reluctant to changing policy along a trajectory and is constrained to follow the same policy at all states, e.g. because of concerns about the consistency of the resulting effective policy. Consequently, we have $\E_{u}\left[R(\pieff(\pia,u))\right] = \theta R(\pia) + (1-\theta) R(\pib)$, so that an optimal recommendation policy $\pi_{\sf alg}^{\star}(\theta)$ may be chosen {\em independent} of the true value of the adherence level $\theta$ and it is equal to the optimal nominal policy for the MDP instance $\M$. 
}

\section{Proof of Theorem \ref{th:model.equiv.informal}}\label{app:model equivalence}
In this section we study the adversarial models of adherence decisions and their equivalence with \ref{eq:definition-ada-mdp}. For concision, we denote 
\begin{align*}
B_{\infty} & := [\theta,1]^{\X \times \N}, \tag{Unconstrained Adversarial} \\
B_{1} & := \{ u \in [\theta,1]^{\X \times \N} \; | \; u_{s,t} = u_{s,t'}, \forall s \in \X, \forall \; t,t' \in \N\}, \tag{Time-invariant Adversarial}\\
 B_{2} & := \{ u \in [\theta,1]^{\X \times \N} \; | \; u_{s,t} = u_{s',t}, \forall s,s' \in \X, \forall \; t \in \N\}, \tag{State-invariant Adversarial} \\
 B_{3} & := \{ u \in [\theta,1]^{\X \times \N} \; | \; u_{s,t} = u_{s',t'}, \forall s,s' \in \X, \forall \; t,t' \in \N\} \tag{Time- and State-invariant Adversarial}
\end{align*}
We will prove the following theorems, \tb{showing the connection between adversarial models of adherence decisions and \ref{eq:definition-ada-mdp}. We then turn to showing strong duality in Appendix \ref{app:strong-duality}.}

\begin{theorem}[Unconstrained Adversarial]\label{th:model.B.infinity}
For a given adherence level $\theta \in [0,1]$, we have the following equality:
\[
\max_{\pia \in \Pi} \:  R\left( \theta \pia + (1-\theta)\pib \right) = \sup_{\pia \in \Pi_{\sf H}} \: \min_{ u \in B_{\infty} } \: R(\pieff(\pia,u)).\]
Additionally, there exists an optimal stationary deterministic policy that is a solution to the right-hand side optimization problem above.
\end{theorem}

\begin{theorem}[Other Adversarial models]\label{th:model.other-adversarial}
\tb{Let $B \subset [\theta,1]^{\X \times \N}$ be either $B_{1}$ ({\bf Time-invariant Adversarial}), $B_{2}$ ({\bf State-invariant Adversarial}), or $B_{3}$ ({\bf Time- and State-invariant Adversarial}).}
For a given adherence level $\theta \in [0,1]$, we have the following equality:
\[
\max_{\pia \in \Pi} \:  R\left( \theta \pia + (1-\theta)\pib \right) = \sup_{\pia \in \Pi_{\sf H}} \: \min_{u \in B }  \: R(\pieff(\pia,u)).\]
Additionally, there exists an optimal stationary deterministic policy that is a solution to the right-hand side optimization problem above.
\end{theorem}
The proofs of Theorem \ref{th:model.B.infinity} and Theorem \ref{th:model.other-adversarial} proceed in several steps.
\begin{itemize}
\item First, we show that the optimization problem with adversarial adherence decisions in $B_{\infty}$:
\begin{equation}\label{eq:app Model Binfinity}
\sup_{\pia \in \Pi_{\sf H}} \min_{u \in B_{\infty}} R(\pieff(\pia,u))
\end{equation} 
admits a robust Bellman equation in Proposition \ref{prop:solving Binfinity}. 
\item In Corollary \ref{cor:equivalence Binfinity and effective policy}, we then show that this robust Bellman equation can be interpreted as the Bellman equation of an alternate MDP, which shows the equivalence between \eqref{eq:app Model Binfinity} and \ref{eq:definition-ada-mdp} as in
\begin{equation}\label{eq:app adherence aware MDP}
\max_{\pia \in \Pi}  R\left(\theta \pia + (1-\theta)\pib\right),
\end{equation}
hence concluding the proof of Theorem \ref{th:model.B.infinity}.
\item 
\tb{ The proof for Theorem \ref{th:model.other-adversarial} follows from $B_{1} \subset B_{\infty}, B_{2} \subset B_{\infty}, B_{3} \subset B_{\infty}$, and from the worst-case $u^{\infty} \in B_{\infty}$ for the {\bf Unconstrained Adversarial} model being $u^{\infty}_{s,t} = \theta, \forall \; (s,t) \in \X \times \N$, which is feasible in $B_{1},B_{2}$ and $B_{3}$.
}
\end{itemize}

\subsection{Proof of Theorem \ref{th:model.B.infinity} (Unconstrained Adversarial)}\label{app:model.B.infinity}
For the sake of conciseness, for a given stationary policy $\pi$ and $\bm{v} \in \R^{\X}$, we define $T^{\pi}(\bm{v}) \in \R^{\X}$ as $T^{\pi}_{s}(\bm{v}) = \sum_{a \in \A} \pi_{sa}  \bm{P}_{sa}^{\top} \left( \bm{r}_{sa} + \lambda \bm{v} \right), \forall \; s \in \X$. Note that for each $s \in \X$, the scalar $T^{\pi}_{s}(\bm{v})$ only depends on $\bm{\pi}_{s} \in \Delta(\A)$ and not on $\bm{\pi}_{s'}$ for $s' \neq s$. The next proposition shows that \eqref{eq:app Model Binfinity} admis a robust Bellman equation.
\begin{proposition}\label{prop:solving Binfinity}
Let $\bm{v}^{\infty} \in \R^{\X}$ satisfying
\begin{equation}\label{eq:v-infinity-app}
     v^{\infty}_{s} = \max_{\bm{\pi}_{s} \in \Delta(\A)} \min_{u \in [\theta,1]}  u \cdot  T^{\pi}_{s}(\bm{v}^{\infty}) + (1-u) \cdot T^{\pib}_{s}(\bm{v}^{\infty}), \forall \; s \in \X.
\end{equation}
Additionally, let $\pi^{\infty}$ be a stationary policy attaining the maximum in the right-hand side in \eqref{eq:v-infinity-app} for each $s \in \X$. Then $\pi^{\infty}$ can be chosen deterministic, and $\pi^{\infty}$ is an optimal solution to \eqref{eq:app Model Binfinity}.
\end{proposition}
\proof{Proof.}
We first note that the vector $\bm{v}^{\infty}$ is well defined and is unique because the following map $f: \R^{\X} \rightarrow \R^{\X}$ is a contraction for the $\ell_{\infty}$-norm:
\[f:\bm{v}\mapsto \left(\max_{\bm{\pi}_{s} \in \Delta(\A)} \min_{u \in [\theta,1]}  u \cdot  T^{\pi}_{s}(\bm{v}) + (1-u) \cdot T^{\pib}_{s}(\bm{v})\right).\]
Since $f$ is a contraction, there exists a unique vector $\bm{v}^{\infty} \in \R^{\X}$ such that $\bm{v}^{\infty}$ is a fixed-point of $f$, i.e., such that $f(\bm{v}^{\infty}) = \bm{v}^{\infty}$.
Let us define $\pi^{\infty}$ as the policy attaining the $\arg \max$ in \eqref{eq:v-infinity-app} and $u^{\star}_{s} \in [\theta,1]$ attaining its worst-case on each state $s \in \X$. We define $u^{\infty} \in [\theta,1]^{\X \times \N}$ with $u^{\infty}_{s,t} = u^{\star}_{s}, \forall \; (s,t) \in \X \times \N$. We will show that $(\pi^{\infty},u^{\infty})$ is an optimal solution to $\sup_{\pia \in \Pi_{\sf H}} \min_{u \in B_{\infty}} R(\pieff(\pia,u))$. To show this, we will show that $\pi^{\infty}$ is an $\epsilon$-optimal policy to $\sup_{\pia \in \Pi_{\sf H}} \min_{u \in B_{\infty}} R(\pieff(\pia,u))$, for any $\epsilon>0$. 
\begin{itemize}
    \item Let $\epsilon>0$.
   Recall that the infinite-horizon return of a policy $\pi$ is defined as
$R(\pi)=  \E^{\pi} \left[ \sum_{t=0}^{+\infty} \lambda^{t}r_{s_{t}a_{t}s_{t+1}} \right].$
 Let $T \in \N$. For any policy $\pi$, we define $R_{T}(\pi)$, {\em the truncated return with terminal reward $\bm{v}^{\infty}$}, as
    \begin{equation}\label{eq:truncated-expected-reward}
     R_{T}(\pi)=\E^{\pi} \left[ \sum_{t=0}^{T-1} \lambda^{t}r_{s_{t}a_{t}s_{t+1}} + \lambda^{T} v^{\infty}_{s_{T}} \right].
     \end{equation}
     Since $\X,\A$ are finite sets, the rewards $r_{s,a,s'}$ are bounded. Therefore, for any $\epsilon>0$, there exists a corresponding $T$ such that $|R_{T}(\pi)-R(\pi)| \leq \epsilon$ for any policy $\pi \in \Pi$. For instance we can take $T$ such that $\lambda^{T} \left( \frac{\max_{s,a} |r_{s,a}|}{1-\lambda}+\| \bm{v}^{\infty}\|_{\infty}\right)<\epsilon.$
 \item We can define the {\em worst-case truncated return with terminal reward $\bm{v}^{\infty}$} as
    \begin{equation}\label{eq:worst-case truncated-expected-reward}
     \min_{u \in B_{\infty}} R_{T}(\pieff(\pi,u))=  \min_{u \in B_{\infty} } \E^{\pieff(\pi,u)}\left[ \sum_{t=0}^{T-1} \lambda^{t}r_{s_{t}a_{t}s_{t+1}} + \lambda^{T} v^{\infty}_{s_{T}} \right].
     \end{equation} 
 Note that the worst-case return and the worst-case truncated return of $\pi^{\infty}$ coincide:
 \[\min_{u \in B_{\infty}} R_{T}(\pieff(\pi^{\infty},u)) = \min_{u \in B_{\infty}} R(\pieff(\pi^{\infty},u)).\] This is by definition of $\pi^{\infty}$ and $\bm{v}^{\infty}$ as the fixed-point of $f$, i.e., as the continuation values of $\pi^{\infty}$.
 \item 
 We claim that for any value of $T \in \N$, the decisions $\pieff\left(\pi^{\infty},u^{\infty}\right), ..., \pieff\left(\pi^{\infty},u^{\infty}\right)$ (repeated $T$ times) are optimal for \eqref{eq:worst-case truncated-expected-reward}. Indeed, the terminal rewards are given by $\left(
 v^{\infty}_{s}\right)_{s \in \X}$, and $f$ is the Bellman operator that relates the worst-case values at period $t \in \{1,...,T\}$ to the worst-case values at period $t-1$. Since $\bm{v}^{\infty}=f(\bm{v}^{\infty})$ and $\pi^{\infty}_{s},u^{\infty}$ is a solution to $f_{s}(\bm{v}^{\infty})$ as a saddle-point program for each $s \in \X$, we conclude that repeating $\pieff\left(\pi^{\infty},u^{\infty}\right)$ $T$-times optimizes the worst-case truncated return.
 \item Overall, we have shown the following inequalities. First, we have shown that the worst-case return and the worst-case truncated return of $\pi^{\infty}$ coincides:
$\min_{u \in B_{\infty}} R_{T}(\pieff(\pi^{\infty},u))  = \min_{u \in B_{\infty}} R(\pieff(\pi^{\infty},u))$ 
 and we have shown that $\pi^{\infty}$ is optimal for the worst-case truncated return:
$
\min_{u \in B_{\infty}} R_{T}(\pieff(\pi^{\infty},u))  \geq \min_{u \in B_{\infty}} R_{T}(\pieff(\pi,u)), \forall \; \pi \in \Pi_{\sf H}.$
Let $\pi^{\star}$ an optimal policy for the worst-case adherence model \eqref{eq:app Model Binfinity}. Then we have
\[
\min_{u \in B_{\infty}} R(\pieff(\pi^{\infty},u)) = \min_{u \in B_{\infty}} R_{T}(\pieff(\pi^{\infty},u)) \geq \min_{u \in B_{\infty}} R_{T}(\pieff(\pi^{\star},u)) \geq \min_{u \in B_{\infty}} R_{T}(\pieff(\pi^{\star},u)) - \epsilon.
\]
where the last inequality follows from the worst-case truncated return approximating the worst-case return up to $\epsilon$.
This shows that for any $\epsilon >0,$ we have
$\min_{u \in B_{\infty}} R(\pieff(\pi^{\infty},u)) \geq \min_{u \in B_{\infty}} R(\pieff(\pi^{\star},u)) - \epsilon$, 
from which we conclude that 
$\min_{u \in B_{\infty}} R(\pieff(\pi^{\infty},u)) \geq \min_{u \in B_{\infty}} R(\pieff(\pi^{\star},u)).$
 Since we have chosen $\pi^{\star}$ as an optimal policy for \eqref{eq:app Model Binfinity}, we can conclude that $\pi^{\infty}$ is an optimal policy for \eqref{eq:app Model Binfinity}. This concludes our proof of Proposition \ref{prop:solving Binfinity}.  \hfill \halmos
\endproof
\end{itemize}

We now have the following corollary, which shows the equivalence (at optimality) between the model with worst-case time-varying adherence and our model of adherence-aware MDP \eqref{eq:app adherence aware MDP}.
\begin{corollary}\label{cor:equivalence Binfinity and effective policy}
For $\bm{v}^{\infty}$ defined as in \eqref{eq:v-infinity-app}, we have
\begin{equation}\label{eq:v-infinity-theta}
     v^{\infty}_{s} = \max_{\bm{\pi}_{s} \in \Delta(\A)} \theta \cdot  T^{\pi}_{s}(\bm{v}^{\infty}) + (1-    \theta) \cdot T^{\pib}_{s}(\bm{v}^{\infty}), \forall \; s \in \X,
\end{equation}
i.e., the worst-case deviation at optimality in \eqref{eq:app Model Binfinity} is attained at $u^{\infty}_{s,t} = \theta$ for all $(s,t) \in \X \times \N$. \tb{In particular, we have the following equality:
\begin{equation}\label{eq:best-response}
\sup_{\pia \in \Pi_{\sf H}}  R(\pieff(\pia,u^{\infty})) = \min_{u \in B_{\infty}}  R(\pieff(\pi^{\infty},u)).
\end{equation}
}
\end{corollary}
\proof{Proof.}
Because the inner objective function is linear in $u$, we have
\begin{align*}
\max_{\bm{\pi}_{s} \in \Delta(\A)} \min_{u \in [\theta,1]}  u \cdot  T^{\pi}_{s}(\bm{v}^{\infty}) + (1-u) \cdot T^{\pib}_{s}(\bm{v}^{\infty}) & = \max_{\bm{\pi}_{s} \in \Delta(\A)} \min_{u \in \{\theta,1\}}  u \cdot  T^{\pi}_{s}(\bm{v}^{\infty}) + (1-u) \cdot T^{\pib}_{s}(\bm{v}^{\infty}) \\
& =  \max_{\bm{\pi}_{s} \in \Delta(\A)} \min \{ \theta \cdot  T^{\pi}_{s}(\bm{v}^{\infty}) + (1-\theta) \cdot T^{\pib}_{s}(\bm{v}^{\infty}), T^{\pi}_{s}(\bm{v}^{\infty})\}.
\end{align*}
Therefore, we want to prove that the minimum 
$\min \{ \theta \cdot  T^{\pi^{\infty}}_{s}(\bm{v}^{\infty}) + (1-\theta) \cdot T^{\pib}_{s}(\bm{v}^{\infty}), T^{\pi}_{s}(\bm{v}^{\infty})\}$ is always attained at $\theta \cdot  T^{\pi^{\infty}}_{s}(\bm{v}^{\infty}) + (1-\theta) \cdot T^{\pib}_{s}(\bm{v}^{\infty})$, i.e., we want to show that 
$\theta \cdot  T^{\pi^{\infty}}_{s}(\bm{v}^{\infty}) + (1-\theta) \cdot T^{\pib}_{s}(\bm{v}^{\infty}) \leq T^{\pi^{\infty}}_{s}(\bm{v}^{\infty}).$
Note that by choosing $\pi = \pib$ in the max-min program \eqref{eq:v-infinity}, we always have
\begin{equation}\label{eq:dominance-v-alg-v-base}
v^{\infty}_{s} \geq T^{\pib}_{s}(\bm{v}^{\infty}), \forall \; s \in \X.
\end{equation}
Now if for some $s \in \X$ we have $\theta \cdot  T^{\pi^{\infty}}_{s}(\bm{v}^{\infty}) + (1-\theta) \cdot T^{\pib}_{s}(\bm{v}^{\infty}) > T^{\pi^{\infty}}_{s}(\bm{v}^{\infty})$,  then  $ T^{\pib}_{s}(\bm{v}^{\infty}) > T^{\pi^{\infty}}_{s}(\bm{v}^{\infty}).$ But since $T^{\pi^{\infty}}_{s}(\bm{v}^{\infty})=v^{\infty}_{s}$, we would obtain $ T^{\pib}_{s}(\bm{v}^{\infty}) > v^{\infty}_{s}$, which is a contradiction with \eqref{eq:dominance-v-alg-v-base}. Therefore, we always have
$\theta \cdot  T^{\pi^{\infty}}_{s}(\bm{v}^{\infty}) + (1-\theta) \cdot T^{\pib}_{s}(\bm{v}^{\infty}) \leq T^{\pi^{\infty}}_{s}(\bm{v}^{\infty}),$ which shows that 
$ v^{\infty}_{s} = \max_{\bm{\pi}_{s} \in \Delta(\A)} \theta \cdot  T^{\pi}_{s}(\bm{v}^{\infty}) + (1- \theta) \cdot T^{\pib}_{s}(\bm{v}^{\infty}), \forall \; s \in \X.$ \tb{Therefore, the worst-case adherence decisions $u^{\infty}$ for $\pi^{\infty}$ can be chosen as $u^{\infty}_{s,t} = \theta$ for any pair $(s,t) \in \X \times \N$, which concludes the proof of Corollary \ref{cor:equivalence Binfinity and effective policy}.} This also shows that we can choose an optimal policy for \eqref{eq:app Model Binfinity} as a stationary deterministic policy, because $\pi^{\infty}$ attains the right-hand side in \eqref{eq:v-infinity-theta}, which maximizes a linear form over the simplex $\Delta(\A)$. \hfill \halmos
\endproof
We can now interpret the equation \eqref{eq:v-infinity-theta} as the Bellman equation of a decision maker which chooses $\pia$ and where the effective policy is $\theta \pia + (1-\theta)\pib$. This is because for any $\bm{v} \in \R^{\X},s \in \X$ and $\pia \in \Pi$, we have \[\theta \cdot  T^{\pia}_{s}(\bm{v}) + (1-    \theta) \cdot T^{\pib}_{s}(\bm{v}) = T^{\theta \pia + (1-\theta)\pib}_{s}(\bm{v})\] which shows that \eqref{eq:app Model Binfinity} is equal to \eqref{eq:app adherence aware MDP} and that the sets of optimal policies of \eqref{eq:app Model Binfinity} and \eqref{eq:app adherence aware MDP} share a common stationary deterministic policy $\pi^{\infty}$, attaining the $\arg \max$ in \eqref{eq:v-infinity-theta}.

\tb{
\subsection{Proof of Theorem \ref{th:model.other-adversarial} (other adversarial models)}\label{app:model.other-adversarial}
We now study the other models of adversarial adherence decisions presented in Theorem \ref{th:model.equiv.informal}. We provide the proof of Theorem \ref{th:model.other-adversarial} below.
\proof{Proof of Theorem \ref{th:model.other-adversarial}.}
Let $B \subset [\theta,1]^{\X \times \N}$ be either $B_{1}$ ({\bf Time-invariant Adversarial}), $B_{2}$ ({\bf State-invariant Adversarial}), or $B_{3}$ ({\bf Time- and State-invariant Adversarial}).

Let $u^{\infty} \in [\theta,1]^{\X \times \N}$ be defined as $u_{s,t}^{\infty} = \theta, \forall \; (s,t) \in \X \times \N$, and $\pi^{\infty}$ be defined as in Proposition \ref{prop:solving Binfinity}. We have
\[\sup_{\pia \in \Pi_{\sf H}} \min_{u \in B} R(\pieff(\pia,u)) \leq \sup_{\pia \in \Pi_{\sf H}}  R(\pieff(\pia,u^{\infty})) = \min_{u \in B_{\infty}} R(\pieff(\pia^{\infty},u)) \leq  \min_{u \in B} R(\pieff(\pia^{\infty},u))\] 
where the first inequality comes from $u^{\infty} \in B$, the equality comes from \eqref{eq:best-response}, and the second inequality comes from $B \subset B_{\infty}$. This shows that $\pi^{\infty}$ is an optimal policy in $\sup_{\pia \in \Pi_{\sf H}} \min_{u \in B} R(\pieff(\pia,u))$ and the two saddle-point formulations are equal: $\sup_{\pia \in \Pi_{\sf H}} \min_{u \in B} R(\pieff(\pia,u)) = \sup_{\pia \in \Pi_{\sf H}} \min_{u \in B_{\infty}} R(\pieff(\pia,u))$. We have proved in Theorem \ref{th:model.B.infinity} that the right-hand side in the previous equation is equal to $\max_{\pia \in \Pi} R(\theta \pia + (1-\theta)\pib)$, which concludes the proof of Theorem \ref{th:model.other-adversarial}.
\hfill \halmos
\endproof
}

\tb{
\subsection{Proof of strong duality for the adversarial models}\label{app:strong-duality}
We now turn to proving that strong duality holds for all the adversarial models considered in Theorem \ref{th:model.equiv.informal}. In particular, we show the following theorem.
\begin{theorem}\label{th:strong-duality}
Let $B \subset [\theta,1]^{\X \times \N}$ be either $B_{\infty}$ ({\bf Unconstrained Aversarial}), $B_{1}$ ({\bf Time-invariant Adversarial}), $B_{2}$ ({\bf State-invariant Adversarial}), or $B_{3}$ ({\bf Time- and State-invariant Adversarial}).
Then
\[\sup_{\pia \in \Pi_{\sf H}} \min_{u \in B} R(\pieff(\pi_{\sf alg},u)) =  \min_{u \in B} \sup_{\pia \in \Pi_{\sf H}} R(\pieff(\pi_{\sf alg},u)).\]
\end{theorem}
\proof{Proof.} Let $B \in \{B_{\infty},B_{1},B_{2},B_{3}\}$.
Since weak duality always holds, we only have to prove that
\[ \min_{u \in B} \sup_{\pi_{\sf alg} \in \Pi_{\sf H}} R(\pieff(\pi_{\sf alg},u)) \leq \sup_{\pi_{\sf alg} \in \Pi_{\sf H}} \min_{u \in B} R(\pieff(\pi_{\sf alg},u)).\]
We have
\[ \min_{u \in B} \sup_{\pi_{\sf alg} \in \Pi_{\sf H}} R(\pieff(\pi_{\sf alg},u)) \leq  \sup_{\pi_{\sf alg} \in \Pi_{\sf H}} R(\pieff(\pi_{\sf alg},u^{\infty})) = \min_{u \in B_{\infty}} R(\pieff(\pi_{\sf alg}^{\infty},u)) \leq \sup_{\pi_{\sf alg} \in \Pi_{\sf H}} \min_{u \in B} R(\pieff(\pi_{\sf alg},u))\]
where the first inequality comes from $u^{\infty} \in B$, the equality comes from \eqref{eq:best-response}, and the second inequality comes from $B \subset B_{\infty}$ and from maximizing over $\Pi_{\sf H}$. Therefore strong duality holds.
\hfill \halmos
\endproof
}

\section{Proof of Theorem \ref{th:hardness-apx-ada-mdp}}\label{app:proof hardness}
We will show that the constrained assortment optimization with a Markov chain-based choice model~\citep{blanchet2016markov,desir2020constrained} can be reduced to \ref{eq:constrained-independent-varying-adversarial}. We first introduce the Markov chain-based choice model below. We follow the lines of \cite{desir2020constrained} here.
\paragraph{Markov chain model.} Let $n \in \N$. The set $\cN= \{1,...,n\}$ represents $n$ items. The no-purchase option is represented by $0$ and we write $\cN_{+} = \cN \cup \{ 0\}$. There are scalars $\nu_{i} \geq 0$ which represents the initial arrival probabilities for every state $i \in \cN_{+}$ and some transition probabilities $\rho_{ij} \in [0,1]$ for all $(i,j) \in \cN_{+}^{2}$.  The return for each item $i$ is written as $\xi_{i} \geq 0$.
The goal of the decision maker is to choose a subset of items $\I \subseteq \cN$ to display to the customers, in order to maximize its expected return $R_{\sf MC}(\I)$, computed as follows:
\begin{itemize}
\item For any state $i$ in the chosen set of items $\I$, the state $i$ is absorbing.
\item A customer arrives in state $i \in \cN_{+}$ with an initial probability $\nu_{i}$. If the state $i$ is non-absorbing, the customer transitions to a different state $j \in \cN_{+}, j \neq i$ with probability $\rho_{ij}$.
\item The process continues until an absorbing state is reached (either in $\I$ or in $\{0\}$).
\item Let $\gamma(i,\I)$ be the probability that item $i \in \I$ is chosen by the customer when the assortment $\I \subset \cN$ is offered. Note that $\gamma(i,\I)$ is equal to the probability that the customer reaches state $i$ before any other absorbing states. Then the return $R_{\sf MC}(\I)$ associated with a chosen subset $\I$ is
\begin{equation}\label{eq:return-MC}
R_{\sf MC}(\I) = \sum_{i \in \I} \gamma(i,\I)\xi_{i}.
\end{equation}
\end{itemize} 
The {\em cardinality assortment } (\carda) problem is the following optimization problem, which solves for an optimal assortment $\I$ with constraints on the number of items selected for display:
\begin{equation}\label{eq:card-assort}\tag{\carda}
\max \; \{ R_{\sf MC}(\I) \; | \; \I \subset \cN, |\I| \leq k\}.
\end{equation}
The authors in \cite{desir2020constrained} prove the following hardness result for \carda, even under some conditions on the parameters $\rho, \xi$ and $\nu$.
\begin{theorem}[Reformulation of Theorem 5, \cite{desir2020constrained}]\label{th:carda-apx-hard-reformulation}
\carda{} is APX-hard, even when $\rho_{i0} = 1/4$ and $\xi_{i}=1, \nu_{i} = 1/n$ for all $i \in \cN$.
\end{theorem}

Our proof of Theorem \ref{th:hardness-apx-ada-mdp} follows from Theorem \ref{th:carda-apx-hard-reformulation}. On the one hand, we can interpret \ref{eq:constrained-independent-varying-adversarial} as an optimization problem where there are two Markov chains, one induced by $\pia$ and one by $\pib$, and where the variable $u_{s}$ decides to follow the Markov chain induced by $\pia$ or the Markov chain induced by $\pib$ at each state $s \in \X$.  On the other hand, \carda{} can be interpreted as follows: given a Markov chain following a transition matrix $\rho$, the decision maker chooses a subset $\I$ (with $|\I| \leq k$) for which the states in $\I$ become absorbing. Based on this interpretation of \ref{eq:constrained-independent-varying-adversarial} and \carda{}, we can reformulate any instance of \carda{} as an instance of \ref{eq:constrained-independent-varying-adversarial} in a straightforward manner. The only technical difficulty is that $R_{\sf MC}(\I)$ a priori does not involve a discount factor, whereas we have defined the objective function in \ref{eq:constrained-independent-varying-adversarial} based on the discounted return \eqref{eq:expected-reward}, which depends on a discount factor $\lambda$. We show how to circumvent this issue in our proof below. In particular, we prove Theorem \ref{th:hardness-apx-ada-mdp} in two steps.
 \paragraph{Step 1: reformulating the objective in \carda.}
We consider the instance of \carda{} from the proof of Theorem 5 in \cite{desir2020constrained}. In this instance, $\rho_{i0} = 1/4$ in all state $i \in \cN$, and for each item $i \in \cN$ there is a subset $\cN_{i} \subset \cN \{ i\}$ such that $|\cN_{i}|=3$ and $\rho_{ij} = 1/4, \forall \; j \in \cN_{i}$. Additionally, the return $\xi_{i}$ is equal to $1$ for each item $i \in \cN$.
 We first note that $R_{\sf MC}(\I)$, defined as $R_{\sf MC}(\I) = \sum_{i \in \I} \gamma(i,\I)\xi_{i}$, is equal to $\E^{\pi} \left[  \sum_{t=0}^{+\infty} \tilde{r}_{s_{t}a_{t}s_{t+1}}  \right]$ for a certain MDP instance $\tilde{\M}$ and a certain policy $\pi$ that represents the subset $\I$ of chosen items. In particular, let us consider the following MDP instance $\tilde{\M}$ with states $\tilde{\X} = \cN_{+}$, actions $\{a_{0},a_{1}\}$ which represent choosing or not an item to display, and where the instantaneous rewards $\tilde{\bm{r}}$ and the transition probabilities $\tilde{\bm{P}}$ are defined as follows:
 \begin{itemize}
 \item All the instantaneous rewards are equal to $0$, except $r_{sa_{1}s} = 1/4, \forall \; s \in \cN$.
 \item $\tilde{P}_{sa_{0}s'} = 1/4$ if $s' \in \cN_{s} \cup \{0\}, \forall \; s \in \cN$,
 \item $\tilde{P}_{sa_{1}s} = 3/4, \tilde{P}_{sa_{1}0}=1/4,\forall \; s \in \cN$,
 \item $\tilde{P}_{0a_{0}0} = \tilde{P}_{0 a_{1} 0} = 1$.
 \end{itemize}
 A stationary deterministic policy $\pi$ is a map $\cN_{+} \rightarrow \{a_{0},a_{1}\}$, and we can construct a policy $\pi$ representing an assortment $\I$ as follows:  $\pi$ chooses action $a_{1}$ in state $s$ if and only if $s \in \I$; otherwise, $\pi$ chooses action $a_{0}$. Finally, the initial probability distribution is just $p_{0,i} = \nu_{i} = 1/n, \forall \; i \in \cN$.
 Let us give some intuition on the values of $\tilde{\bm{r}}$ and $\tilde{\bm{P}}$ given above. In a state where action $a_{0}$ is chosen (which corresponds to a state not included the subset $\I$), the Markov chain induced by $\tilde{\bm{P}}$ evolves exactly as for the kernel $\rho$ defined above and no instantaneous reward is obtained. When action $a_{1}$ is chosen at a state $s$, i.e. when $s$ is chosen to be included in $\I$, the decision maker transitions to the absorbing state $0$ after a number of period that follows a geometric distribution with parameter $\lambda=3/4$, earning an instantaneous reward of $1-\lambda = 1/4$ while remaining in state $s$ and then an instantaneous reward of $0$ while in state $0$.
Overall, we have shown that \carda{} can be reformulated as optimizing an undiscounted return, in contrast to the discounted returns considered in this paper. However, the Markov chain $\tilde{\M}$ has a very particular structure: from any state $s \in \cN$, there is a probability $1/4$ to reach the absorbing state $0$, and no reward is obtained when transitioning to state $0$. We now show in the next step that this can be interpreted as computing a discounted return with a discount factor of $\lambda = 1-1/4 = 3/4$.
\paragraph{Step 2: from undiscounted objective to discounted objective.}
In this section we show that we can reformulate the undiscounted objective function $\E^{\pi} \left[  \sum_{t=0}^{+\infty} \tilde{r}_{s_{t}a_{t}s_{t+1}}  \right]$ as a discounted objective $\E^{\pi} \left[  \sum_{t=0}^{+\infty} \lambda^t r_{s_{t}a_{t}s_{t+1}}  \right]$ for a certain discount factor $\lambda$ and a certain Markov chain $\M$.
We follow the lines of Section 5.3 in \cite{puterman2014markov}, which shows that the discount factor $\lambda \in [0,1)$ can be interpreted as a termination probability. This idea dates back to \cite{derman1970finite}. More precisely, we have the following proposition, which is a reformulation of the results in Section 5.3 in \cite{puterman2014markov}.
\begin{proposition}\label{prop:equivalence discounted to undiscounted}
Let $\M$ be any MDP instance and $\pi$ be a policy. Then the discounted return $R(\pi)=\E^{\pi} \left[ \sum_{t=0}^{+\infty} \lambda^{t}r_{s_{t}a_{t}} \right]$ is equal to the following  undiscounted return $\tilde{R}(\pi) = \E^{\pi}\left[ \sum_{t=0}^{+\infty} \tilde{r}_{s_{t}a_{t}s_{t+1}} \right]$, where $\tilde{\bm{r}} \in \R^{\X_{+} \times \A}, \tilde{\bm{P}} \in \R^{\X_{+} \times \A \times \X_{+}}$ are the instantaneous rewards and the transition probabilities for an MDP instance $\tilde{\M}$ defined over an augmented state space $\X_{+} = \X \cup \{\Delta\}$, defined as follows:
\begin{itemize}
    \item $\tilde{r}_{sas'}=r_{sas'}, \forall (s,a) \in \X \times \A, \forall \; s' \neq \Delta$,
    \item $\tilde{r}_{s a \Delta} = 0, \forall (s,a) \in \X_{+} \times \A$
    \item $\tilde{P}_{sas'} = \lambda P_{sas'}, \forall (s,a,s') \in \X \times \A \times \X$,
    \item $\tilde{P}_{sa\Delta} = 1-\lambda, \forall \; s \in \X$,
    \item $\tilde{P}_{\Delta a \Delta} = 1$.
\end{itemize}
\end{proposition}
Applying Proposition \ref{prop:equivalence discounted to undiscounted} to the MDP instance $\tilde{\M}$ defined in the first step of this proof, we find that the objective function $\E^{\pi} \left[  \sum_{t=0}^{+\infty} \tilde{r}_{s_{t}a_{t}s_{t+1}}  \right]$ can be reformulated as the discounted return in the following MDP instance $\M$: the set of states is $\cN$, the set of actions is $\{a_{0},a_{1}\}$, the discount factor is $\lambda=3/4$, the initial probability distribution $\bm{p}_{0}$ for the MDP instance $\tilde{\M}$ is equal to the probability distribution $\bm{\nu}$ for the instance of \carda, and the instantaneous rewards $\bm{r}$ and the transition probabilities $\bm{P}$ are defined as follows:
\begin{equation}\label{eq:equivalent-mdp-carda}
r_{sa_{0}s'} = 0, r_{sa_{1}s} = 1/4, \forall \; (s,s') \in \cN \times \cN, P_{sa_{0}s'} = 1/3, P_{sa_{1}s} = 1,  \forall s' \in \cN_{s}, \forall \; s \in \cN.
\end{equation}
Overall, we have shown the following proposition.
\begin{proposition}
Let us consider the instance of \carda{} from the proof of Theorem 5 in \cite{desir2020constrained}. Then $R_{\sf MC}(\I)$ can be reformulated as $R(\pi)=\E^{\pi} \left[ \sum_{t=0}^{+\infty} \lambda^{t}r_{s_{t}a_{t}} \right]$ for the MDP instance described in \eqref{eq:equivalent-mdp-carda} with $\pi_{s} = a_{1}$ if and only if $s \in \I$ for $s \in \cN$  and $\lambda = 3/4$.
\end{proposition}
Let $\I \subset \cN$ a subset of displayed items and let $\pi$ be the policy representing $\I$ in the MDP $\M$. Then $\pi_{s} = \pieff(\pia,u) = u_{s} \pi_{{\sf alg} \; s} + (1-u_{s})\pi_{{\sf base}\; s}$, with $\pib$ the policy that chooses $a_{0}$ in all states, $\pia$ the policy that chooses $a_{1}$ in all states, and $u \in \{0,1\}^{\X}$. The cardinality constraint $|\I| \leq k$ can be directly rewritten $\sum_{s \in \X} u_{s} \leq k$, which concludes our proof.

\tb{
\section{Mixed-integer optimization formulation for \ref{eq:constrained-independent-varying-adversarial}}\label{app:MIP-constrained-adamdp}
In this section we provide a mixed-integer optimization formulation for \ref{eq:constrained-independent-varying-adversarial}. We start with the following lemma, which is a direct consequence of a classical contraction lemma, see for instance Lemma 2 in \cite{nilim2005robust} or Lemma 3.1 in \cite{grand2022convex}.
\begin{lemma}\label{lem:contraction-lemma}
Let $\pi \in \Pi$. Then 
$R(\pi) = \min \{ \bm{p}_{0}\trp\bm{v} \; | \; v_{s} \geq \sum_{a \in \A} \pi_{sa}\bm{P}_{sa}\trp\left(\bm{r}_{sa} + \lambda \bm{v}\right), \forall \; s \in \X, \bm{v} \in \R^{\X}\}.$
\end{lemma}
Based on Lemma \ref{lem:contraction-lemma}, for a fixed $\pia \in \Pi$ we can reformulate \ref{eq:constrained-independent-varying-adversarial} as follows:
\begin{equation}\label{eq:MIP-0}
\min \left\{ \bm{p}_{0}\trp\bm{v} \; | \; v_{s} \geq \sum_{a \in \A} u_{s} \pi_{{\sf alg},sa}\bm{P}_{sa}\trp\left(\bm{r}_{sa} + \lambda \bm{v}\right) + (1-u_{s})\pi_{{\sf base},sa}\bm{P}_{sa}\trp\left(\bm{r}_{sa} + \lambda \bm{v}\right) , \forall \; s \in \X, \sum_{s \in \X} u_{s} \leq k, \bm{u}  \in \{0,1\}^{\X},\bm{v} \in \R^{\X}\right\}.
\end{equation}
In the optimization program above, the terms $u_{s} \times \left(\sum_{a \in \A} (\pi_{{\sf alg},sa} - \pi_{{\sf base},sa})\bm{P}_{sa}\trp\bm{v}\right)$ are  bilinear in the variables $(u_{s},\bm{v}) \in \{0,1\} \times \R^{\X}$ for any $s \in \X$. 
However, $u_{s} \in \{0,1\},v_{s} \in [0,r_{\infty}/(1-\lambda)],\sum_{s' \in \X} P_{sas'} = 1$ for any $s,a \in \X$, so we can use classical reformulation techniques to linearize the bilinear constraints. In particular, it is well-known that we can linearize the term $x \times y$ with $x \in \{0,1\}, y \in [L,U]$ with $L \leq U$ by introducing an auxiliary variable $z \in \R$ such that $z \geq Lx, z \leq Ux, z \geq y - (1-x)\max\{|L|,|U|\}, z \leq y + (1-x)\max\{|L|,|U|\}$.
Applying this method with $x = u_{s} \in \{0,1\}$ and $y = \sum_{a \in \A} (\pi_{{\sf alg},sa} - \pi_{{\sf base},sa})\bm{P}_{sa}\trp\bm{v} \in [ - \frac{r_{\infty}}{1-\lambda},2 \frac{r_{\infty}}{1-\lambda}]$ to linearize the bilinear terms appearing in the reformulation \eqref{eq:MIP-0}, we obtain a mixed-integer program for \ref{eq:constrained-independent-varying-adversarial}, only involving a linear objective and linear constraints over continuous and binary variables:
\begin{align*}
\min & \; \bm{p}_{0}\trp\bm{v} \\ 
& v_{s} \geq \lambda z_{s} + \sum_{a \in \A} \left(u_{s} \pi_{{\sf alg},sa} +(1-u_{s})\pi_{{\sf base},sa}\right) \bm{P}_{sa}\trp\bm{r}_{sa} + \lambda \pi_{{\sf base},sa}\bm{P}_{sa}\trp\bm{v}, \forall \; s \in \X,\\
& - 2 \frac{r_{\infty}}{1-\lambda} (1-u_{s}) \leq z_{s} - \sum_{a \in \A} (\pi_{{\sf alg},sa} - \pi_{{\sf base},sa})\bm{P}_{sa}\trp\bm{v} \leq 2 \frac{r_{\infty}}{1-\lambda} (1-u_{s}), \forall \; s \in \X, \\
& - \frac{r_{\infty}}{1-\lambda} u_{s} \leq z_{s} \leq 2 \frac{r_{\infty}}{1-\lambda} u_{s}, \forall \; s \in \X, \\
& \sum_{s \in \X} u_{s} \leq k, \\
& \bm{u}  \in \{0,1\}^{\X}, \bm{v} \in \R^{\X}, \bm{z} \in \R^{\X}.
\end{align*}

}

\tbl{
\section{Detailed computation for Section \ref{sec:bad-mdp-instance}} \label{app:bad-mdp-instance}
We detail the computation of the simple MDP instance we presented in Section \ref{sec:bad-mdp-instance}. 

Considering the policies $\pib$ and $\pia$ represented in Figures \ref{fig:pi-b.tot} and \ref{fig:pi-r-star-one.tot} respectively, we compute the return of the effective policy $\pieff(\pia,\theta) = \theta \pia + (1-\theta) \pib$. For concision, let us denote $r := r_2   = 0.1$. Then, for any $\epsilon$, we have
\begin{align*}
	R\left(\pib\right)  &= \dfrac{\lambda^2}{1-\lambda}, \\
	R\left(\pia\right)  &= r \lambda + \dfrac{\lambda^2}{1-\lambda}, \\
    R\left(\pieff(\pi_{\sf alg},\theta)\right) 
    & = \theta \cdot \left( r \lambda + \theta \cdot \frac{\lambda^2}{1-\lambda} + (1-\theta) \cdot (1+\epsilon) \frac{\lambda^2}{1-\lambda} \right) + (1-\theta) \cdot \left( 0 + \theta \cdot (1+\epsilon) \frac{\lambda^2}{1-\lambda} + (1-\theta) \cdot \frac{\lambda^2}{1-\lambda} \right) \\
    &= \theta \lambda r + \theta R(\pi_{\sf base}) + \theta (1-\theta) \epsilon R(\pi_{\sf base}) + (1 - \theta) R(\pi_{\sf base}) + \theta (1-\theta) \epsilon R(\pi_{\sf base}) \\
    &= \left[ \lambda r + R(\pi_{\sf base}) \right] + (\theta-1) \lambda r + 2 \theta (1-\theta) \epsilon R(\pi_{\sf base}) \\
     &= R(\pia) +[ R(\pia) - R(\pib) ] (\theta-1) + 2 \epsilon R(\pib) \theta (1-\theta).
\end{align*}
In other words, $R\left(\pieff(\pi_{\sf alg},\theta)\right)$ is a second-order polynomial in $\theta$ that equals $R(\pia)$ (resp. $R(\pib)$) for $\theta = 1$ (resp. $\theta=0$). Since we would like  to compare $R\left(\pieff(\pi_{\sf alg},\theta)\right)$ with both $R(\pia)$ and $R(\pib)$, we provide two convenient reformulations:
\begin{align*}
R\left(\pieff(\pi_{\sf alg},\theta)\right) 
   	& = R(\pib) +   \theta \left( [ R(\pia) - R(\pib) ] + 2 \epsilon R(\pib) (1-\theta) \right) \\
	& = R(\pia) + (1-\theta) \left( 2 \epsilon R(\pi_{\sf base})  \theta - [ R(\pia) - R(\pib) ] ) \right),
\end{align*}
and define $\tilde{\theta} - 1:= \dfrac{R(\pia) - R(\pib)}{2 \epsilon R(\pib)} = \dfrac{\lambda r}{2 \epsilon R(\pib)}$. 

\paragraph{Case 1: partial adherence hurts ($\epsilon = -1$).} When $\epsilon = -1 < 0$, the reward of State 5 is strictly less than 1 so $\pia$ is optimal and any deviation from $\pia$ can only deteriorate performance. 
In this case, we have 
\begin{align*}
R\left(\pieff(\pi_{\sf alg},\theta)\right) 
   	& = R(\pib) +   2 \epsilon R(\pib) \theta \left(\Tilde{\theta} - 1+  (1-\theta) \right) = R(\pib) +  2 R(\pib) \theta \left( \theta - \Tilde{\theta}  \right),
\end{align*}
as announced in Section \ref{sec:bad-mdp-instance}.

Furthermore, for any value of $\theta \in [0,1]$, we can find an optimal recommendation policy $\pi\opt_{\sf alg}(\theta)$ via backward induction. 
Let us write $v_{i}$ for the value derived by the DM starting from state $i \in \{1,2,3,4,5\}$. Clearly, $v_{5}=0,v_{4} = 1/(1-\lambda)$. 
In State $3$, $v_{5} < v_{4}$ so the best action is to choose to go to State $4$. Since $\pi_{\sf base}$ also chooses to go to State $4$, we obtain $v_{3} = \lambda/(1-\lambda)$. 
In State $2$, the best action is to choose to State $4$. Since we follow the recommendation policy with probability $\theta$ and the baseline policy with probability $1-\theta$, we have $v_{2} = 0.1 + \theta \lambda \cdot v_{4}+ (1-\theta) \lambda \cdot v_{5} = 0.1 + \theta \frac{\lambda}{1-\lambda}.$
Finally, for the best action in State $1$, we have to choose between going to State $2$ and going to State $3$. If we recommend going to state $i \in \{2,3\}$, then the value derived by the DM from state 1 will be $v_{1} = \theta \lambda \cdot v_{i} + (1-\theta)\lambda v_{3}.$
Hence, the optimal recommendation depends on the comparison between $v_2$ and $v_3$.
If $v_{2} > v_{3}$, we should recommend to go to State $2$ from State $1$ and $v_{1} = \theta \lambda \cdot v_{2} + (1-\theta)\lambda v_{3}$. Otherwise, we recommend State $3$ and $v_1 = v_3$. 
Observe that $v_{2} \geq v_{3} \iff 0.1 + \theta \frac{\lambda}{1-\lambda} \geq \frac{\lambda}{1-\lambda} \iff \theta \geq 1 -  0.1 \cdot \frac{1-\lambda}{\lambda} =: \bar{\theta}.$
All in all, we have the following two cases, 
\begin{itemize}
\item If $\theta \leq \bar{\theta}$, $v_2 \leq v_3$ and an optimal recommendation policy $\pi_{\sf alg}^\star(\theta)$ should recommend the following transitions: $\pi\opt_{\sf alg}(\theta)$ is $1 \rightarrow 3$, $3 \rightarrow 4$. Observe that, as long as $\pi_{\sf alg}^\star(\theta)$ recommends $1 \rightarrow 3$, $\pieff(\pi_{\sf alg}^\star(\theta),\theta)$ will never visit State 2. As a result, $\pi_{\sf base}$ is also an optimal recommendation policy in this case, despite the fact that it prescribes a sub-optimal action at State 2. 
\item If $\theta \geq \bar{\theta}$, $v_2 \geq v_3$ and an optimal recommendation policy $\pi_{\sf alg}^\star(\theta)$ should recommend the following transitions: $1 \rightarrow 2$, $2 \rightarrow 4$, $3 \rightarrow 4$. Unlike in the previous case, $\pieff(\pi_{\sf alg}^\star(\theta),\theta)$ will visit State 3 even if  $\pi_{\sf alg}^\star(\theta)$ does not recommend $1 \rightarrow 3$. Consequently, an optimal recommendation recommends State 4 when at State 3. 
\end{itemize}

\paragraph{Case 2: partial adherence helps ($\epsilon = 1$).} When $\epsilon = 1$, the human DM is currently taking the optimal decision when visiting State 2 (but never visits State 2 in the first place), while the algorithm plays optimally in State 3 but never visits it. In this case, a mixture of $\pia$ and $\pib$ can lead to greater performance than any of the two policy alone. Formally, in this case, 
\begin{align*}
R\left(\pieff(\pi_{\sf alg},\theta)\right) 
	& = R(\pia) + 2 R(\pi_{\sf base})  (1-\theta) \left(  \theta - [\Tilde{\theta} - 1] ) \right),
\end{align*}
which is increasing over the interval $[\Tilde{\theta} - 1, 1]$.

}

\tbl{
\section{Proof of Proposition \ref{prop:pi alg star stationary deterministic}}\label{app:proof prop pi alg star stat deter}
\proof{Proof of Proposition \ref{prop:pi alg star stationary deterministic}.}
Our proof is very similar to the proof of Theorem \ref{th:model.B.infinity} and we only give a sketch here. The gist of the proof is to show that \ref{eq:definition-ada-mdp} admits a Bellman equation, i.e., to show that the stationary deterministic policy $\pi^{\infty}$ as defined in Equation \eqref{eq:v-infinity} is an optimal recommendation policy. We first note that the vector $\bm{v}^{\infty} \in \R^{\X}$, defined as the unique solution to Equation \eqref{eq:v-infinity}, does exist since the following map is a contraction for $\| \cdot \|_{\infty}$:
\[ \bm{v} \mapsto \left( \max_{\bm{\pi}_{s} \in \Delta(\A)}  \theta \cdot  \sum_{a \in \A} \pi_{sa} \bm{P}_{sa}^{\top} \left( \bm{r}_{sa} + \lambda \bm{v} \right) + (1-\theta) \cdot \sum_{a \in \A} \pi_{{\sf base},sa} \bm{P}_{sa}^{\top} \left( \bm{r}_{sa} + \lambda \bm{v} \right) \right)_{s \in \X}.\]
This is a straightforward consequence of $\bm{\pi}_{s} \in \Delta(\A), \bm{\pi}_{{\sf base},s} \in \Delta(\A)$ and $\bm{P}_{sa} \in \Delta(\X)$ for each pair $(s,a) \in \X \times \A$.

We can then show that $\pi^{\infty}$ is $\epsilon$-optimal in \ref{eq:definition-ada-mdp}, for any value of $\epsilon>0$. This is done in the same way as for our proof of Theorem \ref{th:model.B.infinity}, where we build a time $T$ such that the instantaneous rewards obtained after more than $T+1$ periods only accounts for at most $\epsilon$ in the (untruncated) return $R(\cdot)$. We then use a truncated return $R_{T}\left(\cdot\right)$ with continuation value $\bm{v}^{\infty} \in \R^{\X}$ for $\bm{v}^{\infty}$ as defined in Equation \eqref{eq:v-infinity}. The policy $\pi^{\infty}$ is an optimal recommendation policy for $R_{T}\left(\cdot\right)$, so that it is $\epsilon$-optimal for the untruncated return $R(\cdot)$. Since this holds for all $\epsilon>0$, we conclude that $\pi^{\infty}$ as defined in Proposition \ref{prop:solving-adaMDP} is an optimal recommendation policy. Since $\pi^{\infty}$ is stationary and deterministic, this concludes the proof of Proposition \ref{prop:pi alg star stationary deterministic}.
\hfill \halmos
\endproof

}

\section{Proof of Proposition \ref{prop:piece-constant-policy}}
\label{app:prop-piece-constant-policy}
\proof{Proof.}
\begin{enumerate}
    \item The proof uses similar ideas as the proof of {\em Blackwell optimality} for classical MDPs~\citep{feinberg2012handbook} and robust MDPs~\citep{goyal2023robust}, which studies the sensitivity of optimal policies as regards the values of the discount factor $\lambda \in [0,1)$. In particular, we recall the following lemma from \citet{puterman2014markov}.
\begin{lemma}[Lemma 10.1.2, \cite{puterman2014markov}]\label{lem:rational-function} Let $\phi:\I \rightarrow \R$ be a rational function on an interval $\I \subset \R$, that is,  $\phi$ is the ratio of two polynomials and the denominator does not have any zeros in the interval $\I$. Then either $\phi(\theta)=0$ for all $\theta \in \I$, either $\phi(\theta)=0$ for finitely many values of $\theta \in \I$.
\end{lemma}   
We now proceed with the proof of this statement.
From Proposition \ref{prop:pi alg star stationary deterministic}, for any value of $\theta \in [0,1]$, an optimal recommendation policy $\pi_{\sf alg}\opt(\theta)$ can be chosen stationary and deterministic. Since there are finitely many deterministic policies, the map $\theta \mapsto \pi_{\sf alg}\opt(\theta), [0,1] \rightarrow \Pi$ takes a finite number of values. This shows that at least one deterministic policy is visited infinitely often as $\theta \rightarrow 1.$ In particular, let $\left(\theta_{n}\right)_{n \geq 1} \in [0,1]^{\N}$ such that $\theta_{n} \rightarrow 1$ and the same deterministic recommendation policy $\hat{\pi}_{\sf alg}$  is optimal for this sequence of adherence levels: $R(\pieff(\hat{\pi}_{\sf alg},\theta_{n})) \geq R(\pieff(\pi,\theta_{n})), \forall \; \pi \in \Pi, \forall \; n \in \N.$
In particular, for each deterministic recommendation policy $\pi$, let us write $\phi_{\pi}:[0,1] \rightarrow \R$ for the map $\phi_{\pi}(\theta) = R(\pieff(\hat{\pi}_{\sf alg},\theta))  - R(\pieff(\pi,\theta)).$
We know that $
\phi_{\pi}(\theta_{n}) \geq 0, \forall \; n \geq 1.$ We want to show that this inequality $\phi_{\pi}(\theta) \geq 0$ holds for all values of $\theta$ sufficiently close to $1$.
From Lemma 10.1.3 in \cite{puterman2014markov}, we know that $\phi_{\pi}$ is a rational function. From Lemma \ref{lem:rational-function}, we know that either $\phi_{\pi}$ is identically equal to $0$, or it is equal to $0$ for finitely many values of $\theta$ in the interval $[0,1]$. If $\phi_{\pi}$ is identically equal to $0$, then indeed $\phi_{\pi}(\theta) \geq 0$ holds for all $\theta \in [0,1]$. Otherwise, 
$\phi_{\pi}$ is equal to $0$ only for a finite number of values in $[0,1]$, which 
implies that $\phi_{\pi}$ can change sign only finitely many times. 
Since $\phi_{\pi}(\theta_{n}) \geq 0, \forall \; n \geq 1,$ for $\theta_{n} \rightarrow 1$, there exists a threshold $\theta_{\pi} \in [0,1]$, such that $\phi_{\pi}(\theta) \geq 0$, for all $\theta \in [\theta_{\pi},1]$. If we take $\hat{\theta} \geq  \theta_{\pi}$ for  any deterministic policy $\pi$, we find that for all $\theta \in [\hat{\theta},1]$, we have $R(\pieff(\hat{\pi}_{\sf alg},\theta))  \geq R(\pieff(\pi,\theta)), \forall \; \pi \in \Pi.$
This concludes the proof.
\item We can extend the proof of the previous statement to any $\theta \in (0,1)$. In particular, for any $\theta \in (0,1)$, there exists an open interval $\I_{\theta} \subset (0,1)$ containing $\theta$ such that the optimal recommendation policy is constant on $\I_{\theta}^{+}=\I_{\theta} \bigcap [\theta,1]$ and constant on $\I_{\theta}^{-}=\I_{\theta} \bigcap [0,\theta]$. If the optimal recommendation policies on $\I_{\theta}^{-}$ and $\I_{\theta}^{+}$ are different, then they are still both optimal at $\theta$.
We can construct two additional intervals, $\I_{0}=[0,\theta')$ and $\I_{1} = (\theta'',1]$, on which the optimal recommendation policy is constant. Note that $\I_{0},\I_{1}$ are open sets for the subspace topology of $[0,1]$ induced by the classical topology of $\R$.
Overall, we obtain a covering of the compact set $[0,1]$ with open sets $\{ \I_{\theta} \; | \; \theta \in [0,1]\}$: $\bigcup_{\theta \in [0,1]} \I_{\theta} = [0,1]$. From the Heine-Lebesgue covering theorem, there exists a finite number of adherence levels $\theta_{1}',...,\theta_{m}'\in [0,1]$ such that $\bigcup_{i=1}^{m} \I_{\theta'_{i}} = [0,1]$. Finally, from the set $\{ \theta_{i}' \; | \; i=1,...,m\}$, for some $p \in \N$ we can construct $\theta_{1}=0 < \theta_{2} < ... < \theta_{p} =1$ such that the optimal recommendation policy is constant on each of the interval $[\theta_{i},\theta_{i+1}]$ for each $i \in \{1,...,p\}$, with multiple optimal policies at the breakpoints $\theta_{i}$ for $i \in \{1,...,p-1\}$.
\item We show this statement for $\underline{\theta}=1$. Let $\theta \in [0,1]$ and assume that $\pi_{\sf base} = \pi_{\sf alg}\opt(1)$. Note that \[\pieff(\pi_{\sf alg}\opt(1),\theta) = \theta \pi_{\sf alg}\opt(1) + (1-\theta) \pi_{\sf base} = \theta \pi_{\sf alg}\opt(1) +  (1-\theta) \pi_{\sf alg}\opt(1) = \pi_{\sf alg}\opt(1).\]
By definition, $R(\pieff(\pi_{\sf alg}(\theta),\theta)) \geq R(\pieff(\pi_{\sf alg},\theta)), \forall \; \pi_{\sf alg} \in \Pi$. But we have shown in Proposition \ref{prop:monotonicity} that $R(\pieff\opt(\theta)) \leq R(\pieff\opt(1))$ and $R(\pieff\opt(1)) = R(\pieff(\pi_{\sf alg}\opt(1),\theta))$. We conclude that  $\pieff\opt(\theta) = \pieff\opt(1)$ and that $\pi_{\sf alg}\opt(\theta)=\pi_{\sf alg}(1)$. The proof of this statement for $\underline{\theta} <1$ is similar and we omit it for conciseness.
\end{enumerate}
\hfill \halmos
\endproof

\tb{
\section{Proof of Proposition \ref{prop:value-similar}}\label{app:proof prop value similar}

\proof{Proof of Proposition \ref{prop:value-similar}.}

Let $\bs \in \X$ such that $v^{\pib}_{\bs} = v^{\pia\opt(1)}_{\bs}$.  We first prove that $v^{\pieff\opt(\theta)}_{\bs} = v^{\pib}_{\bs}$ for any $\theta \in [0,1]$. We can adapt the proof of Proposition \ref{prop:monotonicity} to show that for any $s \in \X$, the map $\theta \mapsto v^{\pia\opt(\theta)}_{s}$ is non-decreasing. Since we can choose $\pia\opt(0) = \pib$, this shows that
 \begin{equation}\label{eq:maximum-principle}
v^{\pib}_{s} \leq v^{\pieff\opt(\theta)}_{s} \leq  v^{\pia\opt(1)}_{s}, \forall \; s \in \X, \forall \; \theta \in [0,1].
 \end{equation}
Combining \eqref{eq:maximum-principle} with $v^{\pib}_{\bs} = v^{\pia\opt(1)}_{\bs}$, we obtain that $v^{\pieff\opt(\theta)}_{\bs} = v^{\pib}_{\bs}$ for any $\theta \in [0,1]$. Now let $\theta \in [0,1]$. We show that we can choose $\pia\opt(\theta)_{\bs} = \pi_{{\sf base},\bs}$. Recall that there exists a unique vector $\bm{v}^{\infty}_{\theta}$ such that 
\begin{equation}\label{eq:v-infinity-redefinition}
v^{\infty}_{\theta,\bar{s}} = \max_{\bm{\pi}_{\bs} \in \Delta(\A)}  \theta \cdot  \sum_{a \in \A} \pi_{\bs a} \bm{P}_{\bs a}^{\top} \left( \bm{r}_{\bs a} + \lambda \bm{v}^{\infty}_{\theta} \right) + (1-\theta) \cdot \sum_{a \in \A} \pi_{{\sf base},\bs a} \bm{P}_{\bs a}^{\top} \left( \bm{r}_{\bs a} + \lambda \bm{v}^{\infty}_{\theta} \right)
\end{equation}
and $\bm{v}^{\infty}_{\theta} = \bm{v}^{\pieff\opt(\theta)}$.
To show $\pia\opt(\theta)_{\bs} = \pi_{{\sf base},\bs}$, we need to show that 
\begin{equation}\label{eq:cns-optimality-pib-bs}
 \pi_{{\sf base},\bs} \in \arg \max_{\bm{\pi}_{\bs} \in \Delta(\A)}  \theta \cdot  \sum_{a \in \A} \pi_{\bs a} \bm{P}_{\bs a}^{\top} \left( \bm{r}_{\bs a} + \lambda \bm{v}^{\infty}_{\theta} \right) + (1-\theta) \cdot \sum_{a \in \A} \pi_{{\sf base},\bs a} \bm{P}_{\bs a}^{\top} \left( \bm{r}_{\bs a} + \lambda \bm{v}^{\infty}_{\theta} \right).
\end{equation} 
Since $v^{\infty}_{\theta,s} \geq v^{\pib}_{s}, \forall \; s \in \X$, we have 
\begin{equation}\label{eq:intermed-proof-0}
(1-\theta) \cdot \sum_{a \in \A} \pi_{{\sf base},\bs a} \bm{P}_{\bs a}^{\top} \left( \bm{r}_{\bs a} + \lambda \bm{v}^{\infty}_{\theta} \right) \geq (1-\theta) \cdot \sum_{a \in \A} \pi_{{\sf base},\bs a} \bm{P}_{\bs a}^{\top} \left( \bm{r}_{\bs a} + \lambda \bm{v}^{\pib}\right) = (1-\theta) v^{\pib}_{\bs}
\end{equation}
where the equality follows from the fixed-point equation for the value function of a policy.
Similarly, we obtain
\[\max_{\bm{\pi}_{\bs} \in \Delta(\A)}  \theta \cdot  \sum_{a \in \A} \pi_{\bs a} \bm{P}_{\bs a}^{\top} \left( \bm{r}_{\bs a} + \lambda \bm{v}^{\infty}_{\theta} \right) \geq \theta \cdot  \sum_{a \in \A} \pi_{{\sf base},\bs a} \bm{P}_{\bs a}^{\top} \left( \bm{r}_{\bs a} + \lambda \bm{v}^{\infty}_{\theta} \right) \geq \theta v^{\pib}_{\bs}.\]
Overall, we obtain that 
\[ \theta \cdot  \sum_{a \in \A} \pi_{{\sf base}, \bs a} \bm{P}_{\bs a}^{\top} \left( \bm{r}_{\bs a} + \lambda \bm{v}^{\infty}_{\theta} \right) + (1-\theta) \cdot \sum_{a \in \A} \pi_{{\sf base},\bs a} \bm{P}_{\bs a}^{\top} \left( \bm{r}_{\bs a} + \lambda \bm{v}^{\infty}_{\theta} \right) \geq v^{\pib}_{\bs}.\]
But $v^{\pib}_{\bs} = v^{\infty}_{\bs}$, and $v^{\infty}_{\bs}$ satisfies Equation \eqref{eq:v-infinity-redefinition}. Therefore, \eqref{eq:cns-optimality-pib-bs} holds, and we can choose $\pia\opt(\theta)_{\bs} = \pi_{{\sf base},\bs}$.
\hfill \halmos
\endproof
}

\tb{
\section{Bounding the suboptimality of a recommendation policy}\label{app:suboptimality bound}
In this section we show the following proposition, which provides a bound between $\bm{v}^{\pieff(\pia^{\star}(\theta),\theta)}$, the optimal value functions at a given adherence level $\theta$, and $\bm{v}^{\pieff(\pia,\theta)}$, the value function of $\pia$.
\begin{proposition}\label{prop:bound-different-vs-2}
Let $\pia \in \Pi$ be a recommendation policy and $\theta \in [0,1]$. Then we have
\begin{equation}\label{eq:bound-different-vs-2}
\|\bm{v}^{\pieff(\pia^{\star}(\theta),\theta)} - \bm{v}^{\pieff(\pia,\theta)}\|_{\infty} \leq \frac{\theta}{1-\lambda} \cdot \max_{s \in \X} \| \bm{\pi}^{\star}_{{\sf alg}}(\theta)_{s} - \bm{\pi}_{{\sf alg},s} \|_{1} \cdot \| \bm{v}^{\pieff(\pia^{\star}(\theta),\theta)} \|_{\infty}.
\end{equation}
\end{proposition}
Proposition \ref{prop:bound-different-vs-2} bounds the difference between the value function of the optimal recommendation, $\pia\opt(\theta)$ and that of any policy $\pia$. 
We delay the proof of Proposition \ref{prop:bound-different-vs-2} below and we first analyze the bound. Several comments are in order:
\begin{itemize}
    \item Our bound is parametrized by the $\ell_{1}$-norm between the distribution induced by $\pia\opt(\theta)$ and $\pia$ at each state $s \in \X$; note that if both policies are deterministic, we always have $\| \bm{\pi}^{\star}_{{\sf alg}}(\theta)_{s} - \bm{\pi}_{{\sf alg},s} \|_{1} \in \{0,2\}$. This term also reflects the piecewise constant structure of optimal recommendation policies as regards the adherence level, see Proposition \ref{prop:piece-constant-policy}.
    \item Our bound reflects the fact that when $\theta=0$, we have $\pieff(\pia,\theta) = \pib$ for any $\pia$, so that in this case all recommendation policies have the same performances.
    \item The multiplicative factor in $1/(1-\lambda)$ is common for bounds on the difference between two value functions. However, the multiplicative factor $\| \bm{v}^{\pieff(\pia^{\star}(\theta),\theta)} \|_{\infty}$ may also be of order $1/(1-\lambda)$, in which case our bound \eqref{eq:bound-different-vs-2} is not tight.
\end{itemize}
\proof{Proof of Proposition \ref{prop:bound-different-vs-2}.}
Recall the notation $T^{\pi}(\bm{v}) \in \R^{\X}$, defined as $T^{\pi}_{s}(\bm{v}) = \sum_{a \in \A} \pi_{sa}  \bm{P}_{sa}^{\top} \left( \bm{r}_{sa} + \lambda \bm{v} \right), \forall \; s \in \X$. For the sake of clarity, in this proof we use the notation $\pieff(\pia^{\star}(\theta),\theta) = \pieff^{\infty}, \pia^{\star}(\theta) = \pia^{\infty},\pieff(\pia,\theta) = \pieff$. We want to obtain a bound on $\|\bm{v}^{\pieff(\pia^{\star}(\theta),\theta)} - \bm{v}^{\pieff(\pia,\theta)}\|_{\infty} = \|\bm{v}^{\pieff^{\infty}} - \bm{v}^{\pieff}\|_{\infty}$.
By definition, we have the following two fixed-point equations: for all $s \in \X$,
\[ v_{s}^{\pieff^{\infty}} = \theta T_{s}^{\pia^{\infty}}(\bm{v}^{\pieff^{\infty}}) + (1-\theta)T_{s}^{\pib}(\bm{v}^{\pieff^{\infty}}), v_{s}^{\pieff}  = \theta T_{s}^{\pia}(\bm{v}^{\pieff}) + (1-\theta)T_{s}^{\pib}(\bm{v}^{\pieff}). \] 
Therefore, for all $s \in \X$,
$v_{s}^{\pieff^{\infty}} - v_{s}^{\pieff}  = \theta \left( T_{s}^{\pia^{\infty}}(\bm{v}^{\pieff^{\infty}}) - T_{s}^{\pia}(\bm{v}^{\pieff})\right) + (1-\theta)\left( T_{s}^{\pib}(\bm{v}^{\pieff^{\infty}}) - T_{s}^{\pib}(\bm{v}^{\pieff}) \right)$.
Since $\bm{v} \mapsto T^{\pib}_{s}(\bm{v})$ is a contraction, we have, for all $s \in \X$,
\begin{equation}\label{eq:dif-v-ineq-1}
|T_{s}^{\pib}(\bm{v}^{\pieff^{\infty}}) - T_{s}^{\pib}(\bm{v}^{\pieff})| \leq \lambda \|\bm{v}^{\pieff^{\infty}} - \bm{v}^{\pieff} \|_{\infty}.
\end{equation}
We now consider the term $T_{s}^{\pia^{\infty}}(\bm{v}^{\pieff^{\infty}}) - T_{s}^{\pia}(\bm{v}^{\pieff})$. We have, for all $s \in \X$,
\begin{align*}
T_{s}^{\pia^{\infty}}(\bm{v}^{\pieff^{\infty}}) - T_{s}^{\pia}(\bm{v}^{\pieff}) & = T_{s}^{\pia^{\infty}}(\bm{v}^{\pieff^{\infty}}) - T_{s}^{\pia}(\bm{v}^{\pieff^{\infty}})  + T_{s}^{\pia}(\bm{v}^{\pieff^{\infty}}) - T_{s}^{\pia}(\bm{v}^{\pieff}).
\end{align*}
Note that $\bm{v} \mapsto T^{\pia}_{s}(\bm{v})$ is a contraction, therefore we have, for all $s \in \X$,
\begin{equation}\label{eq:dif-v-ineq-2}
|  T_{s}^{\pia}(\bm{v}^{\pieff^{\infty}}) - T_{s}^{\pia}(\bm{v}^{\pieff^{\infty}}) | \leq \lambda \| \bm{v}^{\pieff^{\infty}} - \bm{v}^{\pieff} \|_{\infty}.
\end{equation}
Combining \eqref{eq:dif-v-ineq-1} and \eqref{eq:dif-v-ineq-2}, we obtain that, for all $s \in \X$,
\begin{align*}
v_{s}^{\pieff^{\infty}} - v_{s}^{\pieff} \leq & \theta \cdot \lambda \cdot \| \bm{v}^{\pieff^{\infty}} - \bm{v}^{\pieff} \|_{\infty}  + \theta \cdot   \|T^{\pia^{\infty}}(\bm{v}^{\pieff^{\infty}}) - T^{\pia}(\bm{v}^{\pieff^{\infty}}) \|_{\infty}  + (1-\theta) \cdot \lambda \cdot \| \bm{v}^{\pieff^{\infty}} - \bm{v}^{\pieff} \|_{\infty}.
\end{align*}
This shows that
$\| \bm{v}^{\pieff^{\infty}} - \bm{v}^{\pieff} \|_{\infty} \leq \frac{\theta }{1-\lambda} \cdot \|T^{\pia^{\infty}}(\bm{v}^{\pieff^{\infty}}) - T^{\pia}(\bm{v}^{\pieff^{\infty}}) \|_{\infty}.$

There remains to bound $\|T^{\pia^{\infty}}(\bm{v}^{\pieff^{\infty}}) - T^{\pia}(\bm{v}^{\pieff^{\infty}}) \|_{\infty}$. We have, for all $s \in \X$,
\begin{align*}
T_{s}^{\pia^{\infty}}(\bm{v}^{\pieff^{\infty}}) - T_{s}^{\pia}(\bm{v}^{\pieff^{\infty}})  = \sum_{a \in \A} \left(\pi_{{\sf alg},sa}^{\infty} - \pi_{{\sf alg}, sa} \right)\bm{P}_{sa}^{\top}\left(\bm{r}_{sa} + \lambda \bm{v}^{\pieff^{\infty}}\right) \leq \| \bm{\pi}^{\infty}_{{\sf alg},s} - \bm{\pi}_{{\sf alg},s} \|_{1} \| \left( \bm{P}_{sa}^{\top}\left(\bm{r}_{sa} + \lambda \bm{v}^{\pieff^{\infty}}\right)\right)_{a \in \A}\|_{\infty}.
\end{align*}
Now note that from Corollary \ref{cor:equivalence Binfinity and effective policy}, we have $\| \left( \bm{P}_{sa}^{\top}\left(\bm{r}_{sa} + \lambda \bm{v}^{\pieff^{\infty}}\right)\right)_{a \in \A}\|_{\infty} \leq \| \bm{v}^{\pieff^{\infty}}\|_{\infty}$. 
This shows that $\|T^{\pia^{\infty}}(\bm{v}^{\pieff}) - T^{\pia}(\bm{v}^{\pieff}) \|_{\infty} \leq \left( \max_{s \in \X} \| \bm{\pi}^{\infty}_{{\sf alg},s} - \bm{\pi}_{{\sf alg},s} \|_{1} \right) \cdot \| \bm{v}^{\pieff^{\infty}}\|_{\infty}$. Combining this with \eqref{eq:dif-v-ineq-1} and \eqref{eq:dif-v-ineq-2} concludes the proof of Proposition \ref{prop:bound-different-vs-2}. 
\hfill \halmos
\endproof
}

\section{Proof of Theorem \ref{th:robust-to-nominal}}\label{app:proof-th-robust-theta}
\tbl{
\proof{Proof.}
We want to show that $\sup_{\pi_{\sf alg} \in \Pi_{\sf H}} \min_{\theta \in [ \underline{\theta},\bar{\theta}]} R(\pieff(\pi_{\sf alg},\theta)) = \max_{\pi_{\sf alg} \in \Pi} \min_{\theta \in [ \underline{\theta},\bar{\theta}]} R(\pieff(\pi_{\sf alg},\theta))$
and that $\left(\pi_{{\sf alg}}\opt(\underline{\theta}),\underline{\theta}\right)$ is an optimal solution to the optimization problem in the above equation.
Recall that in Theorem \ref{th:model.other-adversarial} in Appendix \ref{app:model equivalence} we have studied the {\bf Time- and State-invariant Adversarial} model, for which we have shown that $\sup_{\pi_{\sf alg} \in \Pi_{\sf H}} \min_{u \in [\theta,1]} R(\pieff(\pi_{\sf alg},u)) = \max_{\pi_{\sf alg} \in \Pi} \min_{u \in [\theta,1]} R(\pieff(\pi_{\sf alg},u))$ and that $\left(\pi_{{\sf alg}}\opt(\theta),\theta\right)$ is an optimal solution to the optimization problem above. 
Therefore, we can prove Theorem \ref{th:robust-to-nominal} by applying the exact same proof as for Theorem \ref{th:model.other-adversarial} but replacing the interval $[\theta,1]$ by the interval $[ \underline{\theta},\bar{\theta}]$. We omit the proof for the sake of conciseness.
}

\tbl{
\section{Proof of Theorem \ref{th:robust-ada-mdp=robust-mdp}}\label{app:proof-robust-ada=robust-mdp}
We start from the surrogate MDP $\M'$ defined in Section \ref{sec:algorithms}, where the transition probabilities $\bm{P}' \in \left(\Delta(\X)\right)^{\X \times \A}$ and the instantaneous rewards $\bm{r}' \in \R^{\X \times \A}$ are defined in \eqref{eq:surrogate-mdp}. In particular, recall that the Bellman equation of the surrogate MDP \eqref{eq:v-infinity-prime} is $v^{\infty}_{s} = \max_{\bm{\pi}_{s} \in \Delta(\A)}  \sum_{a \in \A} \pi_{sa}\left(r'_{sa} + \lambda \bm{P}'^{\top}_{sa} \bm{v}^{\infty}\right), \forall \; s \in \X.$
From this, we see that the optimization problem \eqref{eq:robust-ada-mdp} is an s-rectangular robust Markov decision process~\citep{iyengar2005robust,wiesemann2013robust}, where the set $\U$ of admissible pairs of instantaneous rewards and transition probabilities $(\bm{r}',\bm{P}')$ is described as 
\begin{equation*}
\begin{aligned}
\U = \{ (\bm{r}',\bm{P}') \; | \; & \bm{r}' \in \R^{\X \times \A}, \bm{P}' \in \left(\Delta(\X)\right)^{\X \times \A}, \pib \in \Gamma, \\
&  \bm{P}_{sa}' := \theta \cdot \bm{P}_{sa} + (1-\theta) \cdot \sum_{a' \in \A} \pi_{{\sf base},sa'}\bm{P}_{sa'}, \forall \; (s,a) \in \X \times \A \\
& r'_{sa}  :=  \theta \cdot \bm{P}_{sa}^{\top}\bm{r}_{sa} + (1-\theta) \cdot \sum_{a' \in \A} \pi_{{\sf base},sa'}\bm{P}^{\top}_{sa'}\bm{r}_{sa'}  , \forall \; (s,a) \in \X \times \A\}.
\end{aligned}
\end{equation*}
By construction, the set $\U$ is s-rectangular, i.e., it can be written $\U = \times_{s \in \X} \; \U_{s}$ with $\U_{s} \subset \Delta(\X)^{\A}$ convex for each $s \in \X$.
From \cite{wiesemann2013robust}, we conclude that an optimal policy for the decision problem \eqref{eq:robust-ada-mdp} can be chosen stationary. Additionally, an optimal policy can be computed efficiently when the sets $\Gamma_{s}$ are described by affine and conic constraints, see corollary 3 in \cite{wiesemann2013robust} for a more precise complexity statement.
}
\end{APPENDICES}

\end{document}